\documentclass[12pt,preprint]{emulateapj}

\begin{document}
\title{$UV$-continuum slopes at $z\sim4-7$ from the HUDF09+ERS+CANDELS
  observations: Discovery of a well-defined $UV$-color magnitude
  relationship for $z\geq4$ star-forming galaxies\altaffilmark{1}}
\author{R. J. Bouwens\altaffilmark{2,3},
  G. D. Illingworth\altaffilmark{3}, P.A. Oesch\altaffilmark{3,7},
  M. Franx\altaffilmark{2}, I. Labb{\'e}\altaffilmark{2},
  M. Trenti\altaffilmark{4}, P. van Dokkum\altaffilmark{5},
  C. M. Carollo\altaffilmark{6}, V. Gonz{\'a}lez\altaffilmark{3},
  R. Smit\altaffilmark{2},
  D. Magee\altaffilmark{3}}

\altaffiltext{1}{Based on observations
  made with the NASA/ESA Hubble Space Telescope, which is operated by
  the Association of Universities for Research in Astronomy, Inc.,
  under NASA contract NAS 5-26555. These observations are associated
  with programs \#11563, 9797.}
\altaffiltext{2}{Leiden Observatory, Leiden University, NL-2300 RA
  Leiden, Netherlands} 
\altaffiltext{3}{UCO/Lick
  Observatory, University of California, Santa Cruz, CA 95064}
\altaffiltext{4}{University of Colorado, Center for Astrophysics and
  Space Astronomy, 389-UCB, Boulder, CO 80309, USA}
\altaffiltext{5}{Department of Astronomy, Yale
  University, New Haven, CT 06520}
\altaffiltext{6}{Institute for Astronomy, ETH
  Zurich, 8092 Zurich, Switzerland}
\altaffiltext{7}{Hubble Fellow}
\begin{abstract}
Ultra-deep ACS and WFC3/IR HUDF+HUDF09 data, along with the wide-area
GOODS+ERS+CANDELS data over the CDF-S GOODS field, are used to measure
UV colors, expressed as the UV-continuum slope $\beta$, of
star-forming galaxies over a wide range in luminosity ($0.1L_{z=3} ^*$
to $2L_{z=3}^{*}$) at high redshift ($z\sim7$ to $z\sim4$).  $\beta$
is measured using all ACS and WFC3/IR passbands uncontaminated by
Ly$\alpha$ and spectral breaks.  Extensive tests show that our $\beta$
measurements are only subject to minimal biases.  Using a different
selection procedure, Dunlop et al. recently found large biases in
their $\beta$ measurements. To reconcile these different results, we
simulated both approaches and found that $\beta$ measurements for
faint sources are subject to large biases if the same passbands are
used both to select the sources and to measure $\beta$.  High-redshift
galaxies show a well-defined rest-frame UV color-magnitude (CM)
relationship that becomes systematically bluer towards fainter UV
luminosities.  No evolution is seen in the slope of the UV CM
relationship in the first 1.5 Gyr, though there is a small evolution
in the zero-point to redder colors from $z\sim7$ to $z\sim4$.  This
suggests that galaxies are evolving along a well-defined sequence in
the $L_{UV}$-color ($\beta$) plane (a ``star-forming sequence''?).
Dust appears to be the principal factor driving changes in the UV
color $\beta$ with luminosity.  These new larger $\beta$ samples lead
to improved dust extinction estimates at $z\sim4$-7 and confirm that
the extinction is essentially zero at low luminosities and high
redshifts. Inclusion of the new dust extinction results leads to (i)
excellent agreement between the SFR density at $z\sim4$-8 and that
inferred from the stellar mass density, and (ii) to higher SSFRs at
$z\gtrsim4$, suggesting the SSFR may evolve modestly (by factors of
$\sim$2) from $z\sim4$-7 to $z\sim2$.
\end{abstract}

\section{Introduction}

One of the biggest frontiers in extragalactic cosmology is to
characterize the early build-up and evolution of galaxies.  It is
important for our understanding of early gas accretion and star
formation in the universe, estimating the rate of early metal
injection into the IGM, and assessing the impact of galaxies on
reionization.  From the observations, we already have a good measure
of how fast galaxies build up through Lyman-Break selections and
luminosity function (LF) studies reaching all the way to $z\sim8$
(e.g., Bouwens et al.\ 2010b; McLure et al.\ 2010; Bunker et
al.\ 2010; Yan et al.\ 2010).  Galaxies show a remarkably uniform
brightening in their LFs from $z\sim8$ to $z\sim4$ (e.g., Bouwens et
al.\ 2011b) and plausibly from $z\sim10$ (Bouwens et al.\ 2011a).
There is even some evidence for very rapid evolution at $z\gtrsim8$
(Bouwens et al.\ 2011a; Oesch et al.\ 2012a).

Despite general constraints on how the galaxy population builds up
with cosmic time, much less is known about how individual galaxies
grow.  Qualitatively, we would expect galaxies to build up
monotonically in mass, metallicity, and dust content as they form
stars.  Quantifying how this build up occurs and with what star
formation history is very challenging however.  The effects of dust,
metal, and age on the colors are very similar and make it very
difficult to disentangle one factor from the others.  Nonetheless,
there is enough information available, i.e., the $UV$-to-optical
colors (Stark et al.\ 2009; Gonz{\'a}lez et al.\ 2010; Labb{\'e} et
al. 2010b), $UV$ colors (e.g., Bouwens et al.\ 2009; Hathi et
al.\ 2008), and high-resolution spectra of high redshift galaxies
(e.g., Stark et al.\ 2010; Vanzella et al.\ 2009), that significant
progress should be made in better characterizing galaxies throughout
the build-up process.

An important piece of the puzzle in deciphering how galaxies build up
is provided by the rest-frame $UV$ spectrum and in particular the $UV$
colors.  The rest-frame $UV$ color provides us with perhaps our best
means for estimating the dust extinction and star formation rate (SFR)
for faint $z>3$ galaxies, given that other techniques for probing the
SFR tend to only detect the most bolometrically luminous galaies
(e.g., Bouwens et al.\ 2009; Smit et al.\ 2012).  The $UV$ colors also
show a systematic dependence on the $UV$ luminosities of star-forming
galaxies (e.g., Bouwens et al.\ 2009; Bouwens et al.\ 2010a) and are
much more amenable to direct measurement than rest-frame UV-optical
colors where mid-IR (IRAC) photometry is necessary.

A significant amount of effort has gone into establishing the
$UV$-continuum slope distribution at high redshift $z>2$ and
determining its dependence upon redshift and luminosity.  The earliest
analyses were at $z\sim2$-3 using either ground-based observations
(Steidel et al.\ 1999; Adelberger \& Steidel 2000) or the deep WFPC2
observations over the HDF North (Meurer et al.\ 1999).  Subsequent
analyses pushed $UV$-continuum slope measurements to $z\sim4$-6 using
Subaru Suprime-Cam, HST ACS, or HST NICMOS observations (Ouchi et
al.\ 2004a; Papovich et al.\ 2004; Stanway et al.\ 2005; Bouwens et
al.\ 2006; Hathi et al.\ 2008).  Bouwens et al. (2009) extended these
previous works by examining the $UV$ slopes as a function of
luminosity over the entire redshift range $z\sim2$-6 -- establishing a
coherent framework for understanding the observational results at that
time.

The availability of both deep and wide WFC3/IR imaging has made it
possible to substantially improve these early measurements of the
$UV$-continuum slope.  These imaging observations allow for accurate
measurements of the $UV$-continuum slopes for large numbers of
$z\sim4$-7 galaxies.  Bouwens et al.\ (2010a) made use of the early
WFC3/IR observations to examine the $UV$-continuum slope distribution
out to $z\sim7$ (see also Oesch et al.\ 2010a; Bouwens et al.\ 2010b;
Bunker et al.\ 2010; Finkelstein et al.\ 2010; Robertson et al.\ 2010;
Dunlop et al.\ 2012; Wilkins et al.\ 2011).  Bouwens et al.\ (2010a)
found that very low luminosity $z\sim7$-8 galaxies in the ultra-deep
HUDF09 WFC3/IR field had $UV$-continuum slopes $\beta$ as steep as
$-2.5$ and plausibly consistent with $-3$ but with uncertainties of
$\sim$0.2-0.3.  Finkelstein et al.\ (2010) also reported on the colors
of ultra-faint $z\sim7$-8 galaxies in the HUDF finding similar steep
$\beta$ values, but with somewhat larger uncertainties.

Since the early WFC3/IR campaign over the HUDF (Bouwens et al.\ 2010b;
Oesch et al.\ 2010a) from the HUDF09 program (GO 11563: PI
Illingworth), the amount of deep, wide-area WFC3/IR observations over
well-known legacy fields has increased dramatically.  At present, we
have ultra-deep WFC3/IR observations over the two HUDF09/HUDF05 fields
(Bouwens et al.\ 2011b; Oesch et al.\ 2007, 2010a), wide-area
($\sim$145 arcmin$^2$) data over the CDF-South GOODS field as a result
of the Early Release Science (Windhorst et al.\ 2011) and CANDELS
(Grogin et al.\ 2011; Koekemoer et al.\ 2011) programs, and even
deeper WFC3/IR observations over the HUDF and HUDF05 fields (Bouwens
et al.\ 2011b).  These observations greatly improve the luminosity
baseline, redshift range, and precision with which we can define the
$UV$-continuum slope distribution for $z\sim4$-7 galaxies.

In this paper, we take advantage of these new observations to
establish the distribution of $UV$-continuum slopes $\beta$ over a
wide range in luminosity and redshift.  These new observations allow
us to determine with great precision how the $UV$-continuum slope
$\beta$ distribution depends upon luminosity, in four distinct
redshift intervals.  This new information puts us in a position to
look for a possible star-forming sequence of galaxies at high redshift
and to characterize its evolution with cosmic time.  The evolution of
such a sequence provides useful information for better understanding
early galaxy build-up.  For example, the slope of a $\beta$-luminosity
relationship constrains how the dust and age of the galaxy population
vary as a function of luminosity.  Scatter in the $\beta$-luminosity
relationship constrains the overall scatter in the stellar populations
or dust extinction of individual galaxies.  Finally, evolution in the
slope and offset of the relation with cosmic time gives us clues as to
possible changes in how galaxies evolve, either in age or dust
extinction.  Independent analyses of the $UV$-continuum slope $\beta$
in $z\sim5$-7 galaxies are provided by Dunlop et al.\ (2012) and
Wilkins et al.\ (2011), but both are based on a much smaller,
shallower set of observations.  A somewhat complementary analysis to
the one described here is given by Gonz{\'a}lez et al.\ (2012) who
quantify the changes in $UV$-optical colors as a function of
luminosity and redshift.

We provide a brief overview of the paper here.  We begin with a brief
summary of the observational data (\S2).  In \S3, we describe the
manner in which we construct our high-redshift samples from the
observational data, measure the $UV$-continuum slope $\beta$
distribution, correct for measurement and selection biases, and
present evidence for well-defined color-magnitude relation for
galaxies in the rest-fame $UV$.  In \S4, we compare the present
$UV$-continuum slope $\beta$ determinations with previous
determinations.  In \S5, we explore the implications of such a
color-magnitude relation for galaxy growth -- using the $UV$-continuum
slopes $\beta$ to infer a luminosity-dependent dust correction for
galaxies.  In \S6, we use these extinction estimates to rederive the
SFR density at $z\gtrsim4$ and then compare these results with what
one infers from the stellar mass density.  Finally, in \S7, we
conclude and provide a summary of our primary results.  The appendices
include a detailed description of many quantitative results and
simulations essential for accurate measurements of the $UV$-continuum
slopes.

Throughout this work, we find it convenient to quote results in
terms of the luminosity $L_{z=3}^{*}$ Steidel et al.\ (1999) derived
at $z\sim3$, i.e., $M_{1700,AB}=-21.07$, for consistency with previous
work -- though we note that the Steidel et al.\ (1999) LF results are
now updated ($M_{1700,AB}=-20.97\pm0.14$: Reddy \& Steidel 2009) but
still consistent with the previous determination.  We present our dust
extinction estimates as the ratio of the bolometric luminosity ($IR$ +
$UV$ luminosity: $L_{IR}$ + $L_{UV}$) to the $UV$ luminosity
($L_{UV}$), i.e., $L_{IR}/L_{UV}+1$.  We refer to the HST F435W,
F606W, F775W, F814W, F850LP, F098M, F105W, F125W, and F160W bands as
$B_{435}$, $V_{606}$, $i_{775}$, $I_{814}$, $z_{850}$, $Y_{098}$,
$Y_{105}$, $J_{125}$, and $H_{160}$, respectively.  Where necessary,
we assume $\Omega_0 = 0.3$, $\Omega_{\Lambda} = 0.7$, $H_0 =
70\,\textrm{km/s/Mpc}$.  We quote all star formation rates and stellar
masses assuming a Salpeter (1955) IMF.  All magnitudes are in the AB
system (Oke \& Gunn 1983).

\begin{deluxetable}{cccc}
\tablecolumns{4}
\tablecaption{A summary of the observational data used to establish the distribution
of $UV$-continuum slopes $\beta$ from $z\sim7$ to $z\sim4$ (see Figure~\ref{fig:obsdata} for the layout of these data within the CDF-South).\label{tab:obsdata}}
\tablehead{
\colhead{} & \colhead{Detection\tablenotemark{a}} & \colhead{PSF FWHM} & \colhead{Areal Coverage}\\
\colhead{Passband} & \colhead{Limits (5$\sigma$)} & \colhead{(arcsec)} & \colhead{(arcmin$^2$)}}
\startdata
\multicolumn{4}{c}{HUDF09 (WFC3/IR HUDF)} \\
$B_{435}$ & 29.7 & 0.09 & 4.7 \\
$V_{606}$ & 30.1 & 0.09 & 4.7 \\
$i_{775}$ & 29.9 & 0.09 & 4.7 \\
$z_{850}$ & 29.4 & 0.10 & 4.7 \\
$Y_{105}$ & 29.6 & 0.15 & 4.7 \\
$J_{125}$ & 29.9 & 0.16 & 4.7 \\
$H_{160}$ & 29.9 & 0.17 & 4.7 \\
\multicolumn{4}{c}{} \\
\multicolumn{4}{c}{HUDF09-1 (WFC3/IR P12)} \\
$V_{606}$ & 29.0 & 0.09 & 4.7 \\
$i_{775}$ & 29.0 & 0.09 & 4.7 \\
$z_{850}$ & 29.0 & 0.10 & 4.7 \\
$Y_{105}$ & 29.0 & 0.15 & 4.7 \\
$J_{125}$ & 29.3 & 0.16 & 4.7 \\
$H_{160}$ & 29.1 & 0.17 & 4.7 \\
\multicolumn{4}{c}{} \\
\multicolumn{4}{c}{HUDF09-2 (WFC3/IR P34)} \\
$B_{435}$\tablenotemark{b} & 28.8 & 0.09 & 3.3 \\
$V_{606}$\tablenotemark{b} & 29.9 & 0.09 & 4.7 \\
$i_{775}$\tablenotemark{b} & 29.3 & 0.09 & 4.7 \\
$I_{814}$\tablenotemark{b} & 29.0 & 0.09 & 3.3 \\
$z_{850}$\tablenotemark{b} & 29.2 & 0.10 & 4.7 \\
$Y_{105}$ & 29.2 & 0.15 & 4.7 \\
$J_{125}$ & 29.5 & 0.16 & 4.7 \\
$H_{160}$ & 29.3 & 0.17 & 4.7 \\
\multicolumn{4}{c}{} \\
\multicolumn{4}{c}{ERS (CDF-S GOODS)} \\
$B_{435}$ & 28.2 & 0.09 & 39 \\
$V_{606}$ & 28.5 & 0.09 & 39 \\
$i_{775}$ & 28.0 & 0.09 & 39 \\
$I_{814}$ & 28.0\tablenotemark{c} & 0.09 & 39 \\
$z_{850}$ & 28.0 & 0.10 & 39 \\
$Y_{098}$ & 27.9 & 0.15 & 39 \\
$J_{125}$ & 28.4 & 0.16 & 39 \\
$H_{160}$ & 28.1 & 0.17 & 39 \\
\multicolumn{4}{c}{} \\
\multicolumn{4}{c}{CDF-S CANDELS Deep} \\
$B_{435}$ & 28.2 & 0.09 & 66 \\
$V_{606}$ & 28.5 & 0.09 & 66 \\
$i_{775}$ & 28.0 & 0.09 & 66 \\
$I_{814}$ & 28.8 & 0.09 & 66 \\
$z_{850}$ & 28.0 & 0.10 & 66 \\
$Y_{105}$ & 28.5 & 0.16 & 66 \\
$J_{125}$ & 28.8 & 0.16 & 66 \\
$H_{160}$ & 28.5 & 0.17 & 66 \\
\multicolumn{4}{c}{} \\
\multicolumn{4}{c}{CDF-S CANDELS Wide} \\
$B_{435}$ & 28.2 & 0.09 & 40 \\
$V_{606}$ & 28.5 & 0.09 & 40 \\
$i_{775}$ & 28.0 & 0.09 & 40 \\
$I_{814}$ & 28.1\tablenotemark{c} & 0.09 & 40 \\
$z_{850}$ & 28.0 & 0.10 & 40 \\
$Y_{105}$ & 28.0 & 0.16 & 40 \\
$J_{125}$ & 28.0 & 0.16 & 40 \\
$H_{160}$ & 27.7 & 0.17 & 40 \\
\enddata
\tablenotetext{a}{$5\sigma$ detection limits for our $z\sim4$-7
  ACS+WFC3/IR selections were measured in a $0.35''$-diameter
  aperture.}
\tablenotetext{b}{Our reductions of the ACS data over the HUDF09-2
  field include both those observations taken as part of the HUDF05
  (82 orbits) and HUDF09 (111 orbits) programs (see Figure~1 from
  Bouwens et al.\ 2011b).  The latter observations add $\sim$0.15-0.4
  mag to the total optical depths.}
\tablenotetext{c}{The depth of the F814W observations vary
  considerably over the ERS and CDF-South CANDELS Wide field, from
  very deep coverage ($\sim$28.8 mag) in some regions to essentially
  no coverage over a small fraction of the area.  Typical depths are
  $\sim$28.0 AB mag (5$\sigma$).}
\end{deluxetable}

\begin{figure}
\epsscale{1.15}
\plotone{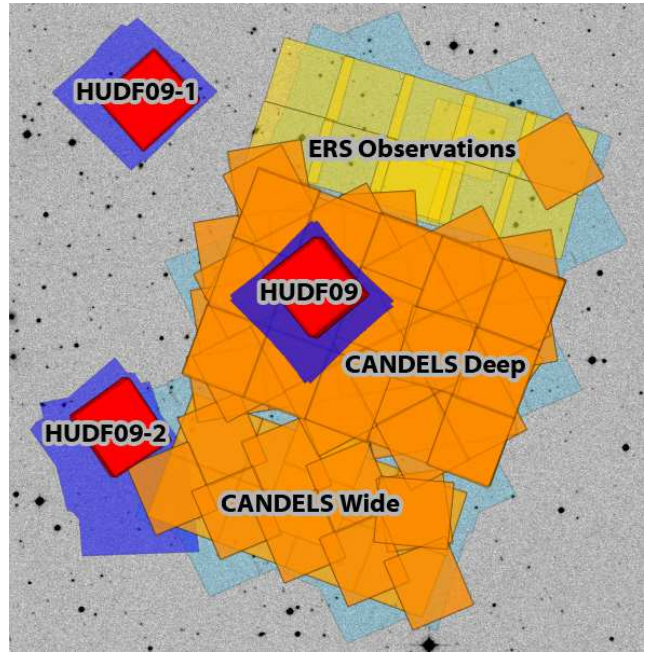}
\caption{Deep WFC3/IR data over the extended CDF-South GOODS field
  that can be used to be establish the $UV$-continuum slope $\beta$
  distribution for star-forming galaxies at $z\geq4$.  Ultra-deep
  WFC3/IR observations are available over the three HUDF09 fields
  HUDF09, HUDF09-1, and HUDF09-2 (red-shaded regions) while moderately
  deep WFC3/IR observations are available over the $\sim$145
  arcmin$^2$ ERS (\textit{yellow regions}) and CANDELS (\textit{orange
    regions}) areas.  The blue and dark blue regions show the position
  of the deep GOODS ACS and ultra-deep HUDF+HUDF05 ACS observations,
  respectively.  A convenient summary of the observational properties
  of each of these fields is provided in
  Table~\ref{tab:obsdata}.\label{fig:obsdata}}
\end{figure}

\section{Observational Data}

Here we utilize two primary data sets to examine the $UV$-continuum
slope $\beta$ distribution of galaxies from $z\sim7$ to $z\sim4$: (1)
the ultra-deep ACS+WFC3/IR observations over the three HUDF09 fields
(Bouwens et al.\ 2010b; Bouwens et al.\ 2011b; Oesch et al.\ 2010a)
and (2) the wide-area ACS+WFC3/IR observations taken over the
CDF-South GOODS field as a result of Early Release Science and CANDELS
programs (Windhorst et al.\ 2011; Giavalisco et al.\ 2004; Grogin et
al.\ 2011; Koekemoer et al.\ 2011).  A brief summary of the properties
of these observations is provided in Table~\ref{tab:obsdata}.
Figure~\ref{fig:obsdata} shows the layout of these observations over
the CDF-South GOODS area.

\subsection{HUDF09 Observations}

The first set of observations we utilize is from the HUDF09 program
(GO11563: PI Illingworth) and involves ultra-deep WFC3/IR observations
over three fields that already have ultra-deep ACS observations.
These fields include the HUDF (Beckwith et al.\ 2006) and two HUDF05
flanking fields (Oesch et al.\ 2007).  These observations are
primarily of use in establishing the distribution of $UV$-continuum
slopes $\beta$ for very faint $z\sim4$-7 galaxies.

At present, the full two years of WFC3/IR observations (all 192 orbits
from the GO11563 program) have been obtained over the three HUDF09
fields.  The WFC3/IR observations over the HUDF include some 111
orbits of observations -- while the HUDF09-1 and HUDF09-2 fields
include 33 orbits and 48 orbits of observations, respectively.  Our
reductions of the WFC3/IR observations over these fields are already
described in Bouwens et al.\ (2011b) and are conducted using standard
procedures.  Particularly important to this process was ensuring that
the geometric distortion and velocity abberation were treated properly
so that the registration between bands and with the ACS observations
was very accurate ($\lesssim$0.01$''$).  Accurate registration is
absolutely essential not only for maximizing the accuracy of our color
measurements, but also for minimizing the effect that misregistration
might have on scatter in these same color measurements.  The depths of
our WFC3/IR observations reach to $\gtrsim$29 AB mag at $5\sigma$ and
are presented in detail in Table~\ref{tab:obsdata}.  The FWHM of the
PSF in the WFC3/IR observations is $\sim$0.16$''$.

Our reductions of the ACS observations over the HUDF09-2 area include
all available data -- including the 82 orbits that were taken during
the execution of the original HUDF05 program and an additional 111
orbits that were taken in parallel with the HUDF09 WFC3/IR
observations over the HUDF.  The total number of orbits per filter
over the HUDF09-2 are 10 orbits F435W, 32 orbits F606W, 46 orbits
F775W, 16 orbits F814W, and 89 orbits F850LP.  The total integration
time is $\gtrsim$50\% of that available for the HUDF, allowing us to
reach within $\lesssim$0.4 mag of the HUDF.  This provides us with
sufficiently high S/N levels required to keep contamination in
$z\gtrsim6$ selections to a minimum.  A more detailed description of
this data set is provided by Bouwens et al.\ (2011b).

\subsection{ERS/CANDELS Observations}

Wide-area WFC3/IR observations allow us to establish the
$UV$-continuum slope $\beta$ distribution for the rarer, higher
luminosity sources.  These observations are available over the
CDF-South GOODS field from the Early Release Science (Windhorst et
al.\ 2011) and CANDELS (Grogin et al.\ 2011; Koekemoer et al.\ 2011)
programs.

The WFC3/IR observations from the ERS program are distributed over the
northern part of the CDF South GOODS field (Windhorst et al.\ 2011:
Figure~\ref{fig:obsdata}).  These observations consist of 60 orbits of
F098M, F125W, and F160W observations distributed over 10 distinct
$2.1'\times2.3'$ pointings, with 2 orbits F098M, 2 orbits F125W, and 2
orbits F160W per pointing.  The WFC3/IR observations extend over
$\sim$43 arcmin$^2$ in total, although only $\sim$39 arcmin$^2$ of
those observations overlap with the ACS GOODS data and can be used
here.

WFC3/IR observations from the CANDELS program are distributed over the
southern two-thirds of the CDF-South GOODS field (106 arcmin$^2$),
with the deepest section planned for the central $\sim$66 arcmin$^2$
portion.  3 orbits of F105W, 4 orbits of F125W, and 4 orbits of F160W
observations are available over the central sections of the CDF-South
(representing all 10 SNe search epochs to February 18, 2012) while
only 1 orbit of F105W, F125W and F160W observations are available over
the southern portion.  The layout of the CANDELS and ERS observations
with the CDF-South GOODS field is shown in Figure~\ref{fig:obsdata}.

Our reductions of the WFC3/IR observations over the CDF-South are
performed in exactly the same manner as for the HUDF09 observations.
See Bouwens et al.\ (2011b) for a detailed description.  As in our
HUDF09 reductions, a $0.06''$ pixel size was used.  For the ACS
observations over the CDF-South, we used the Bouwens et al.\ (2007)
reductions for the F435W, F606W, F775W, and F850LP bands.  These
reductions are comparable to the GOODS v2.0 reduction (Giavalisco et
al.\ 2004) but take advantage of the substantial SNe follow-up
observations over the CDF-South (Riess et al.\ 2007) which add
$\sim$0.1-0.3 mag of depth to the $z_{850}$ band.

We also reduced the new ACS $I_{814}$-band observations over the
CDF-South GOODS fields from the CANDELS+ERS programs to take advantage
of the superb depth available with these data at $\sim$8000\AA$\,$
(typically 13 orbits over the CANDELS-DEEP region, or 0.8 mags deeper
than the GOODS $i_{775}$ band exposures).  The F814W observations are
valuable for controlling for contamination in our $z\sim7$ selections
and for minimizing the uncertainties in our $\beta$ determinations at
$z\sim4$.  We included all F814W observations from the ERS+CANDELS
programs over the CDF-South.  After using public codes (e.g., Anderson
\& Bedin 2010) to correct the raw frames for charge transfer
efficiency defects and row-by-row banding artifacts, we performed the
alignment, cosmic-ray rejection, and drizzling with the ACS GTO apsis
pipeline (Blakeslee et al.\ 2003).

\begin{deluxetable}{cccc}
\tablecaption{Lyman-break samples used to measure the distribution of 
$UV$-continuum slopes $\beta$ as a function of redshift and $UV$ 
luminosity.\label{tab:lumrange}}
\tablehead{
\colhead{} &  \colhead{} & \colhead{Luminosity} & \colhead{\# of} \\
\colhead{Sample\tablenotemark{a}} & \colhead{Field} & \colhead{Range\tablenotemark{b}} & \colhead{Sources}}
\startdata
$z\sim4$ & HUDF09 & $-23<M_{UV,AB}<-16$ & 308 \\
         & ERS/CANDELS & $-23<M_{UV,AB}<-18$ & 1524\\
\\
$z\sim5$ & HUDF09 & $-23<M_{UV,AB}<-17$ & 137 \\
         & ERS/CANDELS & $-23<M_{UV,AB}<-18.5$ & 277 \\
\\
$z\sim6$ & HUDF09 & $-23<M_{UV,AB}<-17$ & 70 \\
         & ERS/CANDELS & $-23<M_{UV,AB}<-19$ & 101 \\
\\
$z\sim7$ & HUDF09 & $-22<M_{UV,AB}<-17$ & 57\tablenotemark{c,d} \\
         & ERS/CANDELS & $-22<M_{UV,AB}<-19$ & 44
\enddata
\tablenotetext{a}{The mean redshift we estimate for these samples is
  3.8, 4.9, 5.9, and 7.0, respectively (Figure~\ref{fig:zdist}).}
\tablenotetext{b}{The faint magnitude limit is set by the 5\,$\sigma$
  depth of the WFC3/IR near-IR observations over our search fields.
  See \S3.2.}
\tablenotetext{c}{These samples include sources from the Bouwens et
  al.\ (2011b) HUDF09 $z\sim7$ samples and the lowest redshift
  galaxies in the Bouwens et al.\ (2011b) HUDF09 $z\sim8$ selection.}
\tablenotetext{d}{To keep contamination in our faint $z\sim7$
  selection to a minimum, we only include faint $J_{125,AB}>28$ sources
  from the HUDF09 fields with the deepest optical data.  We therefore
  only consider sources from the HUDF (HUDF09) and the $\sim$3 arcmin$^2$
  ($\sim$70\%) area in the HUDF09-2 field with deep optical coverage
  from the HUDF09 program (see Figure 1 from Bouwens et al.\ 2011b).}
\end{deluxetable}

\begin{figure*}
\epsscale{1.15}
\plotone{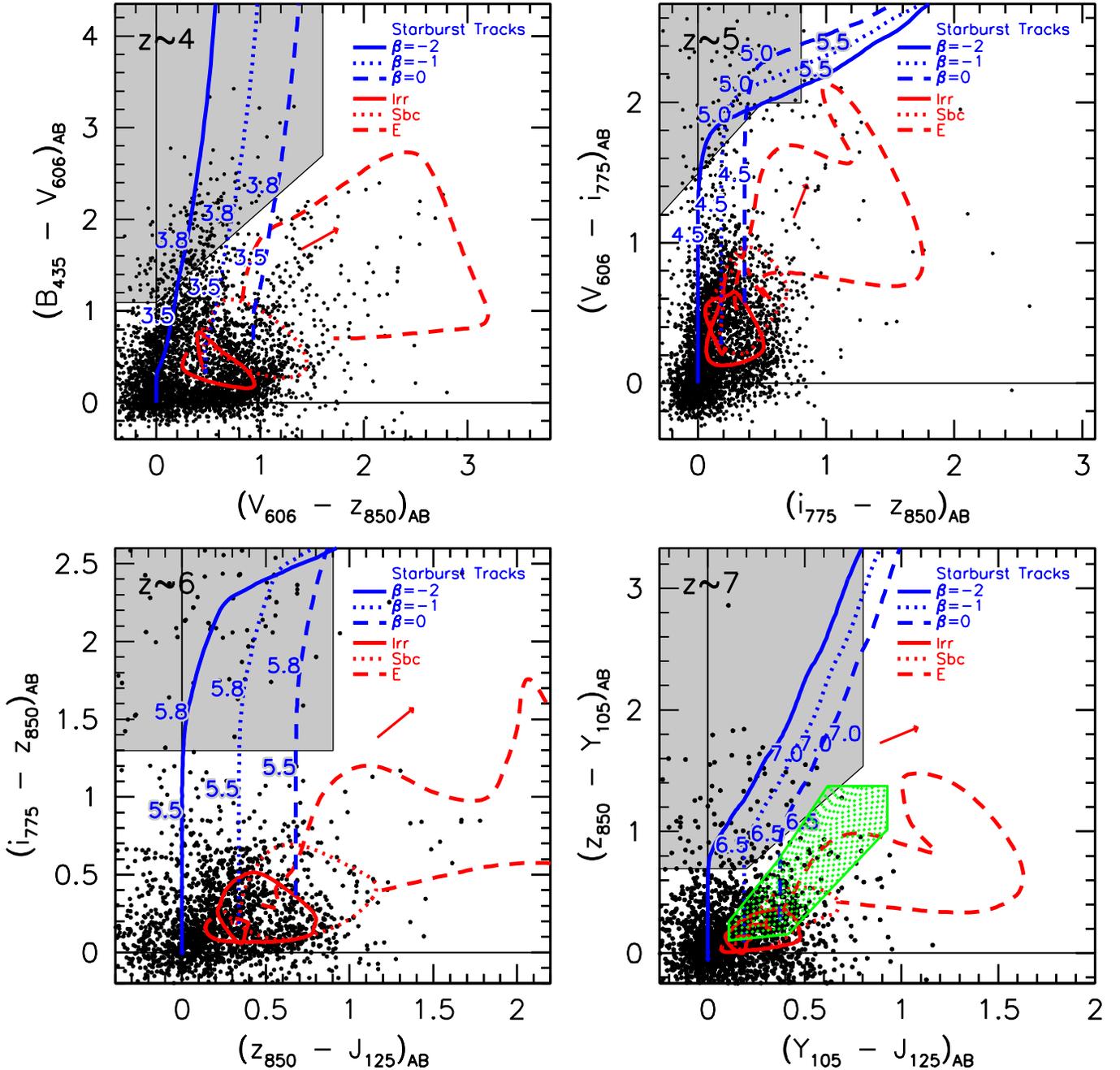}
\caption{The two-color Lyman-Break selection criteria used to select
  our $z\sim4$, $z\sim5$, $z\sim6$, and $z\sim7$ galaxy samples.
  These samples are used to derive the $UV$-continuum slope $\beta$
  distribution as a function of redshift and luminosity (see \S3.2).
  The blue lines show the expected colors versus redshift for
  star-forming galaxies with a range of $UV$-continuum slopes $\beta$,
  while the red lines show the colors expected for low-redshift
  interlopers.  The green hatched region in the $z\sim7$ panel shows
  the position of low-mass stars in
  $z_{850}-Y_{105}$/$Y_{105}-J_{125}$ space.  The black dots show the
  colors of sources found in our HUDF catalog and provide some
  indication of the approximation distribution of sources in
  color-color space.  The two-color selection windows we use to
  identify high-redshift galaxies are indicated in gray.  These
  selection windows allow for the identification of galaxies over a
  wide range of $UV$-continuum slopes $\beta\sim-3$ to 0.5.  The red
  arrows show the Calzetti et al.\ (2000) reddening vectors.  In
  addition to the two colour selection criteria shown here, we also
  utilize a few other criteria in establishing our final samples.  For
  example, we enforce a very stringent optical non-detection criteria,
  especially for our $z\sim7$ samples where we require the optical
  $\chi_{opt} ^2$ be below a certain threshold
  (\S3.2).\label{fig:selcrit}}
\end{figure*}

\begin{figure}
\epsscale{1.15}
\plotone{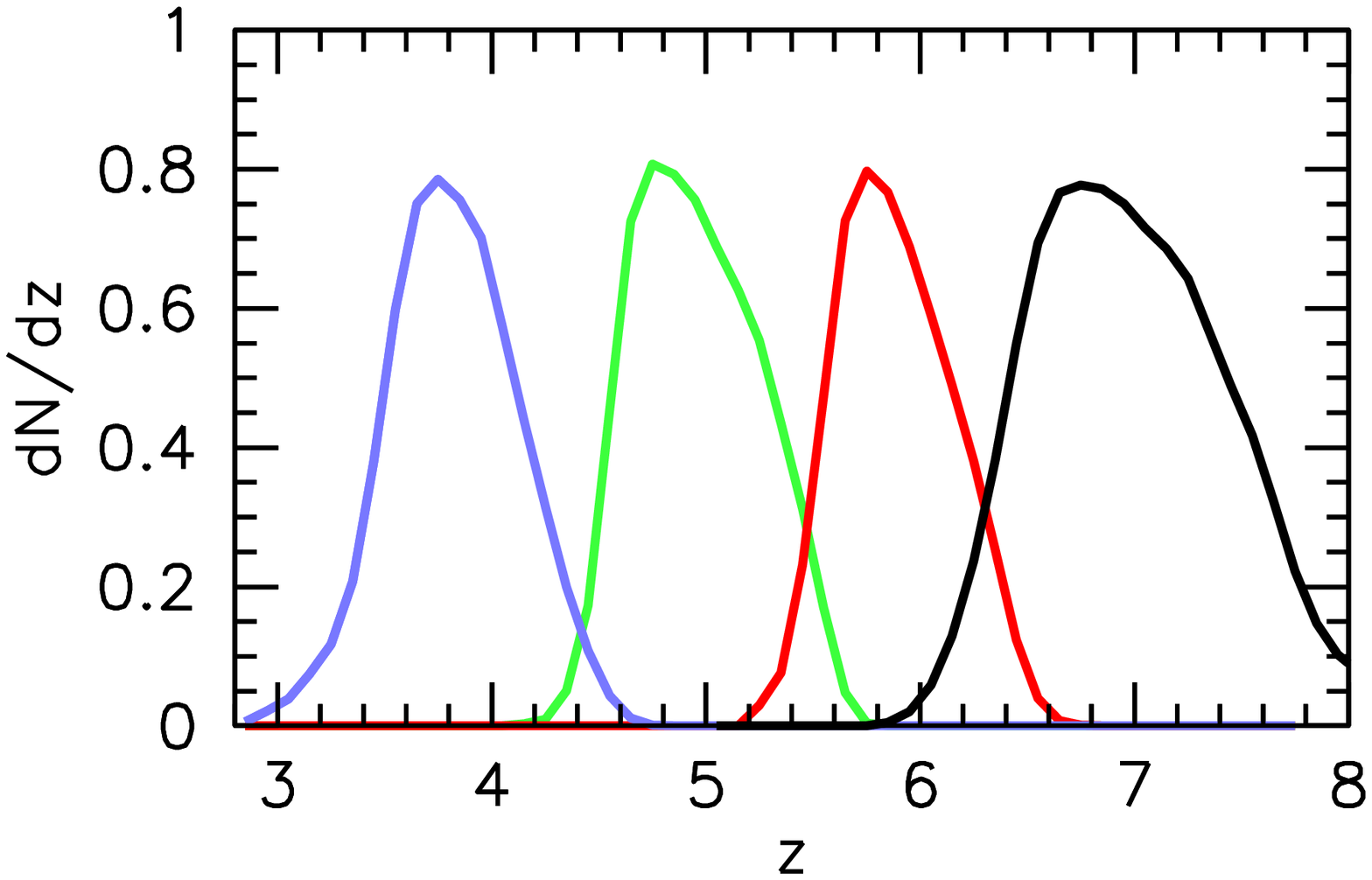}
\caption{The predicted redshift distribution for our $z\sim4$,
  $z\sim5$, $z\sim6$, and $z\sim7$ Lyman-break samples from the HUDF.
  The mean redshifts for these samples are 3.8, 5.0, 5.9, 7.0,
  respectively.  The three lower redshift selections are essentially
  the same as those presented in Bouwens et al.\ (2007: though the
  present selection window for $z\sim6$ galaxies cuts off at
  $z\gtrsim6.2$) while the $z\sim7$ selection is similar to that
  presented in Bouwens et al.\ (2011b) but extends to slightly higher
  redshifts (\S3.2).  This maximizes the size of our $z\sim7$ samples
  while extending our selection to the highest redshift possible
  without suffering significant contamination from Ly$\alpha$ emission
  or IGM absorption.  The redshift distribution for our Lyman-break
  samples from our other search fields are similar to those shown
  here.\label{fig:zdist}}
\end{figure}

\section{Results}

In this section, we describe the procedure we use to establish the
distribution of $UV$-continuum slopes $\beta$ versus $UV$ luminosity
for $z\sim4$-7 galaxies.  We begin with a description of the technique
we use for generating the source catalogs needed for sample selection
and $UV$-continuum slope measurements (\S3.1).  We then describe our
procedure for selecting our high-redshift samples (\S3.2) and for
deriving the $UV$-continuum slope $\beta$ for individual sources
(\S3.3).  In \S3.4, we detail the methods we use to correct the
observed distribution of $UV$-continuum slopes $\beta$ for selection
and measurement biases.  In \S3.5 we combine our $UV$-continuum slope
$\beta$ determinations in the ultra-deep + wide-area data sets to
establish $UV$-continuum slope $\beta$ over a wide-range in redshift
and luminosity.  We then conclude with a discussion of the correlation
we find between the $UV$-slope $\beta$ and the $UV$ luminosity
(\S3.6).

\subsection{Source Catalogs}

The procedure for source detection and photometry are discussed in
many of our previous papers (e.g., Bouwens et al. 2007); we just
summarize the key steps here.  We generate catalogs for our samples
using the SExtractor (Bertin \& Arnouts 1996) software run in double
image mode, with the detection image taken to be the square root of
$\chi^2$ image (Szalay et al.\ 1999: similar to a coadded image of the
WFC3/IR observations) and the measurement image to be a PSF-matched
image from our ACS+WFC3/IR image sets (see below).

We select each of our high-redshift samples (i.e., $z\sim4$, $z\sim5$,
$z\sim6$, and $z\sim7$) from separately-generated source catalogs.
This allows us to take advantage of the fact that $z\sim4$-6 samples
can be selected entirely (or almost entirely) based on the ACS data
(involving generally smaller apertures and therefore reducing the
uncertainties in our color measurements) while $z\sim7$ selections
require PSF-matching to the WFC3/IR data.\footnote{For our $z\sim6$
  selections, some use of the available WFC3/IR data is made.  In
  addition to satisfying a $(i_{775}-z_{850})_{AB}>1.3$ criterion,
  sources in our $z\sim6$ selection must also satisfy a
  $(z_{850}-J_{125})<0.9$ color criterion (\S3.2).}  For our
$z\sim4$-6 catalogs, the first step is to match PSFs by smoothing the
shorter wavelength images to the resolution of the ACS $z_{850}$-band
image.  The square root of $\chi^2$ images we use for the SExtractor
detection images are constructed from the $V_{606}i_{775}z_{850}$,
$i_{775}z_{850}$, and $z_{850}$ images for our $z\sim4$, $z\sim5$, and
$z\sim6$ selections, respectively.  For our $z\sim7$ catalogs, all the
observations are PSF-matched to the WFC3/IR $H_{160}$ band.  The
square root of $\chi^2$ images are constructed from the available
WFC3/IR imaging observations, e.g., $Y_{105}J_{125}H_{160}$ for our
HUDF09 fields, $Y_{098}J_{125}H_{160}$ images for the ERS
observations, and $J_{125}H_{160}$ for the CANDELS observations.

Colors are measured within small scalable apertures using a Kron
(1980) factor of 1.2.  These small-aperture color measurements are
then corrected to total magnitudes in two steps.  First, a correction
is made for the light in a larger scalable aperture, with Kron factor
equal to 2.5.  This correction is made from the square root of
$\chi^2$ image to optimize the S/N.  Secondly, a correction is made
for light outside this large scalable aperture -- using the encircled
energies measured for point sources (typically a $\sim$0.1-0.15 mag
correction).

Photometry was done using the latest WFC3/IR zeropoint calibrations
(March 6, 2012) which differ by $\sim$0.01-0.02 mag from the earlier
zeropoints.  In addition, a correction to the photometry (i.e.,
$E(B-V)=0.009$) was performed to account for foreground dust
extinction from our own galaxy.  This correction was based on the
Schlegel et al.\ (1998) dust map.

\begin{figure}
\epsscale{1.15}
\plotone{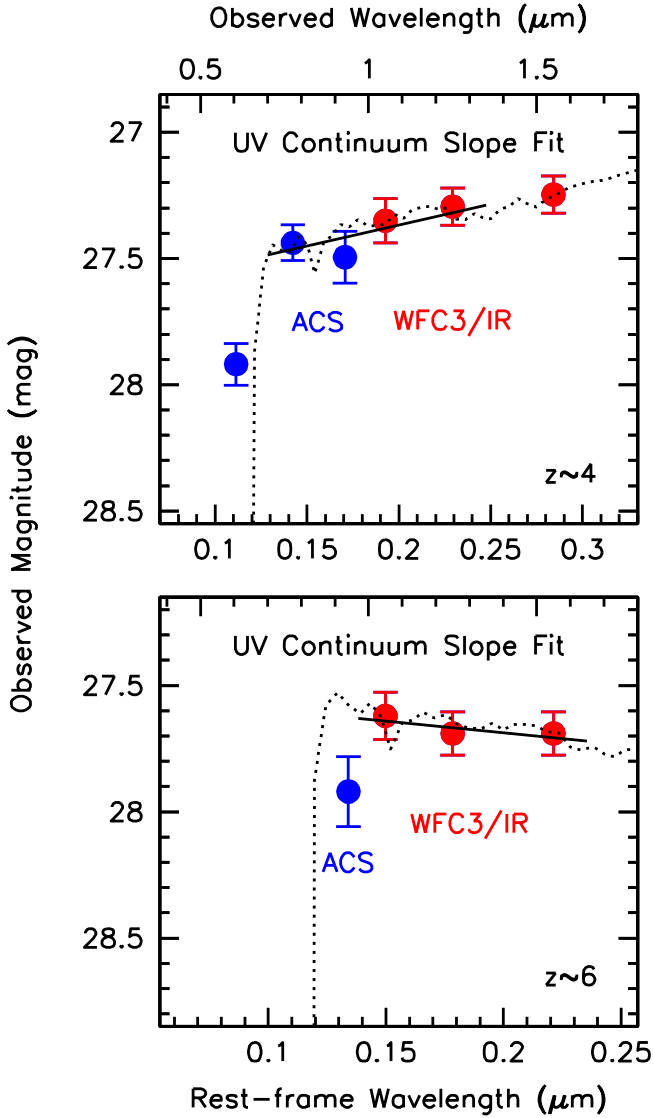}
\caption{(\textit{upper}) Illustration of how we estimate the
  $UV$-continuum slope for a $z\sim4$ galaxy candidate (see \S3.3).
  The blue and red points show the observed magnitudes for the galaxy
  in the ACS and WFC3/IR observations, respectively.  The
  $H_{160}$-band is not used for determining the $UV$-continuum slope
  $\beta$ for our $z\sim4$ samples, since the $H_{160}$-band
  magnitudes for galaxies at the low-redshift end [$z\sim3.0$-3.5] of
  our $z\sim4$ samples will include a contribution from light redward
  of the Balmer break.  The black line shows the $UV$-continuum slope
  we estimate for the source fitting to the $i_{775}$, $z_{850}$,
  $Y_{105}$, and $J_{125}$ band photometry.  The dotted black line is
  a plausible SED from a stellar population model which fits the
  observed photometry and is shown here to show where the major
  spectral features occur (but we emphasize that these SED fits are
  not used to establish the $UV$-continuum slopes).  Using the full
  wavelength baseline provided by both the ACS and WFC3/IR
  observations (Table~\ref{tab:pivotbands}), we are able to establish
  the $UV$-continuum slopes $\beta$ much more accurately than using
  the ACS observations alone.  \textit{(lower)} An illustration of how
  we estimate the $UV$-continuum slope $\beta$ for a $z\sim6$ galaxy.
  The $UV$-continuum slopes $\beta$ for $z\sim6$ galaxies are derived
  using the full flux information in the $Y_{098}$/$Y_{105}$,
  $J_{125}$, and $H_{160}$ bands
  (Table~\ref{tab:pivotbands}).\label{fig:examp}}
\end{figure}

\subsection{Sample Selection}

To select star-forming galaxies at high-redshift, we will use the
well-established Lyman-Break Galaxy (LBG) technique.  This technique
takes advantage of the unique spectral characteristics of
high-redshift star-forming galaxies, which show a very blue spectrum
overall but a sharp cut-off blueward of Ly$\alpha$.  It has been shown
to be very robust through extensive spectroscopic follow-up (Steidel
et al.\ 1996; Steidel et al.\ 2003; Bunker et al.\ 2003; Dow-Hygelund
et al.\ 2007; Popesso et al.\ 2009; Vanzella et al.\ 2009; Stark et
al.\ 2010).

\begin{deluxetable}{ccc}
\tablecaption{Wavebands used to derive the $UV$ continuum slope for
  individual galaxies in our $z\sim4$, $z\sim5$, $z\sim6$, and $z\sim7$
  samples.\tablenotemark{a}\label{tab:pivotbands}}
\tablehead{
\colhead{} & \colhead{Filters used} & \colhead{Mean rest-frame} \\
\colhead{Sample} & \colhead{to derive $\beta$} & \colhead{Wavelength\tablenotemark{b}}}\\
\startdata
\multicolumn{3}{c}{HUDF09 Observations} \\
$z\sim4$ & $i_{775}z_{850}Y_{105}J_{125}$ & 2041\AA \\ 
$z\sim5$ & $z_{850}Y_{105}J_{125}H_{160}$ & 1997\AA \\
$z\sim6$ & $Y_{105}J_{125}H_{160}$ & 1846\AA \\
$z\sim7$ & $J_{125}H_{160}$ & 1731\AA \\ \\
\multicolumn{3}{c}{ERS Observations} \\
$z\sim4$ & $i_{775}I_{814}z_{850}Y_{098}J_{125}$ & 2041\AA \\ 
$z\sim5$ & $z_{850}Y_{098}J_{125}H_{160}$ & 1997\AA \\
$z\sim6$ & $Y_{098}J_{125}H_{160}$ & 1784\AA \\
$z\sim7$ & $J_{125}H_{160}$ & 1731\AA \\ \\
\multicolumn{3}{c}{CDF-S CANDELS Observations} \\
$z\sim4$ & $i_{775}I_{814}z_{850}Y_{105}J_{125}$ & 2041\AA \\ 
$z\sim5$ & $z_{850}Y_{105}J_{125}H_{160}$ & 1997\AA \\
$z\sim6$ & $Y_{105}J_{125}H_{160}$ & 1846\AA \\
$z\sim7$ & $J_{125}H_{160}$ & 1731\AA \\
\enddata
\tablenotetext{a}{These filters probe the $UV$-continuum light of
  sources without contamination from Ly$\alpha$ emission or the
  position of the the Lyman break (being sufficiently redward of the
  1216\AA).  See \S3.3.}
\tablenotetext{b}{Geometric mean}
\end{deluxetable}

We will base our high-redshift samples on selection criteria from
previous work on $z\geq4$ galaxies.  For our $z\sim4$ $B_{435}$ and
$z\sim5$ $V_{606}$ dropout samples, we will apply almost the same
criteria as Bouwens et al.\ (2007), namely,
\begin{eqnarray*}
(B_{435}-V_{606} > 1.1) \wedge (B_{435}-V_{606} > (V_{606}-z_{850})+1.1) \\
\wedge (V_{606}-z_{850}<1.6)
\end{eqnarray*}
for our $B$-dropout sample and
\begin{eqnarray*}
[(V_{606}-i_{775} > 0.9(i_{775}-z_{850})+1.5) \vee (V_{606}-i_{775} > 2))]
\wedge \\ (V_{606}-i_{775}>1.2) \wedge (i_{775}-z_{850}<0.8)
\end{eqnarray*}
for our $V_{606}$-dropout selection (but note we use a
$i_{775}-z_{850}<0.8$ color selection instead of the
$i_{775}-z_{850}<1.2$ selection used by Bouwens et al. 2007 to
minimize contamination).  For our $z\sim6$ $i_{775}$-dropout
selection, we expanded the criteria used in Bouwens et al.\ (2007) to
take advantage of the deep near-IR observations from WFC3/IR to set
limits on the color redward of the break.  Our $z\sim6$
$i_{775}$-dropout criterion is
\begin{displaymath}
(i_{775}-z_{850}>1.3)\wedge(z_{850}-J_{125}<0.9)
\end{displaymath}
Finally, our $z\sim7$ $z_{850}$-dropout criterion is 
\begin{eqnarray*}
(z_{850} - Y_{105} > 0.7)\wedge(Y_{105}-J_{125} < 0.8) \wedge \\
(z_{850}-Y_{105}>1.4(Y_{105}-J_{125})+0.42),
\end{eqnarray*}
or
\begin{eqnarray*}
(z_{850}-J_{125}>0.9)\wedge~~~~~~~~~~~~ \\
(z_{850}-J_{125}>0.8+1.1(J_{125}-H_{160})) \wedge \\
(z_{850}-J_{125}>0.4+1.1(Y_{098}-J_{125})),
\end{eqnarray*}
depending upon whether our search field is from the HUDF09/CANDELS
datasets or the ERS dataset, respectively.  The above selection
criteria closely match those used in our previous study of the
$UV$-continuum slope $\beta$ at $z\sim7$ (Bouwens et al.\ 2010a).
These criteria are slightly more inclusive than those considered by
Bouwens et al.\ (2011b), but this is to allow us to maximize the size
of our samples and to extend our selection to the highest redshift
possible without suffering significant contamination from Ly$\alpha$
emission or IGM absorption.  In cases of a non-detection in the
dropout band, we set the flux in the dropout band to be equal to the
$1\sigma$ upper limit.  To take advantage of the deep $I_{814}$-band
observations over the CDF-South to keep contamination in our $z\sim7$
samples to a minimum, we required that all $z\sim7$ $z_{850}$-dropouts
in our selection to have $I_{814}-J_{125}$ colors greater than 2.0 or
to be undetected in the $I_{814}$ band (at $<1.5\sigma$).  No attempt
is made to select or measure $UV$-continuum slopes for star-forming
galaxies at $z\sim8$ due to the fact that the $J_{125}-H_{160}$ color
(the only available color providing high S/N information on the
rest-frame $UV$ SED at $z\sim8$) is affected by Lyman-series
absorption and Ly$\alpha$ emission at $z>8.1$ (but see Taniguchi et
al.\ 2011).

Figure~\ref{fig:selcrit} provides a convenient illustration of the
approximate range in $UV$-continuum slopes $\beta$ and redshifts
selected by the above two-colour criteria.  

We utilize several additional selection criteria in defining our final
samples.  Sources are required to have SExtractor stellarity
parameters less than 0.8 (i.e., they show evidence of being extended)
to ensure our samples are free of contamination from AGN or low-mass
stars.  Given that $>$97\% of bright sources in Lyman-Break selections
show clear evidence of having extended profiles (i.e., are not
pointlike), this criterion has almost no effect on the overall
composition of our samples.

We also require that sources be undetected in all bands blueward of
the dropout bands.  Sources are rejected if they are detected at
$>$2$\sigma$ in a single band, $>$1.5$\sigma$ in 2 bands, or have a
$\chi_{opt} ^2 > 3$ in the optical bands.  We take $\chi_{opt} ^2 =
\Sigma_{i} \textrm{SGN}(f_{i}) (f_{i}/\sigma_{i})^2$ where $f_{i}$ is
the flux in band $i$ in our smaller scalable apertures, $\sigma_i$ is
the uncertainty in this flux, and SGN($f_{i}$) is equal to 1 if
$f_{i}>0$ and $-1$ if $f_{i}<0$ (Bouwens et al.\ 2011b).  As Bouwens
et al.\ (2011b) illustrate, a $\chi_{opt} ^2$ criterion can be
particularly useful for minimizing contamination in high redshift
samples (see also Oesch et al.\ 2012a,b).

Table~\ref{tab:lumrange} summarizes the properties of the $z\sim4$,
$z\sim5$, $z\sim6$, and $z\sim7$ samples we derive from the
HUDF09+ERS+CANDELS observations.  In total, 308, 137, 70, and 57
$z\sim4$, $z\sim5$, $z\sim6$, and $z\sim7$ galaxies are found in the
HUDF09 fields and 1524, 277, 101, and 44 $z\sim4$, $z\sim5$, $z\sim6$,
and $z\sim7$ galaxies are found in the ERS+CANDELS fields.  The
approximate redshift distributions for our $z\sim4$, $z\sim5$,
$z\sim6$, and $z\sim7$ selections are shown more explicitly in
Figure~\ref{fig:zdist}.  These redshift distributions are as
calculated in Bouwens et al.\ (2007) and Bouwens et al.\ (2011b).

\begin{figure}
\epsscale{1.15}
\plotone{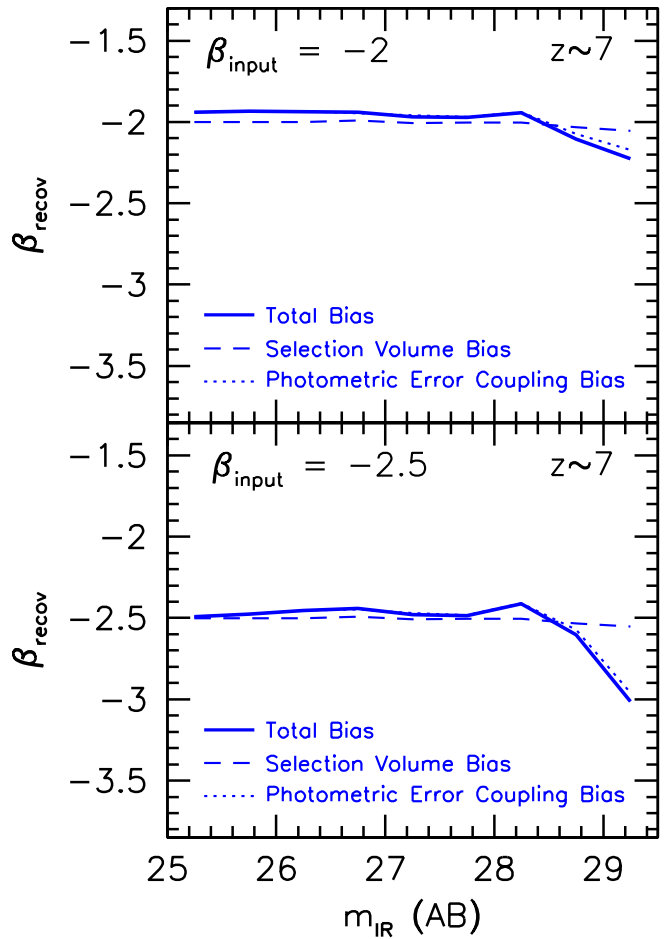}
\caption{The $UV$-continuum slopes $\beta$ expected to be recovered in
  HUDF09-depth $z\sim7$ selections assuming specific input values for
  the $UV$-continuum slope $\beta$ (\S3.4; Appendix B).  The results
  are shown as a function of IR magnitude and based on simulations
  with $>10^5$ galaxies.  The expected biases in the recovered
  $UV$-continuum slopes $\beta$ that result from $\beta$-dependent
  selection effects (\textit{dotted blue lines}:
  Figure~\ref{fig:selvol} and Appendix B.1.1) and from any coupling of
  the photometric errors in the measurement process to source
  selection (\textit{dashed blue lines}: Appendix B.1.2) are presented
  separately.  The solid blue line shows the $\beta$'s we recover
  including both effects.  The results of similar simulations for
  $z\sim4$, $z\sim5$, and $z\sim6$ selections are shown in
  Figure~\ref{fig:bias0c} in Appendix B.  Overall the $UV$-continuum
  slopes $\beta$ we recover are very similar to the input slopes
  ($\Delta \beta \lesssim 0.15$).  One possible exception would the
  faintest luminosity bin in our $z\sim7$ sample, but we emphasize
  that even in this regime our corrected $\beta$ measurements appear
  to be accurate, given their excellent agreement with other largely
  bias-free estimates we make (see \S4.8).  This strongly suggests
  that biases in our derived $UV$-continuum slope $\beta$
  distributions are very small overall.\label{fig:bias_summary}}
\end{figure}

\subsection{Estimating the $UV$-Continuum Slope}

Establishing the $UV$-continuum slope $\beta$ for a source (where
$\beta$ is defined such that $f_{\lambda}\propto \lambda^{\beta}$)
requires that we have at least two flux measurements of a source in
the $UV$-continuum not significantly affected by Ly$\alpha$ or IGM
absorption (unless both the redshift and Ly$\alpha$ flux are already
known).  In practice, this means that we should not make use of the
flux information in the band immediately redward of the break --
since it is frequently contaminated by Ly$\alpha$ emission or
absorption from the IGM.  Assuming we have wavelength coverage in the
$B_{435}V_{606}i_{775}z_{850}Y_{105}J_{125}H_{160}$ bands, we can
obtain good estimates of the $UV$-continuum slope $\beta$ for
high-redshift sources from $z\sim4$ to at least $z\sim7.5$.

There are a few different approaches we could adopt in using the
available flux measurements in the $UV$-continuum to estimate these
slopes.  One approach is to derive the $UV$-continuum slopes from a
fit to all available flux measurements in the rest-frame $UV$.  An
alternate approach is to derive a slope from the flux measurements at
both ends of a fixed rest-frame wavelength range (e.g., as utilized by
Bouwens et al.\ 2009).

Each approach has its advantages.  The first approach uses the full
flux information available on each source and also takes advantage of
a much more extended wavelength baseline.  This results in much
smaller uncertainties in our $\beta$ determinations, particularly at
$z\sim4$ and $z\sim5$ where fluxes in four separate bandpasses are
available.  The simulations we perform in Appendix B.3 suggest factors
of $\sim$1.5 improvement in the uncertainties on $\beta$ at $z\sim4$
and $z\sim5$.  On the other hand, the second approach has the
advantage that all $UV$-continuum slope $\beta$ determinations are
made using a similar wavelength baseline at all redshifts.  As a
result, even if the SED of galaxies in the rest-frame $UV$ is not a
perfect power law (e.g., from the well-known bump in dust extinction
law at 2175\AA: Stecher 1965), we would expect $\beta$ determinations
made at one redshift to show no systematic offset relative to those
made at another, allowing for more robust measurements of evolution
across cosmic time.

After some testing, we adopted the approach that utilizes all the
available flux information to determine the $UV$-continuum slopes
$\beta$.  Figure~\ref{fig:examp} provides an example of such a
determination for a $z\sim4$ galaxy in our HUDF sample and an example
of such a determination for a $z\sim6$ galaxy.  The passbands we will
consider in deriving the slopes include the
$i_{775}z_{850}Y_{105}J_{125}$ bands for $z\sim4$ galaxies, the
$z_{850}Y_{105}J_{125}H_{160}$ bands for $z\sim5$ galaxies, the
$Y_{105}J_{125}H_{160}$ bands for $z\sim6$ galaxies, and the
$J_{125}H_{160}$ bands for $z\sim7$ galaxies.  We include the
$I_{814}$ band in these fits for $z\sim4$ galaxies over the
ERS/CANDELS fields.  We replace the $Y_{105}$ band with the $Y_{098}$
band in these fits for galaxies over the ERS field.  In selecting
these passbands, we explicitly excluded passbands which could be
contaminated by Ly$\alpha$ emission, the Lyman-continuum break, or
flux redward of the Balmer break.  These choices are motivated by our
expected redshift distributions for these samples
(Figure~\ref{fig:zdist}).  Table~\ref{tab:pivotbands} includes a list
of all the bands we use to perform these fits.  The mean rest-frame
wavelengths for our derived $UV$-continuum slopes $\beta$ at $z\sim4$,
$z\sim5$, $z\sim6$, and $z\sim7$ are 2041\AA, 1997 \AA, 1846\AA, and
1731\AA, respectively.

In determining the $UV$-continuum slopes $\beta$ from the flux
information just discussed, we find the $UV$-continuum slope $\beta$
that minimizes the value of $\chi^2$:
\begin{equation}
\chi^2 = \Sigma _{i} \left( \frac{f_i - f_0 \lambda^{\beta}}{\sigma_{i}} \right)^2
\end{equation}
where $f_i$ and $\sigma_i$ are the observed fluxes and uncertainties,
respectively, $\beta$ is the best-fit $UV$-continuum slope, and $f_0$
is the best-fit normalization factor.  The fluxes $f_i \pm \sigma_i$
in the above fits are from photometry of the full ACS+WFC3/IR data set
PSF-matched to the WFC3/IR $H_{160}$-band, using typical Kron-style
apertures.  

\begin{deluxetable}{ccc}
\tablecaption{The biweight mean $UV$-continuum slope $\beta$ and $1\sigma$
  scatter of galaxies, as a function of $UV$
  luminosity.\tablenotemark{a,b}\label{tab:uvslope}} \tablehead{
  \colhead{} & \multicolumn{2}{c}{$UV$-continuum slope $\beta$}
  \\ \colhead{$<M_{UV,AB}>$} & \colhead{Mean\tablenotemark{c}} &
  \colhead{$1\sigma$ Scatter\tablenotemark{d}}} \startdata
\multicolumn{3}{c}{$z\sim4$} \\
$-$21.50 & $-$1.74$\pm$0.04$\pm$0.10 & 0.38\\
$-$20.50 & $-$1.88$\pm$0.02$\pm$0.10 & 0.33\\
$-$19.50 & $-$2.01$\pm$0.02$\pm$0.10 & 0.27\\
$-$18.50 & $-$2.16$\pm$0.03$\pm$0.10 & ---\tablenotemark{e}\\
$-$17.50 & $-$2.24$\pm$0.06$\pm$0.10 & ---\tablenotemark{e}\\
$-$16.50 & $-$2.20$\pm$0.07$\pm$0.10 & ---\tablenotemark{e}\\
\multicolumn{3}{c}{$z\sim5$} \\
$-$21.50 & $-$1.63$\pm$0.08$\pm$0.10 & 0.25\\
$-$20.50 & $-$1.93$\pm$0.05$\pm$0.10 & 0.34\\
$-$19.50 & $-$2.14$\pm$0.04$\pm$0.10 & 0.39\\
$-$18.50 & $-$2.17$\pm$0.06$\pm$0.10 & ---\tablenotemark{e}\\
$-$17.50 & $-$2.35$\pm$0.11$\pm$0.10 & ---\tablenotemark{e}\\
\multicolumn{3}{c}{$z\sim6$} \\
$-$21.50 & $-$1.78$\pm$0.11$\pm$0.14 & ---\tablenotemark{e}\\
$-$20.50 & $-$2.08$\pm$0.08$\pm$0.14 & ---\tablenotemark{e}\\
$-$19.50 & $-$2.30$\pm$0.13$\pm$0.14 & ---\tablenotemark{e}\\
$-$18.50 & $-$2.30$\pm$0.11$\pm$0.14 & ---\tablenotemark{e}\\
$-$17.50 & $-$2.54$\pm$0.17$\pm$0.14 & ---\tablenotemark{e}\\
\multicolumn{3}{c}{$z\sim7$} \\
$-$21.25 & $-$1.89$\pm$0.10$\pm$0.28 & ---\tablenotemark{e}\\
$-$20.25 & $-$2.25$\pm$0.13$\pm$0.28 & ---\tablenotemark{e}\\
$-$19.25 & $-$2.15$\pm$0.12$\pm$0.28 & ---\tablenotemark{e}\\
$-$18.25 & $-$2.68$\pm$0.19$\pm$0.28 & ---\tablenotemark{e}\\
\enddata 
\tablenotetext{a}{The biweight mean $UV$-continuum slopes $\beta$ presented
  here are also shown on Figure~\ref{fig:colmag}.}
\tablenotetext{b}{The slopes presented here have been corrected for
  selection effects and measurement errors (see \S3.4 and Appendix B).
  The tabulated $UV$ luminosities are the geometric mean of the
  measured luminosities in the bands used to establish the $UV$ slope
  (see Table~\ref{tab:pivotbands}).}
\tablenotetext{c}{Both random and systematic errors are quoted
  (presented first and second, respectively).  In \S3.5, we provide a
  brief motivation for our estimates of the approximate systematic
  error in the biweight mean $UV$-continuum slope $\beta$.} 
\tablenotetext{d}{The $1\sigma$ scatter
  presented here has been corrected for photometric scatter using the
  simulations described in Appendix B and therefore should reflect the
  intrinsic $1\sigma$ scatter in the $UV$-continuum slope $\beta$
  distribution.  Typical uncertainties are $\sim$0.1.}
\tablenotetext{e}{It is challenging to establish the intrinsic
  $1\sigma$ scatter in the $\beta$ distribution in this magnitude
  interval to $\lesssim$0.1 accuracy in $\sigma_{\beta}$ either
  because of a large photometric scatter in the individual $\beta$
  measurements or because of a limited number of sources.}
\end{deluxetable}

\begin{figure*}
\epsscale{1.17} \plotone{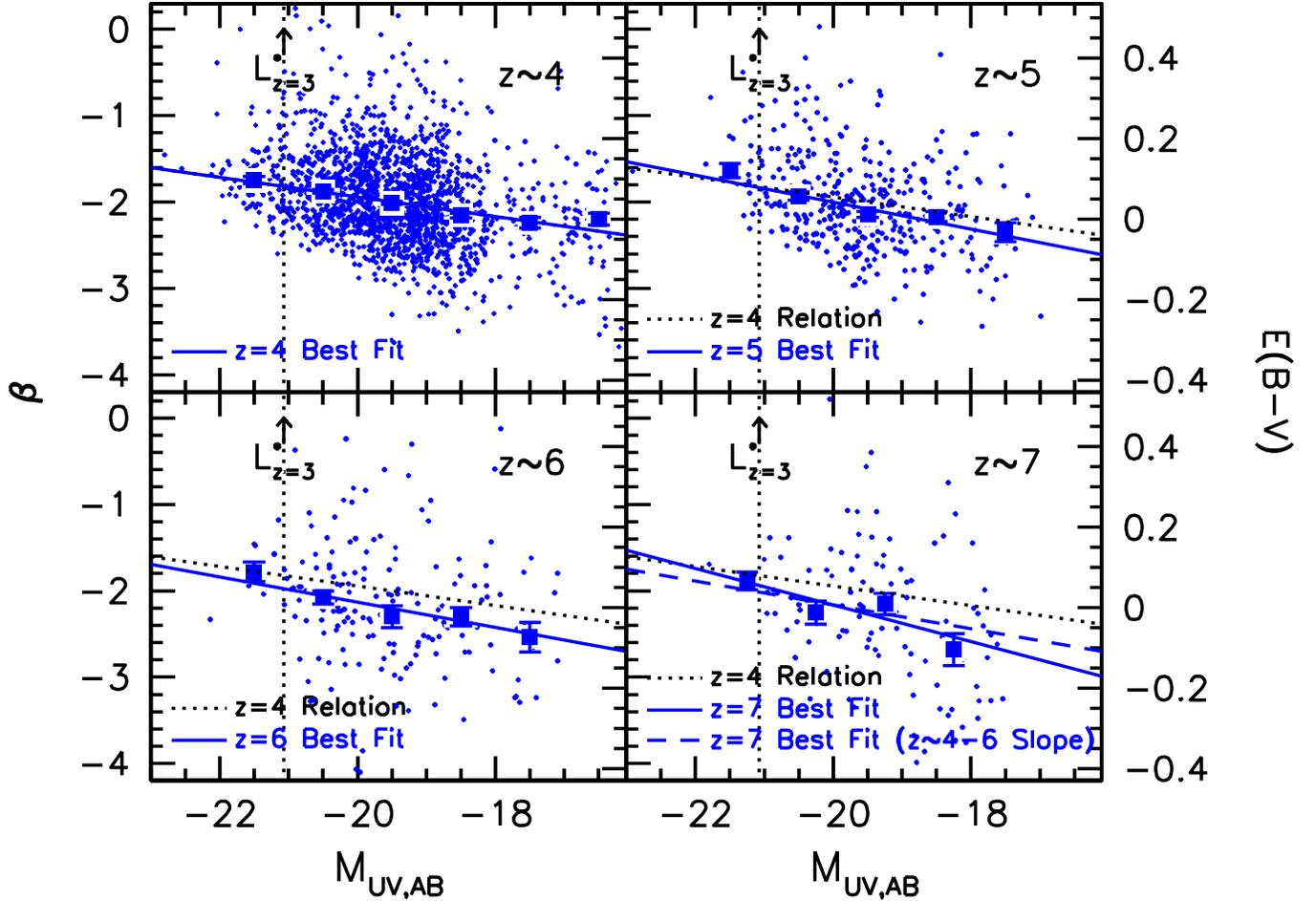}
\caption{Determinations of the $UV$-continuum slope $\beta$
  distribution versus $UV$ luminosity for star-forming galaxies at
  $z\sim4$ (\textit{upper left}), $z\sim5$ (\textit{upper right}),
  $z\sim6$ (\textit{lower left}), and $z\sim7$ (\textit{lower right}).
  The blue points show the $UV$-continuum slope $\beta$ determinations
  for individual sources in our samples.  The large blue squares show
  the biweight mean $UV$-continuum slope $\beta$ determinations in
  each magnitude interval and error on the mean.  We take the absolute
  magnitude of each galaxy in the $UV$ to be equal to the mean
  absolute magnitude of that source in all the HST bands that
  contribute to its $UV$-continuum slope determinations
  (Table~\ref{tab:pivotbands}).  The blue solid line in each panel
  shows the best-fit relationship (see Table~\ref{tab:bestfit} and
  Figure~\ref{fig:bestfit} for the best-fit parameters).  The $z\sim7$
  panel also includes a fit to the binned determinations but keeping
  the slope of the line fixed to the average $d\beta/dM_{UV}$ found
  for our $z\sim4$, $z\sim5$, and $z\sim6$ samples (dashed line). The
  dotted black line in the $z\sim5$, $z\sim6$, and $z\sim7$ panels
  show the best-fit relationship at $z\sim4$ and is included for
  comparison.  The vertical dotted line indicates the absolute $UV$
  magnitude for $L_{z=3}^{*}$ galaxies.  For our $z\sim4$ and $z\sim5$
  samples, the observed dispersion in $UV$-continuum slopes $\beta$
  about the mean relation (\textit{blue line}) is relatively small
  (see also Table~\ref{tab:uvslope} and Figure~\ref{fig:scatter}).
  The intercept to the $UV$-continuum slope $\beta$ versus luminosity
  relationship appears to become slightly bluer towards higher
  redshift.  However, the slope of the $UV$-continuum slope $\beta$
  versus luminosity does not show any significant evolution as a
  function of redshift (see also Figure~\ref{fig:bestfit}).\label{fig:colmag}}
\end{figure*}

This photometry is therefore distinct from that used to select our
$z\sim4$, $z\sim5$, and $z\sim6$ samples -- since sources in our
$z\sim4$, $z\sim5$, and $z\sim6$ Lyman-break samples were selected
based on catalogs where the PSF matching was done to the ACS $z_{850}$
band.  This approach offers two clear advantages.  (1) By selecting
sources based on catalogs where the PSF-matching is done to the
$z_{850}$-band and therefore our photometric apertures are smaller, we
utilise much higher S/N photometry for the selection process,
improving the robustness of our $z\sim4$, 5, and 6 selections.  (2) By
using different photometry (and smaller apertures) to select sources
than we use to estimate $\beta$ (typically involving larger
apertures), we ensure that our $\beta$ measurements are not subject to
exactly the same noise as affects source selection (because the
photometric apertures are different and will have somewhat different
noise characteristics).  This makes our $z\sim4$-6 $\beta$
measurements less susceptible (by $\sim$50\%) to the biases that arise
from a coupling of errors in our $\beta$ measurements with similar
errors in our photometric selection (``photometric error coupling
bias'': Appendix B.1.2).  For our $z\sim7$ Lyman-break sample,
however, the photometry used to estimate $\beta$ is the same as what
we use to select the sources.  See \S3.1 for a description of the
photometry.

In Appendix A, we show that the present approach produces similar
results to that based on $UV$ colors spanning the wavelength range
1600\AA$\,$ to 2200\AA$\,\,$(but with smaller errors).  For typical
sources, the derived $\beta$'s agree to within $\Delta\beta \sim$0.1
which is about as well as we can determine $\beta$ using $UV$ colors
(where possible systematics in $\beta$ estimates are on the order of
$\sim$0.1).  The present $UV$-continuum slope $\beta$ estimates should
therefore be directly comparable with others in the literature
(Bouwens et al.\ 2009; Meurer et al.\ 1999; Ouchi et al.\ 2004a;
Stanway et al.\ 2005; Bouwens et al.\ 2006; Bouwens et al.\ 2010a;
Finkelstein et al.\ 2010) where the wavelength baseline
1600\AA$\,\,$to 2200\AA$\,\,$was typical for $\beta$ determinations.

In deriving $\beta$'s for sources in our samples from the available
photometry, we tested for significant deviations from a pure power-law
shape.  This is important for ensuring that the model we use to
characterize the $UV$-continuum SEDs of sources is meaningful.  For
the average source, we found that the observed photometry and the
best-fit power-law SED differed by less than $\sim0.02$ mag.  Such
deviations are not large enough to have a sizeable impact on our
$\beta$ determinations, shifting the observed $\beta$ determinations
by $\lesssim0.1$.  Such changes are well within the uncertainties we
quote on our $\beta$ determinations (\S3.5).

\subsection{Possible Selection and Measurement Biases}

To establish the actual distribution of $UV$-continuum slopes $\beta$
from the observations, we must account for the effect that object
selection and measurement have on the observed distribution.  These
effects can cause the observed distribution to look very different
from what it is in reality.  Examples of such effects include (1) our
LBG selection criteria preferentially including those galaxies in our
samples with the bluest $UV$-continuum slopes and (2) photometric
noise increasing the spread in the measured $UV$-continuum slopes
$\beta$.

To correct for these selection and measurement biases, we follow a
very similar procedure to what we used in our first major study of the
$UV$-continuum slope $\beta$ distribution at high redshift (Bouwens et
al.\ 2009).  We create mock catalogs of galaxies, generate artificial
images for each source in our catalogs, add these sources to the real
observations, and then reselect the sources and measure their
properties in the same way as with the real observations.  We then
determine the approximate biases that we would expect in the mean
$UV$-continuum slope and $1\sigma$ scatter due to source selection or
photometric scatter.  We compute these biases as a function of
magnitude for each of our data sets.  To ensure that we are able to
determine the relevant biases to high precision, we repeat these
experiments for $>$10$^5$ artificial galaxies.

Appendix B provides a detailed discussion of the simulations we use to
estimate the likely biases in the $UV$-continuum slope $\beta$
distribution and establish the needed corrections.

Figure~\ref{fig:bias_summary} shows the results of these simulations
for $z\sim7$ selections.  The effect of biases related to the LBG
selection itself (Figure~\ref{fig:selvol}) and due to a possible
coupling between the selection and measurement processes
(``photometric error coupling bias'') is shown explicitly.  It is
immediately clear from this figure that we can successfully recover
the input values of the $UV$-continuum slopes $\beta$ and do not
expect especially large biases in the derived slopes.  We include
similar panels for our $z\sim4$, $z\sim5$, and $z\sim6$ selections in
Figure~\ref{fig:bias0c} of Appendix B.

An important check to perform in assessing the overall quality of our
$UV$-continuum slope determinations here is to compare the results
from our deep data with the results from our wide data.  Such a check
is provided in Figure~\ref{fig:colmagq} of Appendix B, and it seems
clear that the results from the HUDF09 and ERS+CANDELS data sets are
in good agreement ($\Delta\beta\lesssim 0.1$-0.2) over the luminosity
range where they overlap and have good statistics.  This suggests that
the $UV$-continuum slope $\beta$ distributions we recover are accurate
and free of any sizeable biases.

\subsection{Establishing the $\beta$ Distribution over a
Wide Luminosity Baseline}

We can utilize the $UV$-continuum slopes $\beta$ measured for
individual sources (\S3.3) to establish the distribution of
$UV$-continuum slopes $\beta$ as a function of luminosity for each of
our $z\sim4$-7 Lyman-Break Galaxy samples.

To establish this distribution while properly accounting for the
relevant selection and measurement biases, we analyze the
$UV$-continuum slope distribution for each data set and Lyman-Break
sample separately.  We determine the mean $UV$-continuum slope and
$1\sigma$ scatter.  We correct the mean and scatter for the relevant
biases (Appendix B and \S3.4).  Typical corrections in the mean
$UV$-continuum slope $\beta$ are $\lesssim$0.1 in general; the
corrections do however reach values of $\Delta\beta \sim 0.1$ near the
selection limit of our shallow and deep probes.

When determining the mean and scatter for each distribution, we use
the biweight mean $C_{BI}$
\begin{displaymath}
C_{BI} = M + \frac{\sum_{|u_i|<1} (x_i - M)(1-u_i
  ^2)^2}{\sum_{|u_i|<1} (1-u_i ^2)^2}
\end{displaymath}
and biweight scale $S_{BI}$
\begin{displaymath}
S_{BI} =
n^{1/2} \frac{\left[ \sum_{|u_i|<1} (x_i - M)^2 (1-u_i^2)^4
  \right]^{1/2}}{|\sum_{|u_i|<1} (1-u_i ^2)(1- 5u_i ^2)|}
\end{displaymath}
where $u_i = (x_i - M)/(c\,\textrm{MAD})$, $\textrm{MAD} =
\textrm{median}(|x_i - M|)$, $M$ is the median of the $x_i$'s, $x_i$
is the data, $n$ is the number of sources summed over, and $c$ is the
``tuning constant'' (Beers et al.\ 1990).  $c$ is taken to be equal to
6 in computing the biweight mean and 9 in computing the biweight
scale.  Use of robust statistics like the biweight mean and scale is
valuable, given that a small fraction of the sources in our samples
may be contaminated by light from nearby sources or may lie outside
the target redshift range.  Biweight mean determinations of $\beta$
are similar to median determinations of $\beta$ (median difference
$\Delta\beta$ $\sim0.01$ with $\sim$0.04 scatter), but somewhat bluer
($\Delta\beta\sim0.1$) than mean determinations (due to the $\beta$
distribution having a somewhat extended tail towards red $\beta$'s and
the biweight mean de-weighting the tail).

For the absolute magnitude of each galaxy in the $UV$, we use the mean
absolute magnitude of that source in all the HST bands that contribute
to its $UV$-continuum slope determinations
(Table~\ref{tab:pivotbands}).  By using the geometric mean for this
luminosity, we aim to avoid giving too much weight to the bluer or
redder bands in defining this luminosity and hence artificially
creating a correlation between the derived $UV$-continuum slope
$\beta$ and $UV$ luminosity.

Figure~\ref{fig:colmag} shows the $UV$-continuum slope $\beta$
distribution versus the magnitude for our $z\sim4$, $z\sim5$,
$z\sim6$, and $z\sim7$ samples.  Scatter in the $\beta$ distribution
is relatively modest for our $z\sim4$ and $z\sim5$ samples, but is
larger for our $z\sim6$ and $z\sim7$ samples.  This is the result of
the fact that our $z\sim4$-5 samples use a larger number of passbands
and larger wavelength baseline to measure the $UV$-continuum slope
$\beta$.

In Table~\ref{tab:uvslope}, we present our corrected determinations of
the biweight mean $UV$-continuum slope and $1\sigma$ scatter as a
function of $UV$ luminosity, for our $z\sim4$, $z\sim5$, $z\sim6$, and
$z\sim7$ samples.  We also include an estimate of the uncertainties in
the biweight mean $UV$-continuum slopes that result from the
measurement uncertainties, intrinsic scatter in the $\beta$
distribution, and small number statistics.

For the systematic error on our $UV$-continuum slope measurements, we
conservatively adopt values of 0.10-0.28.  The dominant component of
this error comes from uncertainties in our photometry.  Allowing for
small errors in the photometric zeropoints, aperture corrections, and
PSF-matching, we estimate a maximum error of 0.05 mag in our flux
measurements.  This translates into possible systematic errors of
0.10, 0.10, 0.14, and 0.28 in the $\beta$ measurements at $z\sim4$,
$z\sim5$, $z\sim6$, and $z\sim7$.

We also allow for small systematic errors in our $UV$-continuum slope
$\beta$ measurements as a result of differences in the wavelength
baseline we use to derive $\beta$ (Table~\ref{tab:pivotbands}).  These
differences would have an effect on the $\beta$'s we derive from the
available photometry, if the $UV$ SEDs differed substantially from a
pure power-law form (i.e., $f_{\lambda} ^{\beta}$).  The tests we
perform in Appendix A suggest the relevant systematic errors are small
$\Delta\beta\lesssim0.13$.  Similar $\beta$'s are found from fits
using the full $UV$-continuum (\S3.3 and Figure~\ref{fig:examp}) as
are found using a smaller wavelength range.

\subsection{Dependence of $\beta$ on Luminosity}

We observe a clear trend in the $UV$-continuum slope $\beta$ as a
function of $UV$ luminosity in all four LBG samples, such that the
$UV$-continuum slope $\beta$ becomes progressively bluer at fainter
luminosities (Figure~\ref{fig:colmag} and \ref{fig:colmagq}).  Such
trends in $\beta$ had already been identified by Bouwens et
al.\ (2009) in their analyses of $z\sim2.5$ and $z\sim4$ galaxy
samples (see also Meurer et al.\ 1999; Labb{\'e} et al.\ 2007; Figure
8 of Overzier et al.\ 2008) and by Bouwens et al.\ (2010a) in their
analyses of $z\sim5$-7 galaxies (see also Wilkins et al.\ 2011).  The
observed luminosity dependence of $\beta$ is thought to be due to a
change in the dust content and perhaps the age of galaxies as a
function of luminosity (Labb{\'e} et al.\ 2007; Bouwens et al.\ 2009).
It is likely to be a manifestation of the well-established
mass-metallicity relationship seen at a wide variety of redshifts
(e.g., Tremonti et al.\ 2004; Erb et al.\ 2006a; Maiolino et
al.\ 2008).

\begin{deluxetable}{cccc}
\tablecaption{Best-fit slopes and intercepts to the $UV$-continuum slope $\beta$ to $UV$ luminosity relationship (\S3.6: see also Figure~\ref{fig:bestfit})\label{tab:bestfit}}
\tablehead{
  \colhead{Dropout} & \colhead{Mean} & \colhead{} & \colhead{} \\
  \colhead{Sample} & \colhead{Redshift} & \colhead{$\beta_{M_{UV}=-19.5}$} & \colhead{$d\beta/dM_{UV}$} }
\startdata
$B_{435}$ & 3.8 & $-$2.00$\pm$0.02$\pm$0.10 & $-$0.11$\pm$0.01\tablenotemark{a} \\
$V_{606}$ & 5.0 & $-$2.08$\pm$0.03$\pm$0.10 & $-$0.16$\pm$0.03\tablenotemark{a} \\
$i_{775}$ & 5.9 & $-$2.20$\pm$0.05$\pm$0.14 & $-$0.15$\pm$0.04\tablenotemark{a} \\
$z_{850}$ & 7.0 & $-$2.27$\pm$0.07$\pm$0.28 & $-$0.21$\pm$0.07 \\
\multicolumn{4}{c}{---------------------------} \\
$U_{300}$\tablenotemark{b} & 2.5 & $-$1.70$\pm$0.07$\pm$0.15 & $-$0.20$\pm$0.04
\enddata
\tablenotetext{a}{While the $d\beta/dM_{UV}$ slopes for our $z\sim4$,
  5, and 6 samples are quite similar overall, these slopes show even
  better agreement if we consider the correlations over the same
  luminosity range, i.e., excluding $\beta$ determinations faintward
  of $-17$ AB mag.  In this case, we find a $d\beta/dM_{UV}$ slope of
  $-$0.13$\pm$0.02 for our $z\sim4$ sample.}
\tablenotetext{b}{From Bouwens et al.\ (2009).}
\end{deluxetable}

The $UV$-continuum slope $\beta$ shows an approximately linear
relationship on the magnitudes $M_{UV}$ of the sources
(Figure~\ref{fig:colmag}), and therefore it makes sense for us to
model this relationship using a first-order polynomial and determine
the best-fit slopes $d\beta / dM_{UV}$ and intercepts $\beta_{M_{UV}=-19.5}$.

\begin{figure}
\epsscale{1.17}
\plotone{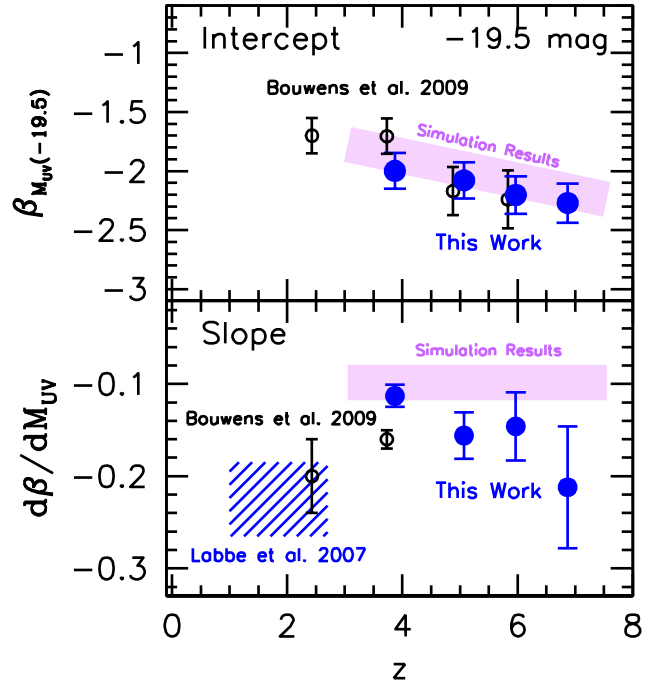}
\caption{Slope and intercept of the $UV$-continuum slope-luminosity
  relationship as a function of redshift (\S3.6: see also
  Table~\ref{tab:bestfit}).  (\textit{upper}) Intercept of the
  $UV$-continuum slope-luminosity relationship, as a function of
  redshift (\textit{large blue circles}).  Previous determinations of
  the intercept to the $\beta$-luminosity relationship are also shown
  (\textit{open black circles}: Bouwens et al.\ 2009).  The shaded
  lavendar region represent the predictions from the cosmological
  hydrodynamical simulations of Finlator et al.\ (2011: \S5.1).  We
  use $-$19.5 AB mag as the intercept, because of the substantial
  $UV$-continuum slope measurements there at all redshifts.  An
  apparent reddening of the $UV$-continuum slope $\beta$ at $-19.5$ AB
  mag ($0.25L_{z=3}^{*}$) with cosmic time is observed, from $z\sim7$
  to $z\sim2.5$.  (\textit{lower}) Slope of the $UV$-continuum
  slope-luminosity relationship, as a function of redshift
  (\textit{large blue circles}).  Previous determinations of this
  dependence on luminosity at $z\sim1-3$ (\textit{blue shaded region}:
  Labb{\'e} et al.\ 2007) and at $z\sim2.5$ and $z\sim4$ (\textit{open
    black circles}: Bouwens et al.\ 2009) are also shown.  The shaded
  lavender region is as in the upper panel.  A very similar dependence
  of $\beta$ on $UV$ luminosity is observed over the entire redshift
  range $z\sim7$ to $z\sim1$.\label{fig:bestfit}}
\end{figure}

In fitting a line to our $UV$-continuum slopes, we use a finer binning
(i.e., 0.5 mag) than shown in Figure~\ref{fig:colmag} to minimize the
impact of the binning scheme on the best-fit slopes and intercept.
The results of our fits to the biweight means are shown as blue lines
on Figure~\ref{fig:colmag}, and it is clear that the mean $\beta$'s
are well fit by the lines.\footnote{There may nonetheless be weak
  evidence in the results from our $z\sim4$ and $z\sim5$ samples that
  the dependence of $\beta$ on luminosity is weaker faintward of $-$18
  mag (see Figure~\ref{fig:colmag}).}  The fits exhibit a similar
dependence on the $UV$ magnitude $M_{UV}$ at each redshift.  The
intercept to the lines is also similar at all redshifts but appears to
evolve monotonically with cosmic time.  The best-fit parameters --
slope and intercept -- we derive for these lines are presented in
Table~\ref{tab:bestfit} and Figure~\ref{fig:bestfit}.  These
determinations are in excellent agreement with previous work, as we
discuss in \S4.

\section{Comparison to Previous Results}

In the present section, we compare the present observational results
on the $UV$-continuum slope $\beta$ with those previously obtained in
the literature (Figure~\ref{fig:mcolor}).  The goal is to assess the
robustness of the present observational results and to give some
perspective on which trends are gaining widespread observational
support.

\subsection{Comparison with Bouwens et al.\ 2009 ($z\sim3$-6)}

\begin{figure}
\epsscale{1.17}
\plotone{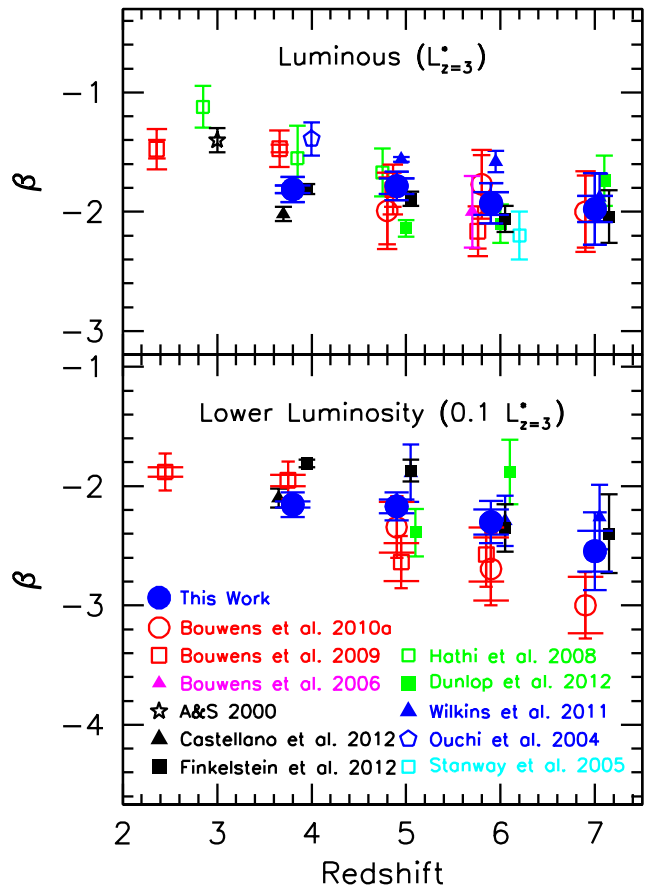}
\caption{Comparison of the biweight mean $UV$-continuum slopes $\beta$
  found here (\textit{solid blue circles}) for luminous ($L^*$/$-$21
  AB mag: \textit{top panel}) and lower luminosity ($0.1
  L_{z=3}^*$/$-18.5$ AB mag: \textit{bottom panel}) galaxies with
  values in the literature.  Generally we only compare with the mean
  $UV$-continuum slope determinations from the literature -- since a
  biweight mean slope is not specified.  However, as described in
  Figure~\ref{fig:bestfit}, the biweight mean values for $\beta$ tend
  to be somewhat bluer ($\Delta\beta\sim0.1$) than the mean values.
  Included on this figure are $UV$-continuum slope determinations from
  Bouwens et al.\ (2010a: \textit{red open circles}), Bouwens et
  al.\ (2009: \textit{red open squares}), Bouwens et al.\ (2006:
  \textit{magenta triangle}), Adelberger \& Steidel (2000:
  \textit{open black star}), Ouchi et al.\ (2004a: \textit{open blue
    pentagon}), Stanway et al.\ (2005: \textit{open cyan square}),
  Hathi et al.\ (2008: \textit{open green squares}), Dunlop et
  al.\ (2012: \textit{solid green squares}), Wilkins et al.\ (2011:
  \textit{blue triangles}), Castellano et al.\ (2012: \textit{solid
    black triangles}), and Finkelstein et al.\ (2012: \textit{solid
    black squares}).  In general, we find good agreement with our
  previous $UV$-continuum slope determinations (Bouwens et al.\ 2009,
  2010a) although the current results are a little redder at lower
  luminosities.  There is a moderate amount of scatter in the
  observational results at lower luminosities (see
  \S4).\label{fig:mcolor}}
\end{figure}

Before the WFC3/IR camera on HST became operational, Bouwens et
al.\ (2009) made use of the ACS+NICMOS observations to quantify the
distribution of $UV$-continuum slopes $\beta$ for star-forming
galaxies over range in redshift ($z\sim2.5$-6) and luminosities.
Bouwens et al.\ (2009) derive $\beta$ directly from the $UV$ colors
(e.g., as in Appendix A).  How do the present $UV$-continuum slope
determinations compare with those from Bouwens et al.\ (2009)?  Both
old and new results are shown in Figure~\ref{fig:mcolor}.  Comparing
the $UV$-continuum slope measurements made on the identical sources in
the two data sets (old ACS+NICMOS data vs. the new WFC3/IR
observations), we find reasonable agreement, with mean offsets
$\beta_{WFC3/IR}-\beta_{ACS+NICMOS}$ of only $-$0.10, 0.04, 0.09 in
the derived $\beta$'s at $z\sim4$, $z\sim5$, and $z\sim6$,
respectively.  The biweight mean $UV$-continuum slopes $\beta$ found
here are also in good agreement with the slopes derived by Bouwens et
al.\ (2009).  At $z\sim4$ and $z\sim5$, the present biweight mean
$\beta$'s are just $\sim$0.26 bluer and $\sim$0.16 redder than the
values found in Bouwens et al.\ (2009); the present biweight mean
$\beta$'s at $z\sim6$ show no average shift at all relative to the
values found by Bouwens et al.\ (2009).  $\Delta\beta\sim0.1$ of the
differences result from our use of the more robust biweight means to
express the central $\beta$'s, so the agreement is quite good overall.

\subsection{Comparison with Bouwens et al.\ 2010a ($z\sim5$-7)}

Bouwens et al.\ (2010a) took advantage of the first-year WFC3/IR
observations over the HUDF and the ERS WFC3/IR observations to
quantify the $UV$-continuum slope $\beta$ distribution to higher
redshifts ($z\sim7$) and lower luminosities.  How do the present
determinations of the $UV$-continuum slope $\beta$ compare with
Bouwens et al.\ (2010a)?  At high luminosities, we find excellent
agreement, with both studies preferring mean $UV$-continuum slopes
$\beta$ of $-2$ (see Figure~\ref{fig:mcolor}).  At lower luminosities,
however, we now find somewhat redder values of the $UV$-continuum
slope $\beta$, i.e., $\Delta\beta\sim0.2-0.4$ than found by Bouwens et
al.\ (2010a).  The observed differences in the mean $UV$-continuum
slope $\beta$ seem to have resulted from both the small number of
sources in previous samples and uncertainties in the photometry of
faint sources in the early HUDF09 data.  We remark that the current
WFC3/IR observations of the HUDF from the HUDF09 WFC3/IR program are
approximately twice as deep as what Bouwens et al.\ (2010a) used.

\begin{figure}
\epsscale{1.15}
\plotone{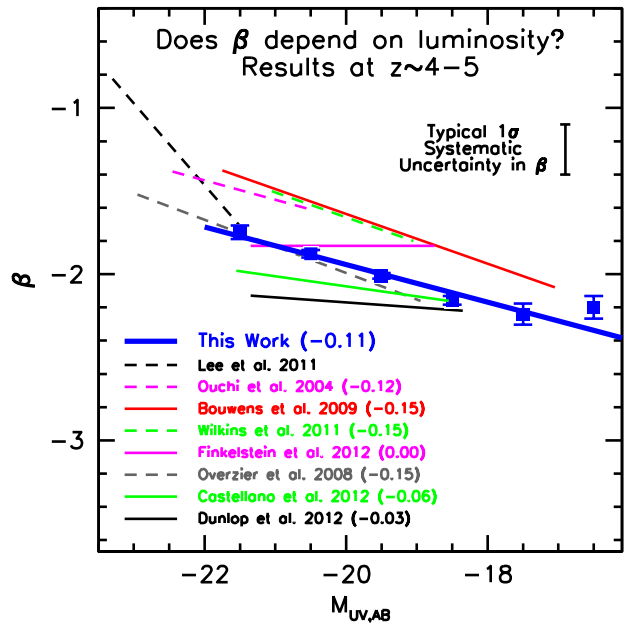}
\caption{The dependence of the $UV$-continuum slope $\beta$ on $UV$
  luminosity $M_{UV}$ at $z\sim4$-5, as determined by different
  analyses in the literature.  The lines show the observed
  $\beta$-$M_{UV}$ relations from Lee et al.\ 2011 (\textit{dashed
    black}), Bouwens et al.\ 2009 (\textit{solid red}), Wilkins et
  al.\ 2011 (\textit{dashed green}), Ouchi et al.\ 2004
  (\textit{dashed magenta}), Finkelstein et al.\ 2012 (\textit{solid
    magenta}), Overzier et al.\ 2008 (\textit{dashed gray}),
  Castellano et al.\ 2012 (\textit{solid green}), Dunlop et al.\ 2012
  (\textit{solid black}), and the present work (\textit{blue}).  The
  blue squares are the $\beta$ determinations at $z\sim4$ from the
  present work.  Error bars are $1\sigma$.  Also included on this
  figure in parentheses is the slope of the $\beta$-$M_{UV}$
  relationship $d\beta/dM_{UV}$ determined in different analyses.  In
  several cases (e.g., Ouchi et al.\ 2004; Wilkins et al.\ 2011) shown
  here, we derived the plotted relations and $d\beta/dM_{UV}$ slopes
  from the individual $\beta$ determinations in those papers.  The
  error bar in the lower left illustrates the approximate systematic
  uncertainties in previous $\beta$ measurements.  Similar luminosity
  dependencies are found at $z\sim1$-3 by Labb{\'e} et al.\ (2007) and
  Bouwens et al.\ (2009: see Figure~\ref{fig:bestfit}).  A large
  number of independent analyses have found evidence for a similar
  correlation between $UV$ luminosity and $\beta$ (see \S4.4).
  Brighter galaxies are consistently found to be redder in their
  $\beta$'s and fainter galaxies are found to be
  bluer.\label{fig:lumdep}}
\end{figure}

\subsection{Comparison with the literature: Does $\beta$ correlate with redshift?}

In the present analysis, we find that the $UV$-continuum slope $\beta$
shows a correlation with the redshift of galaxies, with higher
redshift galaxies being bluer.  This evolution is evident both in
Figure~\ref{fig:colmag} (compare the \textit{solid} lines in the
$z\sim5$, $z\sim6$, $z\sim7$ panels with the dotted lines) and in
Figure~\ref{fig:bestfit}.

How does the correlation we find compare with other studies?  A brief
summary of the evidence for this redshift dependence is provided in
Figure~\ref{fig:mcolor}, and there is a clear trend from bluer
$UV$-continuum slopes $\beta$ at $z\sim6$-7 to redder slopes at
$z\sim2$-4.  Essentially all studies of the $UV$-continuum slope over
the range $z\sim3$-7 (Lehnert \& Bremer 2003; Stanway et al.\ 2005;
Bouwens et al.\ 2006; Bouwens et al.\ 2009; Bouwens et al.\ 2010a;
Wilkins et al.\ 2011; Finkelstein et al.\ 2012) find evidence for this
evolution of $\beta$ with redshift.  The only exception to this is the
recent study of Dunlop et al.\ (2012) who find no dependence (but
Dunlop et al.\ 2012 do note that a comparison of their results with
those at $z\sim3$ does argue for evolution); we discuss the Dunlop et
al.\ (2012) results in \S4.5.

\subsection{Comparison with the literature: Does $\beta$ correlate with $UV$ luminosity?}

As we discuss in \S3.6, we find that the $UV$-continuum slope $\beta$
shows a clear correlation with the rest-frame $UV$ luminosity of
galaxies, with lower luminosity galaxies being bluer at all redshifts.
A good illustration of this correlation is provided in Figure~
\ref{fig:colmag}.  The best-fit relationship (\textit{solid blue
  line}) shows almost exactly the same dependence for each high
redshift sample.  The uniformity of the slope and modest variation as
a function of redshift is also clearly illustrated in
Figure~\ref{fig:bestfit}.  The uniformity extends even to our $z\sim4$
samples.  While $\beta$ in these samples appear to show a somewhat
weaker dependence on luminosity than our other samples, a slightly
steeper dependence is found, i.e., $d\beta/dM_{UV}=-0.13\pm0.02$, if
we use the same luminosity baseline as our $z\sim5$ and $z\sim6$
samples (excluding sources faintward of $-17$ AB mag).

In general, the correlation we find agrees very well with most
previous studies.  Figure~\ref{fig:lumdep} shows different $\beta$
determinations as a function of luminosity at $z\sim4$-5 (Ouchi et
al.\ 2004; Overzier et al.\ 2008; Bouwens et al.\ 2009; Lee et
al.\ 2011; Wilkins et al.\ 2011; Dunlop et al.\ 2012).\footnote{The
  slopes we derive from Labb{\'e} et al.\ (2007) are based upon their
  results at 1700\AA$\,$and 3600\AA$\,$in their Figure 3.  Previously,
  Bouwens et al.\ (2009) had estimated a slope of $\sim-0.25$ in the
  $UV$-continuum slope vs. $M_{UV}$ relationship at $z\sim1$-2.7 from
  the Labb{\'e} et al.\ (2007) results based on a shorter
  1700\AA$\,$and 2200\AA$\,$wavelength baseline (see
  Figure~\ref{fig:bestfit}).  The reason we use a more extended
  wavelength baseline to estimate $\beta$ than Bouwens et al.\ (2009)
  had used is to allow for a fair comparison with the present $\beta$
  results (which use an extended wavelength baseline to estimate
  $\beta$: see \S3.3).}  Despite some dispersion in the precise values
of the $\beta$ measurements and some small variance in the best-fit
slopes $d\beta/dM_{UV}$, nearly all published determinations of the
$UV$-continuum slope $\beta$ at $z\sim4$-5 find bluer values for
$\beta$ at lower luminosities, with the same luminosity dependence.
Two recent analyses that found minimal or no correlation of $\beta$
with luminosity are those of Dunlop et al. (2012) at $z\sim5$-7 and
Finkelstein et al. (2012) at $z\sim4$-5.  We discuss the Dunlop et
al.\ (2012) results in \S4.5 and the Finkelstein et al.\ (2012)
results in \S4.7.

Similar correlations with luminosity are found in the $UV$-continuum
slope results at $z\sim2$-3 (Labb{\'e} et al.\ 2007; Bouwens et
al.\ 2009; Sawicki 2012) and in the $UV$-optical colors (Papovich et
al.\ 2001; Labb{\'e} et al.\ 2007; Gonz{\'a}lez et al.\ 2012).  Again,
two analyses did not find a correlation of $\beta$ with luminosity,
and those are the Adelberger \& Steidel (2000) and Reddy et
al.\ (2008) analyses at $z\sim2$-3.  In these cases, not only was the
luminosity baseline too short to provide much leverage for quantifying
this correlation, but also the luminosity range probed was around
$L^*$ which is where the dispersion in dust properties relative to UV
luminosity is at a maximum (e.g., Figure 13 of Reddy et al.\ 2010).
However, when one adds the Adelberger \& Steidel (2000) samples to the
faint ($\sim$26-27 mag) $z\sim2.5$ samples observed by Bouwens et
al.\ (2009), a strong correlation is present (Figure 3 of Bouwens et
al.\ 2009: see also Figure 5 of Sawicki 2012).  Taken together these
results indicate that there is also a clear trend with luminosity at
$z\sim2$-3.

\subsection{Comparison with Dunlop et al. 2012 ($z\sim5$-7)}

Dunlop et al.\ (2012) use the WFC3/IR observations over the HUDF09 and
ERS observations to quantify the $UV$-continuum slope distribution at
$z\sim5$-7.  They select sources using a photometric redshift
procedure and then measure their $UV$-continuum slopes $\beta$ from
the $UV$ colors.  In both respects, their procedure differs from that
followed here (\S3.2 and \S3.3).  Overall, the individual
$UV$-continuum slope $\beta$ measurements of Dunlop et al.\ (2012) are
in reasonable agreement with the present results (see
Figure~\ref{fig:mcolor}).  The mean $\beta$ they derive for bright
$z\sim5$ galaxies is $\Delta\beta\sim0.3$ bluer and the mean $\beta$
they derive for faint $z\sim6$ galaxies is $\Delta\beta\sim0.4$
redder.  Given the quoted uncertainties on the measurements, the
differences are not particularly significant.

The main differences arise when we look at the trends in the
$UV$-continuum slope $\beta$ with redshift and luminosity.  Dunlop et
al.\ (2012) find a mean $UV$-continuum slope $\beta$ of galaxies equal
to $\sim-2.1$, with no significant dependence on luminosity or
redshift.  This is in contrast to the strong correlation we find of
$\beta$ with both luminosity (Figure~\ref{fig:colmag} and
\ref{fig:bestfit}) and redshift (Figure~\ref{fig:bestfit}).  Their
results are also inconsistent with the trends reported in the
literature (\S4.3-\S4.4; Figure~\ref{fig:bestfit} and
\ref{fig:lumdep}).

We have attempted to understand the source of this difference both by
a qualitative assessment of some issues that we know are important for
obtaining reliable results and by a quantitative assessment using
simulations (see Appendix D and \S4.6).  First, we remark that Dunlop
et al.\ (2012) make no attempt to correct their mean $\beta$'s for the
fact that galaxies with bluer $UV$-continuum slopes $\beta$ are easier
to select than galaxies with redder $UV$-continuum slopes $\beta$.
This effectively biases their $UV$-continuum slopes $\beta$ to bluer
values.  It is a fairly straightforward process to correct for this
issue (e.g., we describe such a correction in Appendix B.1.1 and
Figure~\ref{fig:bias_summary} where we show its effect).\footnote{See
  also Figure 5 of Wilkins et al.\ 2011.}  Second, Dunlop et
al.\ (2012) use an overlapping set of information both to select
sources and to measure the $UV$-continuum slopes $\beta$.  This biases
their results (see Appendix D), though the magnitude of this bias is
mitigated by their consideration of only the brightest sources.

\begin{figure}
\epsscale{1.15} \plotone{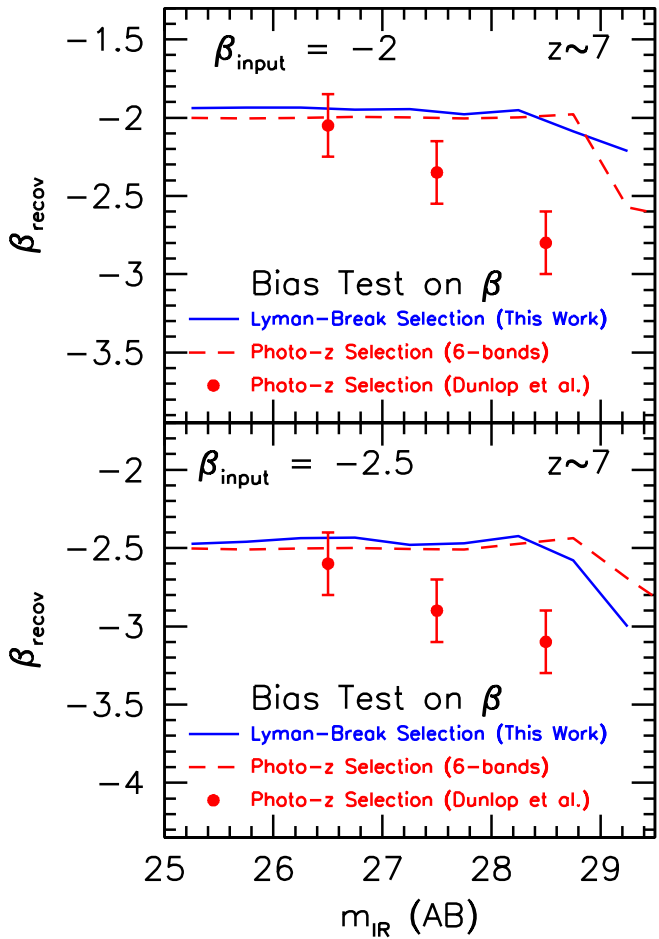}
\caption{Comparison of the recovered $UV$-continuum slopes $\beta$
  versus near-IR magnitude for a $z\sim7$ galaxy population with input
  $UV$-continuum slopes $\beta$ of $-2.0$ and $-2.5$ (see Appendix B
  for details).  Shown are the results from the present Lyman-break
  selection (\textit{solid blue lines}), from a 6-band photometric
  redshift selection (\textit{dashed red lines}), and from the Dunlop
  et al.\ (2012: D11) ``robust'' photometric redshift selections
  (\textit{solid red circles}: from their Figure 8).  The reason our
  measured $UV$-continuum slope measurements $\beta$ are not
  significantly biased towards fainter magnitudes is that we select
  samples using a different part of the rest-frame $UV$ SED (bluest
  two band passes redward of the break) than we use to measure the
  $UV$-continuum slope (second bluest bandpass and redder).  See
  Figure~\ref{fig:examp}.  The situation is similar if one uses a
  6-band photometric redshift selection (Appendix D).  By contrast,
  Dunlop et al.\ (2012) use overlapping information both to select
  their sources and to measure the $UV$-continuum slopes.  Therefore,
  while photometric scatter in the bands we use for selection has
  little effect on our $UV$-continuum slope measurements, these steps
  are tightly coupled in the photometric redshift approach utilized by
  Dunlop et al.\ (2012).  This results in the strong biases shown in
  this figure.  While Dunlop et al.\ (2012) try to minimize the
  magnitude of these biases by restricting their analysis to the
  highest S/N sources, the same inherent biases in their $\beta$
  estimates will remain, but at a lower level.  See \S4.6, Appendix
  B.1.2, and Appendix D.\label{fig:biasdunlop}}
\end{figure}

Third, one further limitation of the Dunlop et al.\ (2012) analysis is
their exclusion of the lowest luminosity sources.  This significantly
reduces the leverage they have to quantify trends in the mean
$UV$-continuum slope as a function of luminosity.  Dunlop et
al.\ (2012) exclude faint sources, because their simulations
suggested to them that $\beta$ could not be measured in an unbiased
way.  However, as we show in Appendix B and D, we are able to obtain
reliable measurements (see also \S4.6).  Not only can we recover the
mean $UV$-continuum slope for our samples to very faint magnitudes
with excellent accuracy, but we can recover these slopes for samples
selected using a photometric redshift procedure, if the information
used for source selection is clearly separated from that used to
measure $\beta$ (see Figure~\ref{fig:biasdunlop}).

As mentioned above, these three issues led us to consider a more
quantitative evaluation of the Dunlop et al.\ (2012) procedure so that
we could better understand what was happening with their measurement
of $\beta$.  This is discussed in more detail in the next section.

\subsection{Can $\beta$ be measured in a largely bias-free way for low S/N sources?}

As discussed above, a key question that has arisen in recent papers is
whether it is possible to determine the $UV$-continuum slope $\beta$
distribution to very low luminosities with small biases.  Dunlop et
al.\ (2012), in particular, have suggested that it is not possible and
have supported this suggestion with a series of simulations where they
add noise to model galaxies and reselect these galaxies with their
photometric redshift code.  Dunlop et al.\ (2012) argue that noise in
the photometry combined with a preference for selecting sources with
blue colors would result in highly biased estimates for the mean
$UV$-continuum slope $\beta$.  Dunlop et al.\ (2012) show that such a
bias towards bluer $UV$-continuum slopes at faint magnitudes is
present in their mock data sets.  Dunlop et al.\ (2012) argue that
similar biases are likely present in Lyman-break selections, without
further substantiating this claim.

We agree that selection biases can affect the distribution of
$UV$-continuum slopes $\beta$.  However, the size of these selection
biases is extremely dependent upon how one selects the galaxies and
measures their $UV$-continuum slopes $\beta$.  As we show in
Figure~\ref{fig:biasdunlop}, our Lyman-Break selections yield much
smaller selection biases overall.  The reason we expect biases in our
selections to be small is that we select galaxies using a different
part of the rest-frame $UV$ SED (the bluest two passbands redward of
the break) than we use to measure $UV$-continuum slopes (the second
bluest passband and redder).  As a result, photometric scatter in the
bands we use to measure the $UV$-continuum slope $\beta$ is largely
independent of similar scatter in the bands we use for selection.
Therefore, we would not expect our $\beta$ measurements to be
significantly biased as a result of the object selection
process.\footnote{This issue is of course in addition to the normal
  selection biases that Lyman-Break samples show against sources with
  red $UV$-continuum slopes $\beta$, but as we show in Appendix B.1.1
  these biases are small for all but the reddest $\beta$'s.}

Achieving similarly small biases to faint magnitudes with a
photometric redshift technique is also possible.  However, one must
again be careful to use different information to select sources from
what one uses to measure the $UV$-continuum slope $\beta$.  To
illustrate this, we consider the situation at $z\sim7$ in the current
HST ACS + WFC3/IR data set.  We have run simulations where we attempt
to measure the mean $UV$-continuum slope $\beta$ for a set of $z\sim7$
galaxies selected using a photometric redshift procedure.  Both the
simulations and results are discussed in Appendix D.  We consider (1)
the case where 5 HST bands are used to select sources and determine
redshifts (this excludes those bands used to measure $\beta$), (2) the
case where 6 bands are used (and so now including one of the two bands
used to measure $\beta$), and (3) the case where all 7 HST bands are
used for selection and redshift determination (and so both bands used
to measure $\beta$ are also included to measure redshifts and select
the sources).  This latter approach is basically what Dunlop et
al.\ (2012) do.

The results are shown in Figure~\ref{fig:dunlop} of Appendix D.  While
$\beta$ measurements show substantial biases when all 7 bands are used
for the photometric redshift estimates (similar to the procedure of
Dunlop et al.\ 2012) $\beta$ measurements made using 5 or 6 bands show
much smaller biases.  This demonstrates that the $UV$-continuum slope
$\beta$ can be measured with very small biases to faint magnitudes.  

\begin{figure}
\epsscale{1.15}
\plotone{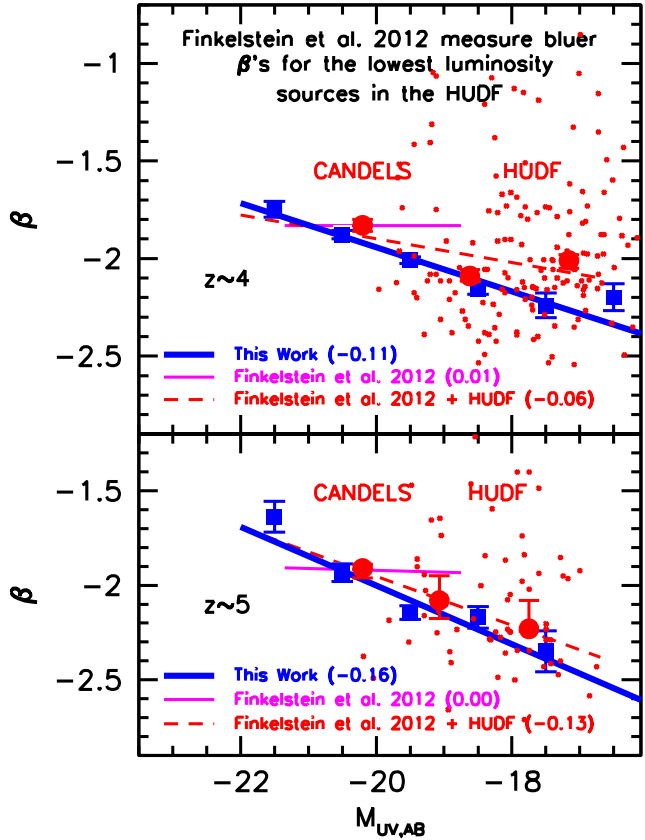}
\caption{\textit{(upper panel)} The approximate median $UV$-continuum
  slopes $\beta$ (\textit{large red circles}) derived by Finkelstein
  et al.\ (2012) for their sample as a whole (based primarily on the
  CANDELS+ERS programs) and two fainter subsamples from the HUDF
  ($-20<M_{UV}<-18$ and $-18<M_{UV}<-16$).  The median $\beta$'s we
  calculate for these two fainter subsamples and the bootstrap
  uncertainties on these medians, i.e., $-2.09_{-0.03}^{+0.03}$ and
  $-2.01_{-0.03}^{+0.04}$, respectively, are based on the $\beta$
  measurements plotted in Figure 5 of their paper (replicated here as
  the small red points).  The solid magenta line shows the trend in
  $\beta$ Finkelstein et al.\ (2012) report in their baseline
  analysis.  The dashed red line shows the trend we find comparing the
  median $\beta$'s Finkelstein et al.\ (2012) measure for their sample
  as a whole (large red circle at $\sim-20$ mag) with the median
  $\beta$'s we calculate for \textit{their} two fainter subsamples in
  the HUDF (large red circles at $\sim-19$ mag and $\sim-17$ mag).
  The solid blue squares and lines are our own $\beta$ determinations
  and are as shown in Figure~\ref{fig:colmag}.  While Finkelstein et
  al.\ (2012) report no correlation between $\beta$ and luminosity in
  their baseline analysis (\textit{magenta line}), we observe quite a
  strong correlation with luminosity, making exclusive use of their
  measurements of $\beta$ for fainter sources in the HUDF to define
  the trend to lower luminosities (\textit{red line}).  The trend
  $d\beta/dM_{UV}$ we find making exclusive use of their $\beta$
  measurements for the fainter HUDF sources, i.e., $-$0.06$\pm$0.02,
  is in much better agreement with what we find, i.e.,
  $-$0.11$\pm$0.01 than it is in their baseline analysis.  The final
  version of Finkelstein et al.\ (2012) also reports this same trend
  making exclusive use of the fainter sources from the HUDF.  The
  median $\beta$'s Finkelstein et al.\ (2012) measure for sources in
  the two fainter-magnitude HUDF subsamples shown here (\textit{two
    large red circles}) are also in very good agreement with our own
  measurements (\textit{blue squares}), particularly at $\sim-19$ mag.
  \textit{(lower panel)} Similar to the upper panel, but comparing
  results from the $z\sim5$ sample of Finkelstein et al.\ (2012) with
  our own results (see \S4.7).  The two fainter subsamples of $z\sim5$
  galaxies from the HUDF are over the magnitude ranges
  $-20.3<M_{UV}<-18.3$ and $-18.3<M_{UV}<-16.3$.  As in the upper
  panel, we note better agreement with the Finkelstein et al.\ (2012)
  $\beta$ measurements, if we restrict our comparison to their $\beta$
  measurements from the HUDF.
\label{fig:lumdepf}}
\end{figure}

\subsection{Comparison with Finkelstein et al. 2012 ($z\sim4$-7)}

In an independent analysis, Finkelstein et al.\ (2012) also use the
recent WFC3/IR observations over the HUDF and CDF-South GOODS field to
quantify the $UV$-continuum slope distribution for star-forming
galaxies at $z\sim4$-8.  Finkelstein et al.\ (2012) split sources into
five different redshift samples using a photometric redshift procedure
and then estimate $\beta$ by finding the model SED which best fits the
photometry of a source and deriving $\beta$ from this model.
Finkelstein et al.\ (2012) find that $\beta$ shows a clear dependence
on redshift, but report only a limited dependence on the $UV$
luminosity.
\vskip0.15cm

\noindent \textit{$\beta$ vs. Luminosity Trends:} While the redshift
dependence Finkelstein et al.\ (2012) find for $\beta$ is in excellent
agreement with what we find (compare the solid black squares and large
blue circles in the upper panel of Figure~\ref{fig:mcolor}), the
luminosity dependence Finkelstein et al.\ (2012) observe would appear
to be considerably weaker.  After all, Finkelstein et al.\ (2012)
report no significant correlation of $\beta$ with $UV$ luminosity in
their baseline analyses of their five redshift samples -- seemingly
different than the clear correlation of $\beta$ with luminosity we
report.

Despite these apparent differences, the overall results from the two
studies are actually in fairly good agreement.  For example, with
regard to the $z\sim6$ and $z\sim7$ samples, the best-fit
$d\beta/dM_{UV}$ values Finkelstein et al.\ (2012) find, i.e.,
$-0.10\pm0.07$ and $-0.20\pm0.11$, respectively, are strikingly
similar to the values we find, i.e., $-0.15\pm0.04$ and
$-0.21\pm0.07$, respectively.

For the $z\sim4$ and $z\sim5$ samples, Finkelstein et al.\ (2012) do
not report a significant correlation between $\beta$ and $UV$
luminosity in what they identify as their baseline analysis (finding
$d\beta/dM_{UV}$ values of 0.01$\pm$0.03 and 0.00$\pm$0.06,
respectively).  However, in this analysis, Finkelstein et al.\ (2012)
only make use of a 2-mag baseline in luminosity (due to their binning
scheme), with their low luminosity anchor point largely coming from
the relatively shallow $\sim$1.6 orbit CANDELS data (with only a small
contribution from the HUDF data).  The CANDELS observations are
clearly poorly suited to determine the trend in $\beta$ to very low
luminosities, given the very low S/N's and potentially large biases
expected for the faintest sources in the CANDELS fields.  We might
expect the situation to change taking advantage of the additional
leverage in luminosity provided by the $\beta$ measurements they
provide for faint sources in the ultra-deep HUDF observations.

We can check this by extracting the HUDF measurements from Figure 5 of
their paper.  By comparing \textit{their} median $\beta$ measurements
for brighter galaxies with \textit{their} median $\beta$ measurements
for fainter galaxies in the HUDF, we find evidence for a significant
correlation with luminosity.  We find $d\beta/dM_{UV}$ trends of
$-$0.06$\pm$0.02 and $-$0.13$\pm$0.04, respectively.  In the final
version of their paper, Finkelstein et al.\ (2012) also note a similar
correlation with luminosity making use of the faintest HUDF sources,
finding $d\beta/dM_{UV}$ trends of $-$0.07$\pm$0.01 and
$-$0.09$\pm$0.03, respectively.  While not in exact agreement with the
trends we derive based on our own $\beta$ measurements, i.e.,
$-$0.11$\pm$0.01 and $-$0.16$\pm$0.03, respectively, the agreement is
much better.  Use of the faint sources in the HUDF is important to
take full advantage of the available leverage in luminosity to
quantify the $\beta$ vs. $M_{UV}$ trend.

In Figure~\ref{fig:lumdepf}, we show the $\beta$ vs. $M_{UV}$ trend
that Finkelstein et al.\ (2012) find in their baseline $z\sim4$ and
$z\sim5$ analyses (\textit{magenta lines}) and the trend we find from
\textit{their} measurements making exclusive use of sources from the
HUDF to constrain $\beta$ to fainter magnitudes (\textit{dashed red
  lines}).  In addition, we show the median $\beta$'s Finkelstein et
al.\ (2012) find for two fainter $z\sim4$ and $z\sim5$ subsamples
within the HUDF (\textit{large solid red circles}: we can extract
$\beta$ measurements for individual sources within the HUDF from their
Figure 5) and the bootstrap uncertainties on these medians.  In both
samples there is a clear trend in the median $\beta$'s towards bluer
values at the lowest luminosities.  It is striking how well the median
$\beta$'s Finkelstein et al.\ (2012) derive from the HUDF agree with
our own $\beta$ measurements, particularly in the luminosity interval
[$-$20 mag, $-$18 mag].  While it is true that the median $\beta$'s
Finkelstein et al.\ (2012) derive for their entire ERS+CANDELS+HUDF09
sample are redder in general than what we find at these luminosities,
these median $\beta$'s receive their largest weight from the shallower
ERS+CANDELS samples and therefore may be subject to possible
selection, measurement, or contamination biases (see the discussion at
the end of this section).
\vskip 0.15cm

\noindent \textit{Expected Trend in Luminosity:} Finkelstein et
al.\ (2012) defend the weak correlation of $\beta$ they report versus
luminosity (particularly as derived in their baseline analysis),
arguing that $\beta$ should show a stronger correlation with stellar
mass than $UV$ luminosity.  We do not dispute this assertion; however,
it would be most surprising if a correlation of $\beta$ with stellar
mass did not also appear as a correlation with $UV$ luminosity.  Given
the correlation found between SFR and stellar mass in high-redshift
galaxies (e.g., Stark et al.\ 2009; Gonz{\'a}lez et al.\ 2011; McLure et
al.\ 2011; Lee et al.\ 2012; Reddy et al.\ 2012b), we would expect the
$UV$ luminosity to be broadly correlated with stellar mass.  A similar
conclusion can be drawn from high-redshift angular correlation
function results.  Higher luminosity galaxies are consistently found
to be more clustered than lower luminosity galaxies (e.g., Ouchi et
al.\ 2004b; Lee et al.\ 2006).  Such would not be the case if $UV$
luminosity was not correlated with mass (in this case halo mass).
Independent of these considerations, we remark that $\beta$ also shows
a clear correlation with $UV$ luminosity in various cosmological
hydrodynamical simulations (e.g., Finlator et al.\ 2011; Dayal \&
Ferrara 2012), and the predicted trends (e.g.,
$d\beta/dM_{UV}\sim-0.10$ is expected in the Finlator et al.\ 2011
simulations) are comparable to what we find
(Figure~\ref{fig:bestfit}).
\vskip 0.15cm

\noindent \textit{Possible Biases in the Finkelstein et
  al. measurements:} Finkelstein et al.\ (2012) have suggested that
the $\beta$-$M_{UV}$ trends we find may be stronger than what they
find due to the fact that we measure the $UV$ luminosity at a
different rest-frame wavelength than they do, and $\beta$ vs. $M_{UV}$
trends may depend on this rest-frame wavelength.  For our $z\sim4$
samples, for example, we measure the rest-frame $UV$ luminosity at
2041\AA (see Table~\ref{tab:pivotbands}) while Finkelstein et
al.\ (2012) measure it at 1500\AA.  We are in full agreement that
$d\beta/dM_{UV}$ will depend on the rest-frame wavelength where the
$UV$ luminosity is derived.  However, the results from Figure 3 in
Labb{\'e} et al.\ (2007) suggest one would find an even stronger
$\beta$ vs. $dM_{UV}$ trend at bluer wavelengths than one would find
at redder wavelengths, which is different from what Finkelstein et
al.\ (2012) find.  It is therefore not clear this explains the
differences.

Instead, one might be concerned that the $\beta$ vs. $dM_{UV}$ trend
Finkelstein et al.\ (2012) find may be biased as a result of the
rest-frame wavelength Finkelstein et al.\ (2012) use to measure the
$M_{UV}$ luminosity.\footnote{Finkelstein et al. (2012) also discuss
  this issue at some length in their paper (as a source of differences
  between the $\beta$ vs. $M_{UV}$ trends we find) and would appear to
  find a similar effect, but given its importance for understanding
  differences between our results, we feel this discussion is worth
  repeating.}  By determining luminosity at the blue end (at 1500\AA)
of the wavelength baseline they use to derive $\beta$, Finkelstein et
al.\ (2012) effectively introduce a coupling between the errors that
affect both their $\beta$ measurements and their determinations of the
$UV$ luminosity $M_{UV}$.  This could be problematic since any errors
in the flux measurements of sources would cause sources to be either
fainter and redder or brighter and bluer, causing the $d\beta/dM_{UV}$
trend derived by Finkelstein et al.\ (2012) to be biased towards too
high of values.  Repeating the determination of $d\beta/dM_{UV}$ at
$z\sim4$-5 based on our own flux measurements but basing $M_{UV}$ on
the flux measurement at the blue end of the wavelength baseline to
derive $\beta$ and using only the wide-area CANDELS+ERS sources, we
estimate that this could bias the derived $d\beta/dM_{UV}$ trend too
high by $\Delta(d\beta/dM_{UV})\sim0.05$.  This bias is analogous to
the photometric error coupling bias we discuss in Appendix D.
However, instead of the coupling being between source selection and
the $\beta$ measurements, it is between the measurement of $\beta$ and
the measurement of the $UV$ luminosity $M_{UV}$.\footnote{Note that
  this same issue does not substantially bias the $\beta$ vs. $M_{UV}$
  trends we find because we take the $UV$ luminosity to be the
  geometric mean of the luminosity measurements that contribute to our
  $\beta$ measurements (\S3.4).  Flux measurements on either end of
  the baseline used to derive $\beta$ bias the $\beta$ vs. $M_{UV}$
  trend in opposite directions and should largely cancel.}
Contamination in the shallower CANDELS+ERS samples (from lower
redshift galaxies) could also be an issue for Finkelstein et
al.\ (2012) in deriving the trend in $\beta$ to lower luminosities
$M_{UV}$.

Given the much smaller flux uncertainties for $z\sim4$-5 sources in
the HUDF (and smaller contamination rates), we would expect the
$\beta$'s Finkelstein et al.\ (2012) measure there to show
significantly smaller biases than $\beta$'s measured for sources in
ERS+CANDELS fields at the same luminosities.  Encouragingly enough,
the median $\beta$'s Finkelstein et al.\ (2012) derive from the HUDF
are in good agreement with our own, particularly in the luminosity
interval [$-$20 mag, $-$18 mag] (\textit{compare the blue squares and
  large red circles in both the upper and lower panels of
  Figure~\ref{fig:lumdepf}}).

Finally, we remark Finkelstein et al.\ (2012) use the same photometry
both to select sources and measure $\beta$.  Given the argumentation
in the previous section and Appendix D (see also Dunlop et al.\ 2012),
we might expect the bias in the derived $\beta$'s to be non-zero.  However,
in practice, given the large number of passbands used to select
sources and measure their redshifts, the bias is likely to be quite
small except at $z\sim7$ (similar to our study: but see \S4.8).
Finkelstein et al. 2012 are aware of this issue and explicitly discuss
it in their paper.

\subsection{How blue are lower luminosity galaxies at $z\sim7$?}

Galaxies with the most extreme $UV$ properties are expected to lie at
very high redshift and have low luminosities -- given the early cosmic
times in which they are observed and likely low masses.  It has
therefore been of considerable interest to establish the
$UV$-continuum slopes $\beta$ for the faintest observable $z\sim7$-8
galaxies.  Early observations of such galaxies in the HUDF gave
tantalizingly steep values of the $UV$-continuum slope $\beta$, i.e.,
$\beta\sim-3$.

In the present study, we find a mean $UV$-continuum slope $\beta$ of
$-2.7\pm0.2$ for faint $z\sim7$ galaxies.  This is slightly redder
than the mean $UV$-continuum slope $\beta$ ($-3.0\pm0.24$) found in
our earlier study of $\beta$ for $z\sim7$ galaxies in the HUDF.  It is
also slightly redder than that ($-3.0\pm0.5$) found by Finkelstein et
al.\ (2010) using the same data.\footnote{Despite an apparent
  difference in the quoted uncertainties on the mean $\beta$, the
  Bouwens et al.\ (2010a) and Finkelstein et al.\ (2010) error
  estimates are actually quite similar, if considered over the same
  magnitude interval and using similarly-sized samples.  Finkelstein
  et al.\ (2010) consider a sample that is half as small and 0.25 mag
  fainter -- which not surprisingly results in a larger quoted
  uncertainty for $\beta$.}  Dunlop et al.\ (2012) have argued that it
is not possible to estimate the $UV$-continuum slope $\beta$ at such
low luminosities, but as we discuss in \S4.6, such measurements are
possible if care is taken to minimize biases by ensuring the
information used for selection is independent of that used to measure
$\beta$.

How robust are our measurements of the mean $UV$-continuum slope
$\beta$ for faint $z\sim7$ galaxies?  While our simulations suggest
the biases are not large, it is useful to check this result by
obtaining an independent estimate of the mean $\beta$.  For this
estimate, we use two completely independent data sets to select
sources and to measure $\beta$.  Source selection is done using the
first-year WFC3/IR observations over the HUDF (18 orbits $Y_{105}$, 16
orbits $J_{125}$, 28 orbits $H_{160}$) from the HUDF09 program while
the measurement of $\beta$ is done using the second-year WFC3/IR
observations over the HUDF (18-orbit $J_{125}$, 25-orbit $H_{160}$
data).  Since the second-year data were not used to select the sample,
we can make an unbiased measurement of the $UV$-continuum slope
$\beta$ for this sample using the new observations.\footnote{We note
  that this measurement could still be affected (at the level of
  $\Delta \beta\sim0.1$) by the selection volume bias (Appendix B.1.1)
  even though the photometric error coupling bias (Appendix B.1.2)
  will be zero.}  For comparison with the Bouwens et al.\ (2010a)
study on $\beta$, we use the same $z\sim7$ sample.  The biweight mean
$\beta$ we derive for our lowest luminosity ($M_{UV,AB}\sim-18.5$)
subsample is $-2.8\pm0.2$.  While this is slightly redder than the
$\beta=-3.0\pm0.2$ we find from the first-year observations (Bouwens
et al.\ 2010a), this completely independent and unbiased estimate does
suggest the $UV$-continuum slopes $\beta$ for lower luminosity
galaxies at $z\sim7$ are very blue.

As one final check on the mean $\beta$ for faint $z\sim7$ galaxies, we
considered one variation on the previous test.  We divided the
$J_{125}$-band data for each HUDF09 field (HUDF09, HUDF09-1, HUDF09-2)
into two disjoint subsets and produced separate $J_{125}$-band
reductions from each.  The $Y_{105}$-band data and first third of the
$J_{125}$-band data were used for the selection of $z\sim7$ sources,
and the $H_{160}$-band data and final two-thirds of the $J_{125}$-band
data were used to measure the $UV$-continuum slopes $\beta$.  As in
the previous test, this is to ensure that the information used for
source selection is completely independent of that used for the
$\beta$ measurements, and therefore the photometric error coupling
bias (discussed in \S4.6 and Appendix B.1.2) must be zero.  The
biweight mean $\beta$ we derive for the faintest $z\sim7$ sample
($M_{UV,AB}\sim-18.3$) based on the three HUDF09 fields (HUDF09,
HUDF09-1, and HUDF09-2) is $-2.7\pm0.2$, again consistent with our
other estimates.

This new determination of the mean $UV$-continuum slope $\beta$ for
very low luminosity $z\sim7$ galaxies is very blue (i.e.,
$\beta\sim-2.7\pm0.2$).  However, this does not appear to be
especially anomalous.  In fact, it appears to be consistent with what
one might expect extrapolating the $z\sim4$-6 $UV$-continuum slope
$\beta$ relationship to $z\sim7$ (see e.g. dashed line in the $z\sim7$
panel to Figure~\ref{fig:colmag}).  The observed correlations of
$\beta$ with both redshift and luminosity are such that one would
expect faint $z\sim7$ sources to be very blue.  This blue $\beta$ is
also not inconsistent with what one can achieve with standard stellar
population modeling (where the $UV$-continuum slope $\beta$ can become
as steep as $\sim$$-$2.7: Schaerer 2003; Bouwens et al.\ 2010a;
Robertson et al.\ 2010).  The observed $UV$-continuum slopes $\beta$
at $z\sim7$ therefore seem to provide no particularly compelling
evidence for exotic stellar populations, i.e., very low metallicity
stellar populations or a high escape fraction (see also Finkelstein et
al.\ 2012).  Bouwens et al.\ (2010a) briefly speculated as to what such
blue $\beta$'s might imply, if future observations confirmed that
$\beta$ was really as blue as $-3$ with small uncertainties.

\section{Discussion}

\begin{figure}
\epsscale{1.15} \plotone{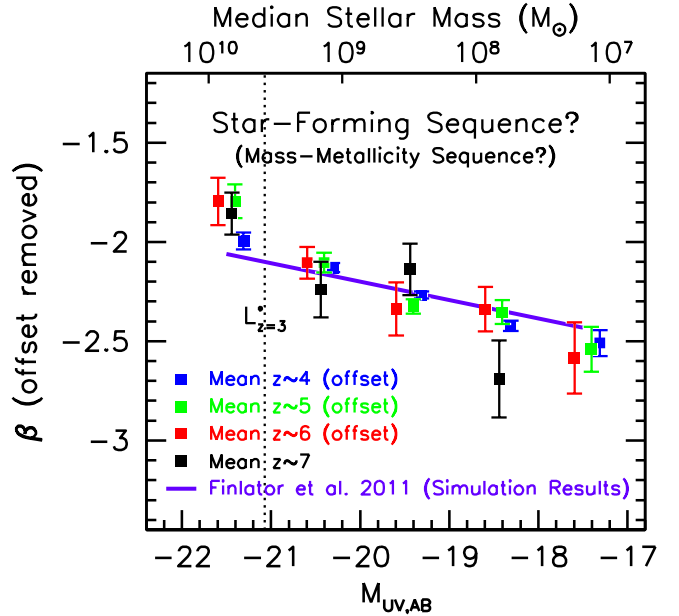}
\caption{Biweight mean $UV$-continuum slope $\beta$ versus the
  absolute magnitude in the rest-frame $UV$.  The mean $UV$-continuum
  slope results at $z\sim4$ (\textit{blue}), $z\sim5$
  (\textit{green}), $z\sim6$ (\textit{red}), and $z\sim7$
  (\textit{black}) from Figure~\ref{fig:colmag} are summarized here
  and compared with the $UV$-continuum slopes $\beta$ expected by the
  Finlator et al.\ (2011) cosmological hydrodynamical simulations at
  $z\sim6$-7 (\textit{solid purple line}).  The top axis gives the
  median stellar mass Gonz{\'a}lez et al.\ (2011) found for galaxies
  at a given rest-frame $UV$ luminosity.  Outside of the range
  $-21<M_{UV,AB}<-18.5$, these stellar masses are an extrapolation of
  the trends found by Gonz{\'a}lez et al.\ (2011).  The mean slopes
  $\beta$ found at $z\sim4$, $z\sim5$, and $z\sim6$ are offset to
  bluer values by a uniform $\Delta \beta \sim 0.1$-0.3 offset to show
  how similar the dependence of $\beta$ on luminosity is at redshifts
  $z\sim4$, $z\sim5$, and $z\sim6$ where the $UV$-continuum slopes are
  the most well defined.  The existence of some change (or offset) in
  $\beta$ versus redshift is plausible -- as both the age and dust
  properties of galaxies could easily be a function of cosmic time.
  The dependence of $\beta$ on luminosity at $z\sim7$ appears to be
  stronger, but is consistent with the other redshifts at $1\sigma$.
  The Finlator et al.\ (2011) results are shifted $\Delta\beta\sim0.10$
  bluer to better illustrate their similarity with the observed
  trends.  Table~\ref{tab:bestfit} and Figure~\ref{fig:bestfit}
  provide our best-fit determinations of how the $UV$-continuum slope
  $\beta$ depends on luminosity at $z\sim4$-7.  The similar luminosity
  dependencies strongly argue that the observed luminosity dependence
  is real and indicative of a sequence in star-forming galaxies at
  $z\sim4$-7 (\S5.1).\label{fig:starformseq}}
\end{figure}

\subsection{Sequence in SF Galaxies at High Redshift}

In the previous section, we presented evidence that the $UV$-continuum
slopes $\beta$ of star-forming galaxies at high redshift were
distributed along a well-defined sequence in $UV$ luminosity.  The
intrinsic scatter in $\beta$ along the sequence is small
($\sigma_{\beta} \sim 0.34$), and the dependence of $\beta$ is such
that galaxies become bluer towards lower luminosities.  Such a
sequence is particularly prominent in our $z\sim4$ sample, but all of
our higher redshift selections ($z\sim4$, $z\sim5$, $z\sim6$,
$z\sim7$) show strong evidence for such a sequence as well
(Figure~\ref{fig:colmag}).

The fact that we observe the same dependence of $\beta$ on $UV$
luminosity in each of four Lyman-Break selections suggests the trend
we are recovering from the observations is real.  Such a
color-magnitude relationship was already evident in many studies of
galaxies over the redshift range $z\sim2$-5 -- though the clearest
evidence was presented by Bouwens et al.\ (2009) for $z\sim2.5$ and
$z\sim4$ samples and Labb{\'e} et al.\ (2007) for $z\sim1.05$,
$z\sim1.8$, and $z\sim2.7$ samples.  See \S4.4.  The present work
confirms these trends and extends them to $z\sim5$, $z\sim6$, and
$z\sim7$.

Consistent with these trends, the $UV$-continuum slopes $\beta$ we
measure for the most luminous (and presumably most massive) galaxies
have similarly red values for $\beta$ of $-1.6$ to $-1.9$ in all four
redshift samples examined here (Table~\ref{tab:uvslope}: similar
values were also found by Lee et al.\ 2011 and Willott et al.\ 2012)
while the lowest luminosity galaxies probed have relatively similarly blue
values for $\beta$ of $-2.2$ to $-2.7$ for all four samples.

The existence of well-defined sequences for both star-forming and
evolved galaxies at lower redshift is now very well established.  In
the local universe ($z\sim0.1$), for example, Salim et al.\ (2007)
find that galaxies fall along a well-defined sequence in SFR versus
stellar mass.  Similar star-forming sequences were found by Noeske et
al.\ (2007) and Martin et al.\ (2007) at somewhat higher redshifts,
from $z\sim0.2$ to $z\sim1$ (with a 0.3 dex scatter in the SFRs).
Daddi et al.\ (2007) find evidence for such a sequence in star-forming
galaxies at $z\sim2$, and Elbaz et al.\ (2011) show that such a
sequence exists for even more luminous systems from recent Herschel
observations.  While the present color-magnitude sequence we observe
is not a SFR vs. stellar mass sequence, the existence of such a
sequence suggests that galaxies build up in a relatively well-defined
way, versus cosmic time.

Theoretically, we would expect such a sequence due to the build-up in
metals and dust anticipated to occur as galaxies grow in luminosity
and mass.  One useful illustration of this can be seen in some recent
work by Dav{\'e} et al. (2006) and Finlator et al.\ (2011) who use
smooth particle hydrodynamics to model the evolution of galaxies to
$z\sim6$.  In Figure 6 of Dav{\'e} et al. (2006), for example, we see
a clear mass-dependence in the metallicity of galaxies, with $\sim$0.3
dex change in metallicity per $\sim$1 dex change in mass.  

Figure 7 of Finlator et al.\ (2011) shows the expected trends in
$UV$-continuum slopes $\beta$ as a function of luminosity, including
the effects of starlight, dust, and emission lines.  Finlator et
al.\ (2011) predict mean $UV$-continuum slopes of $\sim$$-$2.02 for
luminous ($M_{UV,AB}\sim-20.5$) $z\sim7$ galaxies and $\sim$$-$2.28
for lower luminosity ($M_{UV,AB}\sim-18.5$) $z\sim7$ galaxies --
equivalent to an approximate slope to the $\beta$ - $M_{UV}$
relationship of just $\sim$$-$0.13.  A fit to the $\beta$ vs. $M_{UV}$
relationship for all the $z\sim4$, $z\sim5$, $z\sim6$, and $z\sim7$
sources in the Finlator et al.\ (2011) simulations yield
$d\beta/dM_{UV}$ slopes of $\sim$$-$0.10, $\sim$$-$0.08,
$\sim$$-$0.13, and $\sim$$-$0.09, respectively (Finlator 2011, private
communication).  The mean slope to this relationship $d\beta/dM_{UV}$
of $\sim-0.10$ is in excellent agreement with what is observed
(Figure~\ref{fig:bestfit} and Table~\ref{tab:uvslope}).  To better
illustrate this, we include a comparison of the $UV$-continuum slopes
$\beta$ observed with that predicted from the Finlator et al.\ (2011)
simulations (Figure~\ref{fig:starformseq}).

Rest-frame optical studies of $z\sim4$-6 galaxies with Spitzer IRAC
have provided evidence for a similar sequence at $z\geq4$ for
star-forming galaxies (Stark et al.\ 2009; Labb{\'e} et al.\ 2010b;
Gonz{\'a}lez et al.\ 2011).  Typical star-forming galaxies at $z\sim5$
have rest-frame $UV$-optical colors of $\sim$0.6 mag, with a scatter
of 0.5 dex (Gonz{\'a}lez et al.\ 2011, 2012).  These $UV$-optical
colors appear to show a slight dependence on luminosity, in the sense
that brighter galaxies are redder and fainter galaxies are bluer
(Gonz{\'a}lez et al.\ 2012).  So far there is no evidence for
evolution in the UV-optical colors over the redshift range $4<z<6$
(Stark et al.\ 2009; Labb{\'e} et al.\ 2010b; Gonz{\'a}lez et
al.\ 2011).  Both trends in the $UV$-optical colors parallel those
found in the $UV$-continuum slopes.  Together these findings suggest
that the evolution of galaxies at high-redshift may be self-similar
(see also Gonz{\'a}lez et al.\ 2012).

\begin{figure}
\epsscale{1.15}
\plotone{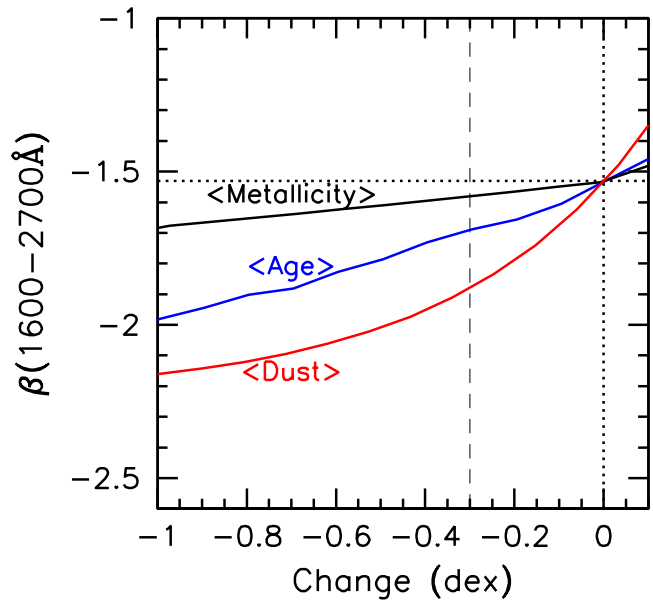}
\caption{Sensitivity of the $UV$-continuum slope $\beta$ (wavelength
  baseline 1600\AA$\,\,$to 2700\AA) to changes in the mean
  metallicity, age, or dust extinction of a galaxy population.  For
  our fiducial model (where $\beta\sim-1.5$), we assume $t = 70$ Myr,
  $\tau=10$ Myr, $[Z/Z_{\odot}]=-0.7$, $E(B-V)=0.15$, and a Salpeter
  IMF (where the star formation history is parametrized as
  $e^{-t/\tau}$) from the Papovich et al.\ (2001) fits to $z\sim2.5$
  $U$-dropouts from the WFPC2 HDF North.  In modifying our fiducial
  model to have younger ages, we make changes to both $t$ and $\tau$.
  Factor of 2 (0.3 dex) changes in the mean metallicity, age, or dust
  content of galaxies result in 0.07, 0.15, 0.35 changes in the
  $UV$-continuum slope $\beta$.  Similar to Figure 7 from Bouwens et
  al.\ (2009) but for $UV$-continuum slopes derived over a wider
  wavelength baseline.  This wavelength baseline is appropriate given
  our procedure for determining $\beta$ using flux information over a
  wide wavelength baseline (\S3.3; Figure~\ref{fig:examp};
  Table~\ref{tab:pivotbands}).  It seems clear that changes in the
  mean dust content of galaxies at high-redshift likely have the
  biggest effect on the $UV$-continuum slope $\beta$ and setting up
  trends with luminosity and possibly redshift
  (\S5.2).\label{fig:dbeta}}
\end{figure}

\begin{figure}
\epsscale{1.15}
\plotone{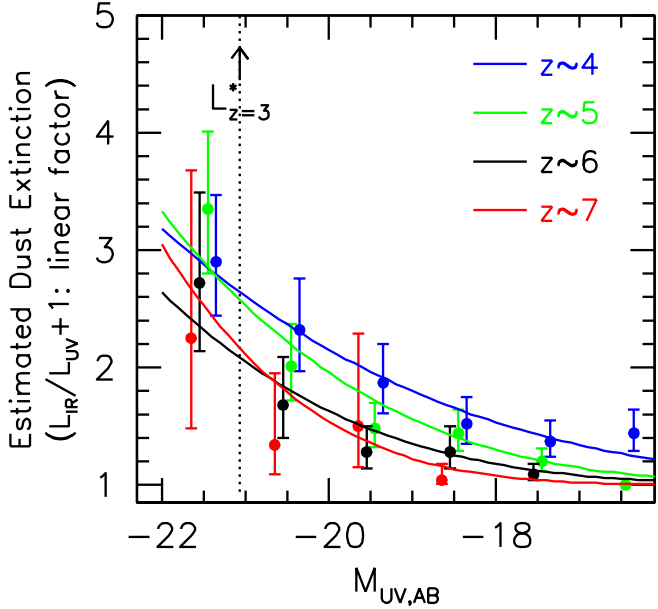}
\caption{Estimated dust extinction ($L_{IR}/L_{UV}+1$) versus $UV$
  luminosity at redshifts $z\sim4$ (\textit{blue}), $z\sim5$
  (\textit{green}), $z\sim6$ (\textit{black}), and $z\sim7$
  (\textit{red}: see \S5.3).  The dust extinction is estimated based
  on the measured $UV$-continuum slopes (Table~\ref{tab:uvslope} and
  Figure~\ref{fig:colmag}) and adopting the Meurer et al.\ (1999)
  IRX-$\beta$ relationship.  The solid circles and lines correspond to
  the results using the biweight mean $UV$-continuum slopes and linear
  fits to the biweight means (Figure~\ref{fig:colmag}).  A smaller but
  similar dust extinction would be derived based on the $z\sim0$
  Overzier et al.\ (2011) Lyman Break Analogue sample.  The
  uncertainties we estimate for the dust extinction almost exclusively
  derive from the assumed systematic errors in $\beta$, i.e., $\Delta
  \beta\sim0.10$-0.28.  The typical dust extinction inferred for
  luminous galaxies is much larger than it is for lower luminosity
  galaxies.  The dependence of the dust extinction on redshift is not
  as large as it is on $UV$ luminosity, but the dust extinction for
  higher redshift galaxies is lower than it is for lower redshift
  galaxies at the same luminosity.\label{fig:dust}}
\end{figure}

\begin{figure}
\epsscale{1.15}
\plotone{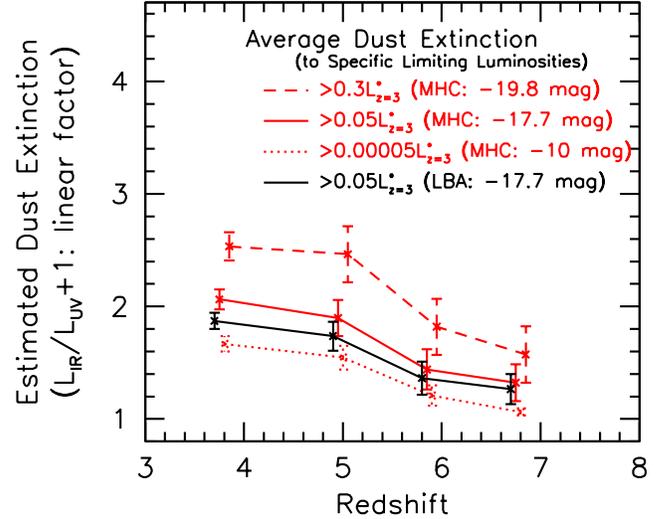}
\caption{Dust extinction ($L_{IR}/L_{UV}+1$) versus redshift (\S5.3).
  These factors are calculated from the $UV$-continuum slope $\beta$
  distribution using two different IRX-$\beta$ relationships: the
  Meurer et al.\ (1999) relationship (red lines) and the one derived
  from Lyman-Break Analogues (Overzier et al.\ 2011: black line).  The
  total correction factors for the luminosity density are integrated
  down to three different limits for the Meurer et al.\ (1999)
  IRX-$\beta$ relationship corresponding to 0.3 $L_{z=3}^{*}$ (the
  approximate limiting luminosity for searches over the GOODS fields),
  0.05 $L_{z=3}^{*}$ (the approximate limiting luminosity for searches
  over the HUDF09 data), and 0.00005 $L_{z=3}^{*}$ (the approximate
  lowest luminosity we might expect galaxies to form: e.g., Read et
  al.\ 2006; Dijkstra et al.\ 2004); these are shown with the dashed,
  solid, and dotted lines, respectively.  The corrections are weighted
  according to luminosity based on the $z\sim4$, $z\sim5$, $z\sim6$,
  and $z\sim7$ LFs from Bouwens et al.\ (2007) and Bouwens et
  al.\ (2011b).  The dust correction factors are also presented in
  Table~\ref{tab:effdust}.  This figure is similar to Figure 4 of
  Overzier et al.\ (2011) but updated to include our latest results on
  the $UV$-continuum slope $\beta$ distribution at $z\gtrsim4$.  Our
  estimated uncertainties on the dust extinction almost exclusively
  derive from the assumed systematic errors on $\beta$, i.e., $\Delta
  \beta\sim0.10$-0.28.  The correction factors we find show a slight
  dependence on redshift, but not nearly as strong as what Bouwens et
  al.\ (2009) inferred using the ACS+NICMOS observations (Figure 8 of
  Bouwens et al.\ 2009).\label{fig:dustext}}
\end{figure}

\begin{deluxetable}{cccc}
\tablecaption{The effective dust extinction (at $\sim$1600\AA)
  estimated for the LBG population integrated down to various $UV$
  luminosities (\S5.3: see also Figure~\ref{fig:dustext}).\label{tab:effdust}}
\tablehead{\colhead{} &
  \multicolumn{3}{c}{Effective Extinction} \\ 
\colhead{} &
  \colhead{$>0.3L_{z=3}^{*}$} & \colhead{$>0.05L_{z=3}^{*}$} &
  \colhead{$>0.00005L_{z=3}^{*}$} \\
\colhead{Sample} &
  \colhead{($<$$-19.8$ mag)} & \colhead{($<$$-17.7$ mag)} & \colhead{($<$$-10$ mag)}}
\startdata 
\multicolumn{4}{c}{Using Meurer et  al.\ (1999) Relationship\tablenotemark{a,b}} \\ 
$z\sim 4$ & 2.5$_{-0.1}^{+0.1}$$_{-0.4}^{+0.4}$ & 2.1$_{-0.1}^{+0.1}$$_{-0.3}^{+0.3}$ & 1.7$_{-0.1}^{+0.1}$$_{-0.2}^{+0.2}$ \\
$z\sim 5$ & 2.5$_{-0.2}^{+0.3}$$_{-0.4}^{+0.4}$ & 1.9$_{-0.2}^{+0.2}$$_{-0.3}^{+0.3}$ & 1.5$_{-0.1}^{+0.1}$$_{-0.2}^{+0.2}$ \\
$z\sim 6$ & 1.8$_{-0.2}^{+0.2}$$_{-0.3}^{+0.4}$ & 1.4$_{-0.2}^{+0.2}$$_{-0.2}^{+0.2}$ & 1.2$_{-0.1}^{+0.1}$$_{-0.1}^{+0.1}$ \\
$z\sim 7$ & 1.6$_{-0.2}^{+0.3}$$_{-0.4}^{+0.6}$ & 1.3$_{-0.1}^{+0.2}$$_{-0.3}^{+0.4}$ & 1.1$_{-0.0}^{+0.0}$$_{-0.1}^{+0.1}$ \\
\multicolumn{4}{c}{----------------------------------------------------------------------} \\
$z\sim 3$ & 4.8$_{-0.8}^{+1.0}$$_{-1.3}^{+1.7}$ & 3.5$_{-0.6}^{+0.7}$$_{-0.9}^{+1.3}$ & 2.4$_{-0.3}^{+0.4}$$_{-0.6}^{+0.9}$ 
\enddata

\tablenotetext{a}{The effective dust extinctions given here are the
  multiplicative factors, i.e., $L_{IR}/L_{UV}+1$, needed to correct the
  observed $UV$ luminosity densities at $\sim$1600\AA$\,\,$to their
  intrinsic values, after integrating to specific limiting
  luminosities (specified at the top of each column).  These
  extinctions are estimated using Meurer et al.\ (1999) IRX-$\beta$
  relationship and integrating over the distribution of $UV$-continuum
  slopes observed (Table~\ref{tab:uvslope}).  Notice that the dust
  extinctions are much lower when integrated to very low luminosities
  (see also Reddy \& Steidel 2009; Bouwens et al.\ 2009).}
\tablenotetext{b}{Both random and systematic errors are quoted
  (presented first and second, respectively).}
\end{deluxetable}

\subsection{Sequence in SF Galaxies: Interpreting $\beta$ vs. luminosity trends}

In the previous section, we briefly discussed the well-defined
sequence in $\beta$'s we observed versus luminosity in our $z\sim4$-7
samples as another instance of a ``star-forming'' sequence for
galaxies.  How shall we interpret the changes we observe in the mean
$UV$-continuum slope $\beta$ of galaxies on this sequence versus their
$UV$ luminosity?

Given the approximate correlation of galaxy mass with $UV$ luminosity
(e.g., Ouchi et al.\ 2004b; Lee et al.\ 2006, 2009; Stark et
al.\ 2009) and gradual build-up of galaxies in mass, we would expect
the mean properties of galaxies to change gradually as a function of
their $UV$ luminosity.  While we can imagine many properties of
galaxies driving changes in the $UV$-continuum slope $\beta$ as a
function of $UV$ luminosity, i.e., dust, age, metallicity, AGN
content, changes in the dust content of galaxies would likely have the
largest effect.  Figure~\ref{fig:dbeta} provides a simple illustration
of this.  0.3 dex changes in the dust content have a much larger
effect on the $UV$-continuum slope $\beta$ than similarly-sized
changes in the age, metallicity, or the stellar IMF.

Moreover, given the mass-metallicity relationship observed at
$z\sim0$-4 (e.g., Tremonti et al.\ 2004; Erb et al.\ 2006a; Maiolino
et al.\ 2008), we would expect galaxy metallicity -- and hence dust
content -- to show a strong correlation with mass (and luminosity).
Higher luminosity galaxies would largely be redder because of their
greater dust content while lower luminosity galaxies would be bluer
due to a scarcity of dust.  Such a correlation of dust content with
mass has been explicitly shown (e.g., Figure 18 from Reddy et al. 2010
vand Figure 5 from Pannella et al. 2009).  We would, of course, also
expect changes in the metallicity and age of star-forming galaxies to
contribute to the observed trends in $\beta$; however their effect on
the $UV$-continuum slope would likely be much smaller in general.  See
also Bouwens et al.\ (2009: \S4.4) and Labb{\'e} et al.\ (2007).

\subsection{Sequence in SF Galaxies: Dust Extinction}

Based upon the above tests and discussion, we will assume that dust
extinction is the dominant variable in setting up these observed
trends.  We therefore use our results on the $UV$-continuum slope
$\beta$ distribution to estimate a mean dust extinction for
high-redshift galaxies.  We will make use of well-known IRX-$\beta$
relationships known to work well at $z\sim0$ (e.g., Meurer et
al.\ 1999; Burgarella et al.\ 2009; Overzier et al.\ 2011) and
$z\sim2$ (e.g., Reddy \& Steidel 2004, Reddy et al.\ 2006a, 2010,
2012a; Daddi et al.\ 2007).  The canonical $z=0$ IRX-$\beta$
relationship (Meurer et al.\ 1999) is
\begin{equation}
A_{1600} = 4.43 + 1.99\beta.
\label{eq:mhc}
\end{equation}
where $A_{1600}$ is the dust extinction at 1600\AA.  The Meurer et
al.\ (1999) approach is functionally equivalent to correcting for dust
extinction based upon the Calzetti et al.\ (2000) dust curve.

Of course, use of the Meurer et al.\ (1999) IRX-$\beta$ relation at
$z\geq3$ has not been without controversy, and there has been
suggestions that the dust extinction in high redshift galaxies may be
either higher or lower than that implied by the Meurer et al.\ (1999)
relationship.  Certainly we might expect some change given that it is
likely that AGB stars -- thought to be the principal sites for the
formation of dust -- will not be present in the universe until the
universe is at least 1 Gyr old, and therefore the dust that existed in
the first 1 Gyr of the universe must have formed in another way, e.g.,
in the winds of supernovae (e.g., Maiolino 2006; Maiolino et
al.\ 2008).  The attenuation curve for dust from AGB stars may be very
different from dust of other origin (e.g., from SNe).

Dust obscuration would be higher in the high redshift universe if the
attention curve were flatter than Calzetti et al.\ (2000) while the
obscuration would be lower if the attenuation curve were steeper than
Calzetti et al.\ (2000), i.e., much more like that from the SMC.  A
flatter (steeper) attenuation curve implies a higher (lower) dust
extinction for a given UV slope.  Arguments for its being flatter come
from efforts to derive the dust properties of QSOs (Gallerani et
al.\ 2010) while arguments for its being steeper follow from studies
of very young galaxies at $z\sim2$-3.  Both Reddy et al.\ (2006b) and
Siana et al.\ (2008, 2009) find that dust corrections implied by the
$UV$-continuum slopes $\beta$ of young galaxies are much too large for
the Meurer et al.\ (1999) IRX-$\beta$ relation to apply.  Chary \&
Pope (2011) argue for low dust extinction in high-redshift galaxies
based upon extragalactic background light stacking results.  It has also
been argued that the Carilli et al.\ (2008) stacking results of the
radio emission in $z\sim3$ galaxies from COSMOS also suggest lower
values for the dust extinction, but Reddy et al.\ (2012a) dispute this,
arguing that the Carilli et al.\ (2008) results are consistent with
previous results (which support the Meurer et al.\ 1999 IRX-$\beta$
relationship in the mean).

The above arguments aside, there is circumstantial evidence that dust
obscuration in high-redshift galaxies is likely at least as large as
implied by the Meurer et al.\ (1999) relationship (see also discussion
in \S6.3).  Perhaps the strongest piece of evidence for substantial
dust extinction is provided by the large number of high mass
($\sim$1-3$\times10^{10}$ $M_{\odot}$) galaxies found at $z\sim5$-6 in
the GOODS fields (Eyles et al.\ 2005; Yan et al.\ 2005,
2006).\footnote{Nonetheless, it is worth noting that early estimates
  of the stellar masses for $z\sim6$ galaxies may have been somewhat
  too high as a result of their neglecting to account for the effect
  of rest-frame optical emission lines on the observed IRAC fluxes.}
Building up these stellar masses by $z\sim6$ requires SFRs of
$\sim$20-30 $M_{\odot}$ / yr, even assuming constant SFRs for 1 Gyr.
Without dust extinction, the progenitors to these massive galaxies
would need to be very luminous indeed, i.e., $-22$ AB mag (Yan et
al.\ 2006).  However, such galaxies are not observed at $z\gtrsim7$ in
the requisite numbers (e.g., Bouwens et al.\ 2011b), arguing that the
progenitors must be moderately dust obscured or have existed in a
smaller form (i.e., having subsequently merged).

One other possible uncertainty in the dust corrections regard the
possible impact of scatter in the IRX-$\beta$ relationship.  Smit et
al. (2012) found that scatter in this relationship could potentially
have a modest effect, i.e., $\sim$0.09 dex, on the estimated dust
extinction, but it depended in detail on the precise cross
correlations between dust extinction, $\beta$, UV luminosity, and the
SFR.  Since the effect of scatter in the IRX-$\beta$ relation on the
mean dust extinction is not at all clear (and plausibly consistent
with no net change), we will not consider a correction at this time.

Utilizing Eq.~\ref{eq:mhc} and the observed distribution of
$UV$-continuum slopes (Table~\ref{tab:uvslope}), we can estimate the
mean extinction corrections ($L_{IR}/L_{UV}+1$) as a function of $UV$
luminosity.  We have plotted the results in Figure~\ref{fig:dust} for
the four different redshift intervals considered here.  As in other
work (Bouwens et al.\ 2009; Reddy \& Steidel 2009; Sawicki 2012), we
find that the typical dust extinction in galaxies increases
systematically as a function of the $UV$ luminosity.  Note that in
calculating these mean extinction factors we integrate over the full
$UV$-continuum slope distribution.  This distribution is approximated
as a normal distribution with the biweight means given in
Table~\ref{tab:uvslope}) and $1\sigma$ scatter of 0.34 (the median
scatter presented in Table~\ref{tab:uvslope}).  We take $A_{1600}=0$
when $A_{1600}<0$ in Eq.~\ref{eq:mhc} above.

To determine the actual extinction corrections that are appropriate
for real samples, we must weight these extinction corrections
according to the $UV$ LFs determined at $z\sim4$, $z\sim5$, $z\sim6$,
and $z\sim7$ integrated to specific limiting luminosities.  We will
make use of the $z\sim4$, $z\sim5$, $z\sim6$, and $z\sim7$ LFs of
Bouwens et al.\ (2007) and Bouwens et al.\ (2011b).  The results are
presented in Table~\ref{tab:effdust} and Figure~\ref{fig:dustext}.  We
also include the dust extinction we would derive using an alternate
IRX-$\beta$ relationship $A_{1600} = 4.01 + 1.81\beta$ derived by
Overzier et al.\ (2011).  Overzier et al.\ (2011) derive this
relationship based upon a small sample of $z\sim0$ galaxies with
similar properties to $z\sim2$-3 Lyman-Break galaxies.

The extinction corrections given in Table~\ref{tab:effdust} and
Figure~\ref{fig:dustext} are an update to the estimates we previously
provided in Bouwens et al.\ (2009) based upon the best $UV$-continuum
slope $\beta$ estimates available at that time.  Relative to the
Bouwens et al.\ (2009) extinction estimates, the most significant
change is in the extinction estimates we find at $z\sim4$ which are
now $\sim$1.5-2$\times$ lower (see also Castellano et al.\ 2012).
This is the direct result of the somewhat bluer $UV$-continuum slopes
$\beta$ we find using the present ACS+WFC3/IR photometry.  Such
photometry extends over a wider wavelength baseline than is available
using ACS alone (as employed by Bouwens et al.\ 2009) and therefore
allows for much more accurate estimates of $\beta$.

\begin{figure}
\epsscale{1.15}
\plotone{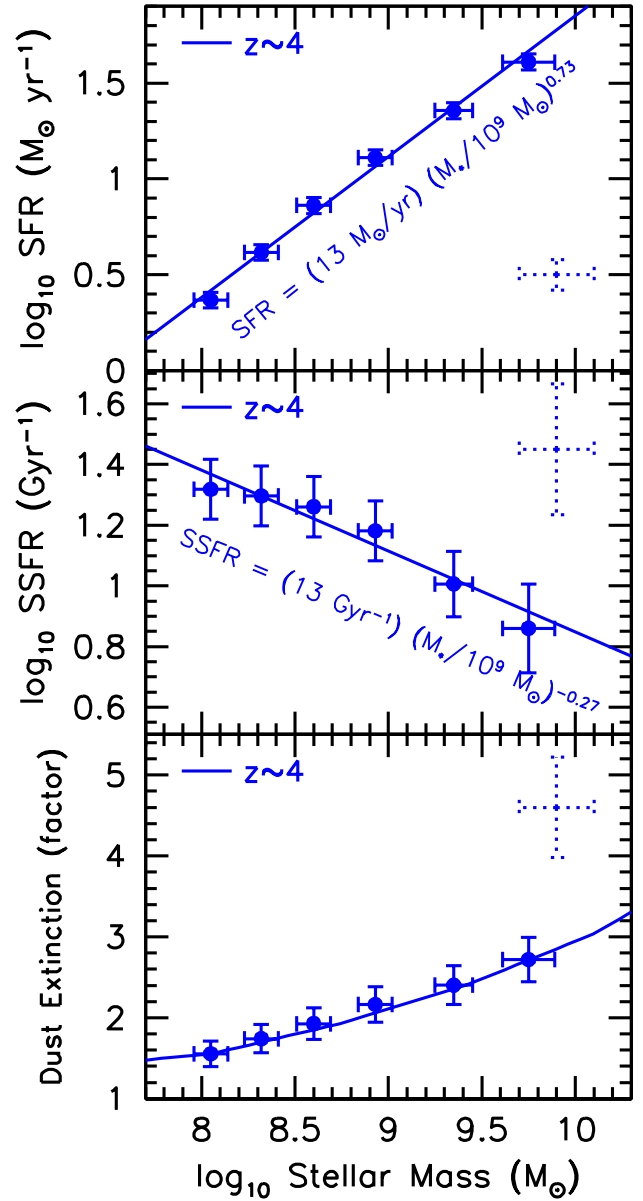}
\caption{\textit{(upper}) The relationship between the star formation
  rate and stellar mass of galaxies at $z\sim4$ (\S5.5).  The points
  correspond to the mean SFRs and median stellar masses for galaxies
  with absolute $UV$ magnitudes ranging from $-$21.0 mag and $-$18.5
  mag, in 0.5 mag intervals.  The stellar masses are from Gonz{\'a}lez
  et al.\ (2011) and the SFRs are derived from $UV$ luminosities
  adopting the dust corrections tabulated in Table~\ref{tab:effdust}.
  The error bars shown are random errors.  The dotted error bar
  included to the lower right on the plot is representative of the
  typical systematic error that likely applies to the SFR and stellar
  mass estimates.  The blue line is a linear fit to the points, with
  $\textrm{SFR} \sim (13_{-5}^{+7} \textrm{M}_{\odot}/\textrm{yr})
  (M_{*}/10^{9} \textrm{M}_{\odot})^{0.73\pm0.32}$.  The relationship
  between SFR and stellar mass shown here is expected to be
  representative for galaxies that are luminosity selected.  The
  stellar masses may be up to a factor of 2 higher using M/L ratios
  from a mass-based selection (but the exact correction depends upon
  uncertain details of the SF histories: see discussion in \S5.5).
  \textit{(middle)} The relationship between the specific star
  formation rate and the stellar mass.  The best-fit SFR-stellar mass
  relationship from the top panel is presented here in terms of the
  SSFR (\textit{black line}: $\textrm{SSFR} \sim (13_{-5}^{+7}
  \textrm{Gyr}^{-1}) (M_{*}/10^{9}
  \textrm{M}_{\odot})^{-0.27\pm0.32}$).  The error bars are as in the
  top panel.  The SSFRs may be up to a factor of 2 lower using M/L
  ratios from a mass-based selection (see discussion in \S5.5).
  \textit{(lower)} Extinction correction ($L_{IR}/L_{UV}+1$) we apply
  as a function of stellar mass.  The extinction correction is based
  on the $UV$-continuum slope $\beta$ distribution observed and the
  Meurer et al.\ (1999) IRX-$\beta$ relationship.  The solid line is
  derived using the best-fit $\beta$ and M/L ratio versus luminosity
  relationship at $z\sim4$ (Gonz{\'a}lez et al.\ 2011).  The error
  bars are as in the top panel.
\label{fig:sfrmass}}
\end{figure}

\subsection{Sequence in SF Galaxies: Interpreting $\beta$ vs. redshift trends}

The above discussion highlights several clear trends that are present
in the properties of star-forming galaxies at $z\sim4$-7 as a function
of their $UV$ luminosity and presumably as a function of their stellar
mass.  The focus of this discussion was the $UV$-continuum slopes
$\beta$ of galaxies -- and by inference -- their overall dust
extinction.

However, we also noted that the observed $\beta$'s in the apparent SF
sequence showed a modest dependence on the redshift of the sources.
At face value, this suggests that the dust extinction in galaxies must
increase, from high redshift to low redshift (as per the discussion in
\S5.2-\S5.3).  We might expect such an evolution based on the gradual
build-up of both metals and mass in star-forming galaxies with cosmic
time, as seen in the evolution of the mass-metallicity relationship
(e.g., Tremonti et al.\ 2004; Erb et al.\ 2006a; Maiolino et
al.\ 2008; Mannucci et al.\ 2009; Laskar et al.\ 2011), or based on
the evolution in the observed correlation between dust extinction and
bolometric luminosity from $z\sim2$ to $z\sim0$ (e.g., Reddy et
al.\ 2006b, 2010; Buat et al.\ 2007).  Such a change in the mean dust
content of galaxies is also supported by the simulations of Finlator
et al.\ (2011) who predict essentially the same evolution in mean
$\beta$ with redshift (Figure~\ref{fig:bestfit}), and almost all of
the change in $\beta$ ($\gtrsim$75\%) comes from a change in the dust
content (Finlator 2011, private communication).

Of course, it is always possible that the observed evolution in the
$UV$-continuum slopes $\beta$ might be due to a change in the overall
dust composition or extinction curve with cosmic time, as might occur
if the dust composition depended on the age of the stellar population
in a galaxy.  Indeed, we might expect some change in the dust
composition of galaxies as a result of the fact that dust from SNe
would be expected to form much earlier in the lifetime of a galaxy
than dust from AGB stars (e.g., Maiolino et al.\ 2004; Maiolino 2006;
Reddy et al.\ 2006b; Gallerani et al.\ 2010; Finkelstein et al.\ 2012).
While it seems clear that such changes may affect the colors of very
young star-forming galaxies, it is not clear how important they are in
driving the trends we observe with cosmic time.  After all, even
without considering such changes in dust composition, the detailed
cosmological hydrodynamical simulations of Finlator et al.\ (2011) are
successful in reproducing the approximate evolution in $\beta$ we
observe.

\subsection{Sequence in SF Galaxies: SFR - Stellar Mass Relationship}

In the previous sections, we saw that the overall shape of the SED for
star-forming galaxies -- both in the UV continuum slope and the
UV-optical colors (Gonz{\'a}lez et al. 2012) -- exhibits a very
similar dependence on luminosity at all redshifts that we examined
$z\sim4$, 5, 6, and 7 -- suggesting there is a standard star-formation
sequence for galaxies at high redshift.  

\begin{figure}
\epsscale{1.15}
\plotone{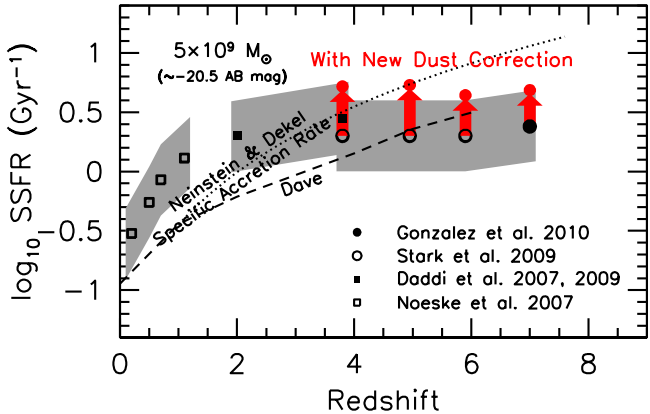}
\caption{Specific Star Formation Rate (SFR / $M_{*}$) for luminous,
  intermediate mass ($\sim5\times10^9$ $M_{\odot}$) galaxies versus
  redshift (\textit{solid blue circles}: \S5.6).  The star formation
  rates are estimated using the canonical $UV$ luminosity-to-SFR
  conversion factors (Madau et al.\ 1998; Kennicutt 1998) and the
  extinction factors given in Figure~\ref{fig:sfrmass}.  Previous
  estimates of the specific star formation rate from Noeske et
  al.\ (2007: \textit{open black squares}), from Daddi et al.\ (2009:
  \textit{solid black squares}), from Stark et al.\ (2009:
  \textit{open black circles}), and Gonz{\'a}lez et al.\ (2010:
  \textit{solid black circle}) are also shown.  The gray shaded
  regions provide an illustration of the rough systemic uncertainties
  in the SSFRs quoted in the literature.  The red arrows show the
  changes in the SSFRs we would derive including the current estimates
  of the dust extinction at $z\sim4$-7.  Earlier measurements of the
  $UV$-continuum slopes $\beta$ at $z\sim5$-7 (Bouwens et al.\ 2009;
  Stanway et al.\ 2005) were sufficiently blue that no dust
  corrections were previously applied in estimating the SSFRs at high
  redshift.  The SSFRs at $z\sim4$-7 may be up to a factor of 2 lower
  using M/L ratios from a mass-based selection (but the exact
  correction depends upon uncertain details of the SF histories: see
  also discussion in \S5.5 and Appendix B of Reddy et al.\ 2012b).
  The SSFR seems to evolve much more gradually with cosmic time in the
  observations than in many theoretical models, e.g., Neinstein \&
  Dekel (2008: \textit{dotted line}) and Dav{\' e} (2008:
  \textit{dashed line}), but see however Krumholz \& Dekel (2011).
  The approximate SSFR-scaling in typical models scales as the
  specific accretion rate of gas (proportional to $(1+z)^{2.5}$:
  Neinstein \& Dekel 2008).\label{fig:ssfr}}
\end{figure}

In this section, we derive an approximate relationship between the SFR
and stellar mass of galaxies that reside on the ``star-forming
sequence.''  To do this, we utilize the observed $UV$ luminosities,
$UV$-continuum slopes, and M/L ratios inferred from the observations.
We transform the $UV$ luminosities into SFRs by using the canonical
Kennicutt (1998) and Madau et al.\ (1998) $UV$ luminosity-to-SFR
conversion factor.  A dust correction is made at different $UV$
luminosities using the Meurer et al.\ (1999) IRX-$\beta$ relationship
and the $UV$-continuum slope $\beta$ distribution determined in this
paper.  As we have seen, these dust corrections are quite significant
($\sim$3$\times$) at higher luminosities around $L_{z=3}^{*}$, but are
essentially zero at lower luminosities ($\lesssim0.1L_{z=3}^{*}$:
$\gtrsim-18.5$ AB mag).

Stellar masses can be calculated using the luminosity-dependent M/L
ratios derived by Gonz{\'a}lez et al.\ (2011).  Gonz{\'a}lez et
al. (2011) derived these M/L ratios utilizing the HST optical + HST
near-IR + Spitzer IRAC photometry for a large sample of $z\sim4$
galaxies within the CDF-South ERS field (including those sources that
are not individually detected in the IRAC observations).  Gonz{\'a}lez
et al.\ (2011) find that the M/L ratios scale as $(M/L_{UV})\propto
L_{UV} ^{0.7}$.  Gonz{\'a}lez et al.\ (2011) find a steeper dependence
on luminosity than found by Stark et al.\ (2009), where the M/L ratio
scale as $(M/L_{UV})\propto L_{UV} ^{0.175}$.  The $z\sim4$ results
should be fairly indicative of the results at higher redshift given
the lack of clear evolution in the M/L ratio from $z\sim7$ to $z\sim4$
for galaxies at a fixed UV luminosity ($M_{UV}\sim-22$ to $-18$ mag:
Stark et al.\ 2009; Labb{\'e} et al.\ 2010a,b; Gonz{\'a}lez et
al.\ 2011).

The resulting SFRs and stellar masses for galaxies in various $UV$
luminosity bins are shown in the top panel of Figure~
\ref{fig:sfrmass}.  Fitting a line to these points in log-log space,
we find that the SFR varies as $(13_{-5}^{+7} \textrm{M}_{\odot}
\textrm{yr}^{-1}) (M_{*}/10^{9} \textrm{M}_{\odot})^{0.73\pm0.32}$.
Without any dust correction, we find that the SFR varies as
$(6_{-2}^{+3} \textrm{M}_{\odot} \textrm{yr}^{-1}) (M_{*}/10^{9}
\textrm{M}_{\odot})^{0.59\pm0.32}$.  The equivalent results for the
SSFR and dust extinction are shown in the lower two panels of
Figure~\ref{fig:sfrmass} (see also \S5.6).

Interestingly the SFR vs.  stellar mass relationship we derive
including the effects of dust extinction is much more linear than we
would derive without it.  Without any dust correction, the results of
Stark et al.\ (2009) and Gonz{\'a}lez et al.\ (2011) imply that
$\textrm{SFR}\propto M^{0.85}$ and $\textrm{SFR}\propto M^{0.59}$,
respectively.  However, correcting for dust extinction, these
relationships become a much more linear $\textrm{SFR}\propto M^{1.05}$
and $\textrm{SFR}\propto M^{0.73}$, respectively.  An approximately
linear proportionality, i.e., SFR vs. $M_{*}$, is exactly what is
expected from cosmological hydrodynamical simulations (e.g., Dav{\'e}
et al.\ 2006; Finlator et al.\ 2011; Dayal \& Ferrara 2012).  Indeed,
it points to a scenario where galaxies build up exponentially with
time along a well-defined star-forming sequence (e.g., Stark et
al.\ 2009; Papovich et al.\ 2011; Lee et al.\ 2011).

We emphasize that the SFRs and stellar masses shown in
Figure~\ref{fig:sfrmass} (and also quoted for the ``star-forming
sequence'') are representative values for a
\textit{luminosity}-selected sample.  They were derived using the
\textit{median} M/L ratios found in specific bins of $UV$ luminosity
(Gonz{\'a}lez et al.\ 2011).  In general, one would expect different
results for the M/L ratios using \textit{mass} rather than
\textit{luminosity}-selected samples.  Different results are also
expected using \textit{mean} rather than \textit{median} M/L ratios.
In particular, a mass selection would yield higher values for the M/L
ratios (though the size of the effect will depend substantially on the
scatter in the M/L ratios).  Also use of \textit{mean} rather than
\textit{median} M/L ratios for these calculations would increase the
quoted masses.  The reason is that \textit{medians} will not account
for the large amounts of mass in the tail of the distribution that
extends to high masses.  Both effects work in the same sense and would
tend to increase the masses of galaxies at each point on the sequence.
Overall, by correcting for these effects, we would expect somewhat
higher stellar masses for galaxies on the sequence and somewhat lower
values of the SSFRs.  The precise corrections depend, of course, upon
the duty cycle for star formation and scatter in the M/L ratios, but
factor of $\sim$2 corrections would not be surprising.  Reddy et
al.\ (2012b) include an extended discussion of the effect of source
selection on observed SFR vs. stellar mass relations in their Appendix
B and Figure 26.

\subsection{Sequence in SF Galaxies: Evolution in the SSFR}

The results of the previous section allow us to update previous
estimates of the specific star formation rate at high redshift to
include a correction for the dust extinction.  The specific star
formation rate -- the star formation rate divided by the stellar mass
-- has been of considerable interest recently due to the evidence that
the specific star formation rate may not evolve very rapidly at high
redshift (Stark et al.\ 2009; Gonz{\'a}lez et al.\ 2010), in
significant contrast to that expected from theory (e.g., Bouch{\'e} et
al.\ 2010; Dav{\'e} 2010; Dutton et al.\ 2010; Weinmann et al.\ 2011;
but see Krumholz \& Dekel 2011).

One shortcoming of these early SSFR determinations at high redshift
(Stark et al.\ 2009; Gonz{\'a}lez et al.\ 2010) was that no dust
correction was applied (Gonz{\'a}lez et al.\ 2010).  Such an approach
seemed appropriate at the time, given the very blue $UV$-continuum
slopes $\beta$ observed for galaxies in the redshift range $z\gtrsim5$
(Bouwens et al.\ 2009) and large uncertainties on those $UV$-continuum
slopes (e.g., Bouwens et al.\ 2009).  However now that sufficiently
deep, wide-area near-IR data are available we can establish the
$UV$-continuum slopes $\beta$ and approximate dust corrections more
accurately.

Our new estimates of the dust extinction at $z\sim4$-7 allow us to
correct previous SSFR estimates at $z\sim4$-7.  The results are
presented in Figure~\ref{fig:ssfr}.  Typical corrections result in a
factor of $\sim$2-3 increase in the SSFR.

We emphasize that the above corrections to the SSFR at $z\gtrsim4$ are
very schematic in nature.  A proper determination of the SSFR in this
regime requires a fairly extensive, self-consistent analysis of
the observations.  In addition to the issue of dust extinction, other
issues that need to be considered are (1) the selection of the samples
by mass in contrast to selection by luminosity and (2) emission-line
contamination of our broadband flux measurements (e.g., Schaerer \& de
Barros 2010).

Already there are many theoretical predictions about how the SSFR
should evolve with cosmic time (e.g., Bouch{\'e} et al.\ 2010;
Weinmann et al.\ 2011; Krumholz \& Dekel 2011).  How well do these
expectations agree with our corrected SSFRs?  The evolution of the
SSFR in many models follows the specific accretion rate $\dot{M}/M$,
which scales as $(1+z)^{2.5}$ (\textit{dotted line} on
Figure~\ref{fig:ssfr}: Neinstein \& Dekel 2008; Weinmann et
al.\ 2011).  This implies a factor of $\sim$10 decrease in the
specific star formation rate from $z\sim7$ to $z\sim2$.  By
comparison, our revised estimates of the dust extinction at $z\sim4$-6
imply a SSFR that decreases by a factor of $\sim$3 to $z\sim2$.  While
this still does not match the evolution predicted by standard models,
the agreement is better.  Are there additional ingredients that might
lead to further changes?  One possiblity that has been discussed
includes accounting for metallicity dependencies in one's SFR
prescription (e.g., Krumholz \& Dekel 2011) or changes to the stellar
IMF (Dav{\'e} 2010; Schaye et al.\ 2010).

\begin{figure}
\epsscale{1.15}
\plotone{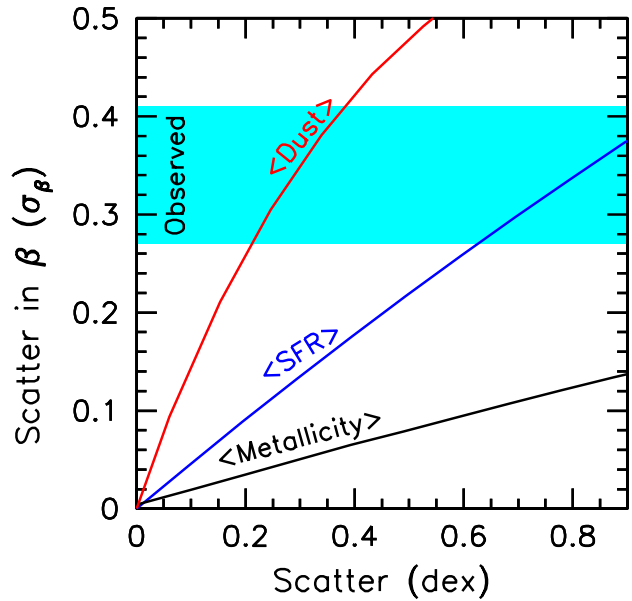}
\caption{Scatter in the $UV$-continuum slope $\beta$ distribution
  $\sigma_{\beta}$ resulting from scatter in the dust extinction,
  metallicity, or instantaneous SFRs (see \S5.7).  The typical scatter
  in the intrinsic $\beta$ distribution (after subtracting the
  contribution due to photometric uncertainties) is $\sim$0.34, as
  indicated by the cyan-shaded region (see Table~\ref{tab:uvslope}).
  In estimating the approximate scatter in $\beta$ that would result
  from variations in the dust, metallicity, or star formation rate, we
  adopt a base stellar population model an $E(B-V)$ of $\sim0.15$, a
  $[Z/Z_{\odot}]$ of $-0.7$, a Salpeter IMF, and a constant star
  formation history.  In considering variations in SF history (or
  instantaneous SFRs of galaxies), we break up the star formation
  history of each galaxy in ten 50 Myr-segments, treat each 50 Myr
  segment in the SF history as independent, and randomly choose a SFR
  for each segment from a log-normal distribution.  It is evident from
  this figure that one can approximately match the observed scatter in
  the $UV$-continuum slope distribution, by allowing for $\sim$0.3 dex
  variations in the dust content or $\sim$0.9 dex in the instantaneous
  SFR.  Substantially larger variations in dust content or
  instantaneous SFRs would introduce a larger scatter in the
  $UV$-continuum slope $\beta$ distribution -- which is inconsistent
  with the observations.\label{fig:scatter}}
\end{figure}

\subsection{Sequence in SF Galaxies: Small Scatter in the $\beta$ Distribution}

The distribution of $UV$-continuum slopes $\beta$ shows a remarkably
small intrinsic scatter $\sigma_{\beta}$ for a fixed $UV$ luminosity.
The typical scatter $\sigma_{\beta}$ observed is just $\sim$0.34
(Table~\ref{tab:uvslope}).  A similar scatter was found by Labb{\'e}
et al.\ (2007) and Bouwens et al.\ (2009) for galaxy samples at
$z\sim1$-2.7 and $z\sim4$, respectively.  Such a small scatter allows
us to set upper limits on variations in the star formation histories
and dust content of galaxies (assuming that changes in one variable do
not offset changes in other variables).

Figure~\ref{fig:scatter} illustrates the impact that scatter in dust,
age, or metallicity would have on scatter in the $\beta$ distribution.
For the typical galaxy, we assume an $E(B-V)$ of $\sim$0.15, a
metallicity $[Z/Z_{\odot}]$ of $-0.7$, a Salpeter IMF, and an
approximately constant star formation history with some scatter in the
star formation rate.  These parameters are fairly representative for
what has been found in stellar population models of luminous
$z\sim2$-4 galaxies (e.g., Papovich et al.\ 2001; Shapley et al. 2001,
2005; Erb et al. 2006b; Reddy et al.  2006b).

\begin{deluxetable*}{cccccc}
\tablewidth{13cm}
\tabletypesize{\footnotesize}
\tablecaption{$UV$ Luminosity Densities and Star Formation Rate Densities to $-17.7$ AB mag (0.05 $L_{z=3} ^{*}$: see \S6.1-\S6.2).\tablenotemark{a}\label{tab:sfrdens}}
\tablehead{
\colhead{} & \colhead{} & \colhead{$\textrm{log}_{10} \mathcal{L}$} & \colhead{$\textrm{log}_{10}$ SFR density} \\
\colhead{Dropout} & \colhead{} & \colhead{(ergs s$^{-1}$} & \colhead{($M_{\odot}$ Mpc$^{-3}$ yr$^{-1}$)} \\
\colhead{Sample} & \colhead{$<z>$} & \colhead{Hz$^{-1}$ Mpc$^{-3}$)} & \colhead{Uncorrected} & \colhead{Corrected} & \colhead{Incl. ULIRG\tablenotemark{b}}}
\startdata
$B$ & 3.8 & 26.38$\pm$0.05 & $-1.52\pm$0.05 & $-1.21\pm0.05$ & $-1.12\pm0.05$ \\
$V$ & 5.0 & 26.08$\pm$0.06 & $-1.82\pm$0.06 & $-1.54\pm0.06$ & $-1.51\pm0.06$ \\
$i$ & 5.9 & 26.02$\pm$0.08 & $-1.88\pm$0.08 & $-1.72\pm0.08$ & $-1.71\pm0.08$ \\
$z$ & 6.8 & 25.88$\pm$0.10 & $-2.02\pm$0.10 & $-1.90\pm0.10$ & $-1.90\pm0.10$ \\
$Y$ & 8.0 & 25.65$\pm$0.11 & $-2.25\pm$0.11 & $-2.13\pm0.11$ & $-2.13\pm0.11$ \\
$J$\tablenotemark{d} & 10.3 & 24.1$_{-0.7}^{+0.5}$ & $-$3.8$_{-0.7}^{+0.5}$ & $-$3.8$_{-0.7}^{+0.5}$ & $-$3.8$_{-0.7}^{+0.5}$\\
$J$\tablenotemark{d} & 10.3 & $<$24.2\tablenotemark{c} & $<-$3.7\tablenotemark{c} & $<-$3.7\tablenotemark{c} & $<-$3.7\tablenotemark{c}
\enddata
\tablenotetext{a}{Integrated down to 0.05 $L_{z=3}^{*}$.  Based upon
  LF parameters in Table 2 of Bouwens et al.\ (2011b: see also Bouwens
  et al.\ 2007) (see \S6.1).  The SFR density estimates assume
  $\gtrsim100$ Myr constant SFR and a Salpeter IMF (e.g., Madau et
  al.\ 1998).  Conversion to a Chabrier (2003) IMF would result in a
  factor of $\sim$1.8 (0.25 dex) decrease in the SFR density estimates
  given here.}
\tablenotetext{b}{See \S6.2}
\tablenotetext{c}{Upper limits here are $1\sigma$ (68\%
  confidence).}
\tablenotetext{d}{$z\sim10$ determinations and limits are from Oesch
  et al.\ (2012a: see also Bouwens et al.\ 2011a) and assume 0.8
  $z\sim10$ candidates in the first case and no $z\sim10$ candidates
  (i.e., an upper limit) in the second case.}
\end{deluxetable*}

\begin{figure*}
\epsscale{1.17}
\plotone{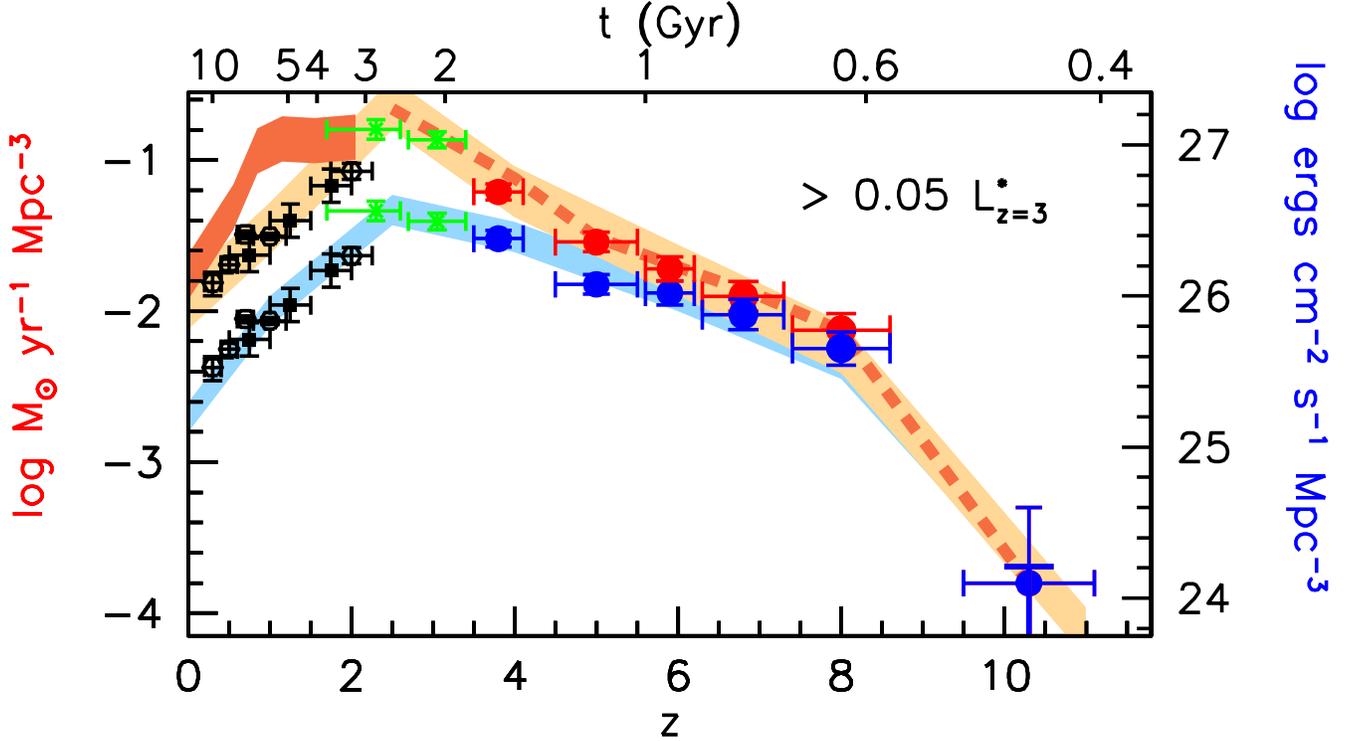}
\caption{Star formation rate density as a function of redshift
  (\S6.1-6.2: see also Table~\ref{tab:sfrdens}).  The lower set of
  blue points and blue-shaded region give the UV luminosity density
  (\textit{right axis}) and hence the SFR density before correction
  for dust extinction.  The upper set of red points and red-shaded
  region give the SFR density (\textit{left axis}), after applying our
  estimated dust corrections (see Table~\ref{tab:effdust}).  The $UV$
  luminosity density determinations are integrated down to $-17.7$ AB
  mag (0.05 $L_{z=3}^{*}$) and taken from Schiminovich et al.\ (2005:
  \textit{open black squares}) at $z\leq1$, Oesch et al.\ (2010c:
  \textit{solid black squares}; see also Hathi et al.\ 2010) at
  $z\sim1$-2.5, Reddy et al.\ (2009: \textit{green crosses}) at
  $z\sim2$-3, Bouwens et al.\ (2007, 2011b) at $z\geq4$, and Oesch et
  al.\ (2012a) at $z\sim10$.  The dark red shaded region and dark
  dashed line include the contribution from IR bright sources
  (Magnelli et al.\ 2009, 2011; Daddi et al.\ 2009).  The contribution
  from the IR bright population needs to be explicitly included in the
  SFR density estimates -- since dust corrections do not typically
  recover the total SFRs for especially luminous $>10^{12}$
  $L_{\odot}$ galaxies (e.g., Reddy et al.\ 2008).  The IR bright
  population contributes very little to the SFR density at $z\gtrsim4$
  (e.g., Bouwens et al.\ 2009).  Conversion to a Chabrier (2003) IMF
  would result in a factor of $\sim$1.8 (0.25 dex) decrease in the SFR
  density estimates given here.\label{fig:sfzulirg}}
\end{figure*}

We begin by using the observed scatter in the $UV$-continuum slope
$\beta$ distribution to set constraints on scatter in the ages (or
star formation history) of star-forming galaxies at high redshift.
One potentially promising approach is to consider star-formation
history for galaxies with stochastic variations in the SFR over 50 Myr
intervals (a typical time scale over which one might imagine the SFR
in a galaxy might be correlated).  Assuming similar variations in the
SFRs of $z>4$ galaxies to that observed at $z\sim0$-1 (where a scatter
of $\sim$0.3 dex in observed the SFR-stellar mass relationship: Noeske
et al.\ 2007) results in just a 0.14 scatter in $\beta$.  We computed
this scatter by (1) running a simulation with 1000 input galaxies, (2)
dividing up the star formation history for each galaxy into ten 50-Myr
segments, (3) randomly selecting a star formation rate for each 50-Myr
segment from a log-normal distribution with 0.3 dex scatter, (4)
computing the resultant $\beta$'s for each galaxy based on its star
formation history, and (5) calculating the scatter in the derived
$\beta$ distribution for the simulated galaxies.  For simplicity, each
galaxy in the simulation is taken to have an age of 500 Myr.

The predicted scatter in $\beta$, i.e., 0.14, is considerably less
than we observe.  We can of course increase the predicted scatter in
$\beta$ by considering star formation histories with larger variations
in the SFRs.  For example, a 0.9 dex scatter in the SFRs translates
into a 0.34 dex scatter in $\beta$ -- which is a good match to
intrinsic scatter in $\beta$ ($\sigma_{\beta}$).  This is similar to
the scatter found by Gonz{\'a}lez et al.\ (2011) in modelling the
distribution of M/L ratios for $z\sim4$ star-forming galaxies.  In the
above modeling, no account is made for changes to the total luminosity
of galaxies, as a result of a stochastic star formation history.

Scatter in the dust content can add significantly to the scatter in
the $UV$-continuum slope $\beta$, but the magnitude of the scatter
will depend directly on how dusty galaxies are in the luminosity range
one is considering.  If the dust extinction is low, for example, dust
has very little impact on the $\beta$ observed, and therefore small
multiplicative changes to the dust extinction factor would have
similarly little impact.  On the other hand, if the dust extinction is
non-negligible, multiplicative changes to the total dust extinction
factor would have a big impact on the value of $\beta$ one observes.
For the fiducial luminous galaxy at $z\sim2.5$, with an
$E(B-V)\sim0.15$ and a Calzetti et al.\ (2000) dust law, a $\sim$0.3
dex scatter in the dust extinction would result in an observed scatter
$\sigma_{\beta}$ of $\sim$0.34 (Figure~\ref{fig:scatter}).

Scatter in the metallicity adds very little to scatter in the
$UV$-continuum slopes (Figure~\ref{fig:scatter}), and therefore one
cannot use the observed scatter in the $UV$-continuum slope $\beta$
distribution to set strong limits on scatter in galaxy
metallicity.

In summary, the observed scatter in the $UV$-continuum slopes $\beta$
distribution allows us to set upper limits on variations in dust
content and instantaneous SFR of galaxies (assuming that changes in
one variable do not offset changes in other variables).  Scatter in
the dust extinction of galaxies appear to be $\lesssim$0.3 dex and
scatter in the instantaneous SFR is $\lesssim$0.9 dex.

\section{Star Formation Rate Density at High Redshift}

The availability of WFC3/IR data over both the ultra-deep HUDF09
fields and wide-area fields has allowed us to establish the
$UV$-continuum slopes to great accuracy over a wide range in redshift
and luminosity.  In the previous section, we used these $UV$-continuum
slope distributions to estimate the mean dust extinction in
star-forming galaxies at high redshift.

Here we utilize these new estimates of the dust extinction to revisit
our determinations of the SFR density at high redshift.  In \S6.1, we
begin by first determining the SFR density from those galaxies that
make up our high-redshift LBG selections.  In \S6.2, we include the
contribution from ultra-luminous ($L_{bol}>10^{12}$ $L_{\odot}$)
IR-bright galaxies (\S6.2).  We include this contribution explicitly
since dust corrections tend to underestimate the SFRs for the the most
luminous, IR-bright galaxes (e.g., Reddy et al.\ 2009) and since such
galaxies are not typically well-represented in rest-frame UV, LBG-type
selections.  Finally, in \S6.3, we compare our total SFR density
estimates with what we would infer from current measures of the
stellar mass density.

\subsection{SFR Density at High Redshift}

In this subsection, we determine the SFR density using our current
estimates of the dust extinction.  As in previous work (Bouwens et
al.\ 2009; Bouwens et al.\ 2011b), we base our SFR density
determinations on our most recent LF determinations at $z\sim4$-8
(Bouwens et al.\ 2007; Bouwens et al.\ 2011b) and search results at
$z\sim10$ (Bouwens et al.\ 2011a; Oesch et al.\ 2012a).  Luminosity
densities are derived by integrating these LFs down to $-$17.7 AB mag
(0.05 $L_{z=3} ^{*}$) which is the limit to which we probe the LFs at
both $z\sim7$ and $z\sim8$.  These luminosity densities are then
converted into SFR densities using the canonical Madau et al.\ (1998)
and Kennicutt et al.\ (1998) relation:
\begin{equation}
L_{UV} = \left( \frac{\textrm{SFR}}{M_{\odot} \textrm{yr}^{-1}} \right) 8.0 \times 10^{27} \textrm{ergs}\, \textrm{s}^{-1}\, \textrm{Hz}^{-1}\label{eq:mad}
\end{equation}
where a $0.1$-$125\,M_{\odot}$ Salpeter IMF and a constant star
formation rate of $\gtrsim100$ Myr are assumed.  Finally, for our dust
extinction estimates, we will use those from Table~\ref{tab:effdust}
calculated using the Meurer et al.\ (1999) IRX-$\beta$ relationship.

Our latest UV luminosity density and SFR density estimates are
summarized in Table~\ref{tab:sfrdens} and presented in
Figure~\ref{fig:sfzulirg}.  The new estimates are in broad agreement
with previous estimates (Bouwens et al.\ 2009), but we find a lower
SFR density at $z\sim4$, as expected, given the lower dust extinction
we infer at these redshifts.  The change is significant, with the SFR
density decreasing by a factor of $\sim$1.5-2 at this redshift (see
also Castellano et al.\ 2012).

\begin{figure*}
\epsscale{1.17}
\plotone{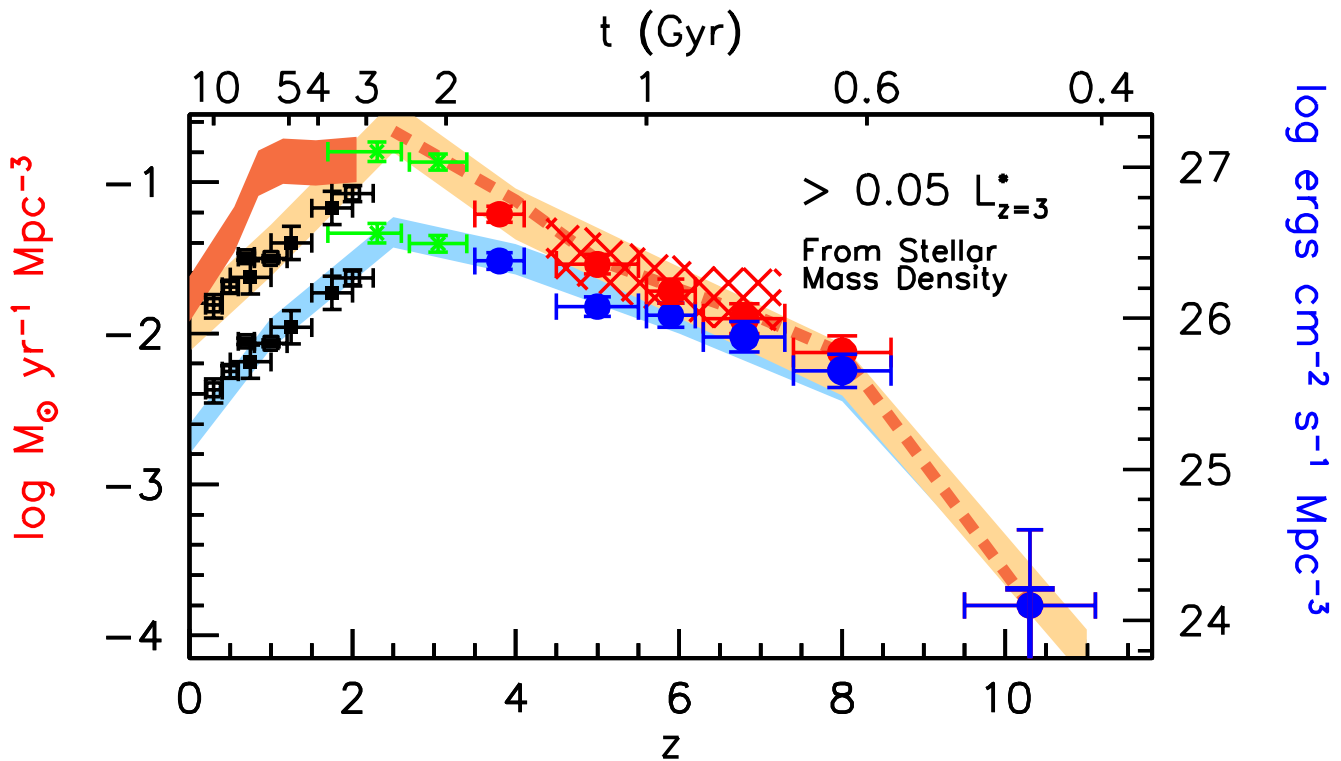}
\caption{A comparison of the derived SFR density with that implied by
  stellar mass density determinations in the literature
  (\textit{hatched red region}: Labb{\'e} et al.\ 2010b; Gonz{\'a}lez
  et al.\ 2011: \S6.3).  Published estimates of the stellar mass
  density at $z\sim8$ (Labb{\'e} et al.\ 2010b) are very uncertain at
  present and therefore not used to infer a SFR density at $z>8$.  The
  SFR density determinations are as in Figure~\ref{fig:sfzulirg} and
  Table~\ref{tab:sfrdens}.  Conversion to a Chabrier (2003) IMF would
  result in a factor of $\sim$1.8 (0.25 dex) decrease in the SFR
  density estimates given here.  Good agreement is observed between
  the SFR density and that implied by the stellar mass density.  The
  agreement is much better than in Bouwens et al.\ (2011b), as
  expected given the improvements in the $UV$-continuum slope $\beta$
  determinations at $z\gtrsim4$ (and therefore likely dust
  corrections).  \textit{Note that we actually require a dust
    correction to obtain good agreement between the SFR density and
    that implied by the stellar mass density.}  \label{fig:sfzmass}}
\end{figure*}

\subsection{SFR Density Estimates including IR Luminous galaxies}

Dusty, infrared-luminous galaxies also potentially contribute quite
meaningfully to the SFR density at high redshift.  However, it can be
quite challenging to account for this contribution on the basis of
optical/near-IR surveys with the Hubble Space Telescope.  Not only can
it be difficult to identify such sources in these surveys (due to
their faintness in the UV or red colors which cause them to be
excluded from LBG selections), but it is now well established that the
dust extinction for the most dusty, infrared luminous ($>$10$^{12}$
$L_{\odot}$) galaxies at $z\sim1$-3 cannot be accurately estimated
using the observed $UV$-continuum slopes $\beta$ and IRX-$\beta$
relationship (e.g., Reddy et al.\ 2006b; Elbaz et al.\ 2007).

To accurately account for the SFR density in this population, a better
approach is to include that population explicitly by utilizing a
luminosity function in the mid-IR/far-IR and integrating down to
$10^{12}$ $L_{\odot}$ (e.g., Reddy et al.\ 2008; Reddy \& Steidel
2009; Bouwens et al.\ 2009).  We replicate that approach here
integrating published mid-IR LFs to $10^{12} L_{\odot}$, converting
the total IR luminosity to SFR using the canonical relation in
Kennicutt et al.\ (1998), and then adding the inferred SFR densities
to the dust-corrected $UV$ SFR densities.  The IR luminosity functions
we utilize are Magnelli et al.\ (2009, 2011) to $z\sim2$ and Daddi et
al.\ (2009) at $z\sim4$.  Magnelli et al.\ (2011) utilize the full set
of deep 24$\mu$m and 70$\mu$m observations over the two GOODS fields
from the FIDEL program.  Since the $70\mu$m and $24\mu$m data that
Magnelli et al.\ (2011) utilise are significantly deeper than that
used by Caputi et al. (2007) and allow for a self-consistent
correction to the total bolometric luminosity, the Magnelli et
al.\ (2011) SFR density estimate represents a noteworthy improvement
on the Caputi et al.\ (2007) estimates we previously utilized at
$z\sim2$ (Bouwens et al.\ 2009).

Including the contribution from IR bright galaxies, we present our
total SFR density estimates in Table~\ref{tab:sfrdens} and
Figure~\ref{fig:sfzulirg}.  Interestingly, but not surprising, this
correction makes very little difference to the SFR density derived at
very high redshifts $z>4$, but adds modestly to the SFR density at
late cosmic times ($z\leq3$).  A small contribution of ULIRGs to the
SFR density at high redshifts is expected given their position at the
very end of the extended build-up process whereby galaxies gradually
acquire higher and higher masses in gas, dust and stars (e.g., Bouwens
et al.\ 2009).

\subsection{SFR Densities Implied by Stellar Mass Density Measurements}

As in our previous work (Bouwens et al.\ 2011b), we can compare our
SFR density estimates with that implied by recent stellar mass density
determinations (e.g., Stark et al.\ 2009; Gonz{\'a}lez et al.\ 2011;
Labb{\'e} et al.\ 2010a,b; Gonz{\'a}lez et al.\ 2011).  In doing so,
we consider the integrated SFR density and stellar mass density to the
same luminosity limits 0.05 $L_{z=3}^{*}$ for self
consistency.\footnote{The importance of using a consistent limit for
  this comparison was made by Reddy \& Steidel (2009) in regards to
  claims that SFR density determinations at $z\sim2$-4 might not be
  consistent with stellar mass determinations in the same redshift
  range.}

We use the following formula to infer an approximate SFR density
(SFRD) at $z\gtrsim4$ from the observed stellar mass density (SMD):
\begin{equation}
\textrm{SFRD}(z_i,z_j) = \frac{\textrm{SMD}(z_i) -
\textrm{SMD}(z_j)}{\textrm{time}(z_i)-\textrm{time}(z_j)}(1-\epsilon)^{-1}
f_{LE}
\end{equation}
where $z_i<z_j$ are the redshifts of adjacent Lyman-break samples and
$\epsilon$ is the gas recycling factor.  The $1-\epsilon$ factor
accounts for the recycling of gas mass from high-mass stars back into
the interstellar medium through SNe explosions.  Recycling results in
only a fraction of the stars formed being locked up in stellar mass,
i.e., $dM_{*}/dt = (1-\epsilon)\textrm{SFR}$ where $\epsilon=0.3$
appropriate for a Salpeter IMF (e.g., Bruzual \& Charlot 2003).  The
$f_{LE}$ factor in the above equation accounts for the fact that new
galaxies enter our magnitude-limited samples at all redshifts simply
as a result of galaxy growth.  Since these galaxies (and their stellar
mass) would not have been included in the magnitude-limited sample
just above them in redshift, these sources would cause us to
overestimate the SFR density (see \S7.4 of Bouwens et al.\ 2011b).
Accounting for this latter effect reduces the inferred SFR density by
a factor of 1.3; therefore, we take $f_{LE}$ to be $1/1.3$.  

The SFR densities implied by several recent stellar mass density
determinations (e.g., Stark et al.\ 2009; Gonz{\'a}lez et al.\ 2010;
Labb{\'e} et al.\ 2010a,b; Gonz{\'a}lez et al.\ 2011) are presented in
Figure~\ref{fig:sfzmass}, and there is remarkably good agreement over
the redshift range $z\sim4$-6.  This is a useful consistency check and
suggests that the dust extinctions we are inferring at $z\sim4$-7 are
reasonable and fit into a consistent picture.  The general agreement
we observe also points towards no clear evolution in the stellar
initial mass function (IMF) to high redshift ($z\geq4$: see also
Bouwens et al.\ 2011b; Papovich et al.\ 2011) since any changes in the
IMF would have an effect on the SFR density or stellar mass densities
we compute from the observed light and likely result in a mismatch.

The observed agreement also suggests that biases in our stellar mass
density estimates at high redshift, e.g., due to the contamination of
the rest-frame optical light probed by IRAC with strong emission lines
(Schaerer \& de Barros 2010), are not huge.  Of course, we cannot
totally rule out modest levels of contamination by emission lines,
particularly to the SEDs of $z\sim6$-7 galaxies, and in fact there may
be some evidence in the stacked SEDs of $z\sim5$-7 galaxies that
emission lines do have some effect on the IRAC fluxes (Gonz{\'a}lez et
al.\ 2012).

\section{Summary}

The recent availability of ultra-deep WFC3/IR observations over the
HUDF, CDF-South GOODS, and the two HUDF05 fields has allowed us to
measure $UV$-continuum slopes for large samples of star-forming
galaxies at $z\sim4$-7.  Such measurements can be made both for
samples of $z\sim4$-7 galaxies to very faint levels (i.e., $-17$ AB
mag) and also for much brighter samples (i.e., $-21$ AB mag).  Use of
the Lyman-break selection technique allows us to divide these sources
by redshift into four distinct redshift samples ($z\sim4$, $z\sim5$,
$z\sim6$, and $z\sim7$) and thus to quantify the changes in the
$UV$-continuum slope $\beta$ with cosmic time from $\sim$0.7 Gyr to
1.8 Gyr after the Big Bang.

This is the first time the $UV$-continuum slope $\beta$ distribution
can be derived with such small uncertainties for a large sample of
$z\sim4$-7 galaxies ($\sim$2500 galaxies).  Full use of the flux
information in the $UV$-continuum is made in determining the
$UV$-continuum slope $\beta$ for individual sources (\S3.3), except of
course those bands contaminated by emission from Ly$\alpha$ or
affected by the position of Lyman or Balmer breaks (at $\sim$1216$\AA$
or $\sim$3600$\AA$, respectively).  This resulted in $>1.5\times$
smaller uncertainties in our measurements of $\beta$ than obtained
using other techniques (see Appendix B.3).  It is essential to keep
the uncertainties in our $\beta$ measurements to a minimum if we are
to accurately characterize the scatter in the $\beta$ distribution.

Care was taken to minimize the effect of source selection and
photometric scatter on our results (Appendix B).  Such effects can
significantly bias determinations of the $UV$-continuum slope
distribution (see e.g., Dunlop et al.\ 2012), so it is crucial to
utilize techniques that minimize the bias.  We estimate the bias using
extensive Monte-Carlo simulations where we added artificial sources to
the observations and then select and measure their properties in the
same way as the real observations.  We demonstrate that we can recover
the distribution of $UV$-continuum slopes $\beta$ to very faint
magnitudes with very small biases ($\lesssim$0.1: see
Figure~\ref{fig:bias_summary} and \ref{fig:bias1}).  The very small
biases found here appear to be in significant contrast to the large
biases found by techniques that use similar information both to select
sources and measure their $UV$-continuum slopes $\beta$ (see
Figures~\ref{fig:biasdunlop} and \ref{fig:dunlop}).  In particular, as
we demonstrate through extensive simulations (Appendix D), a coupling
between source selection and $\beta$ measurement seems to have
produced the large biases reported by Dunlop et al.\ (2012: \S4.6)
towards measuring blue slopes (since sources with blue slopes show a
greater likelihood to be at $z\gtrsim5$).  In addition, we find that
we can select galaxies with $UV$-continuum slopes as red as 0.5
(Figure~\ref{fig:selcrit} and \ref{fig:selvol}), so the $UV$-continuum
slope distributions we derive should be valid over a wide range in
$UV$-continuum slope $\beta$ (i.e., $-$3.5 to 0.5).  Small corrections
were made based on the selection biases found in our simulations.

Using the above procedure, we accurately establish the distribution of
$UV$-continuum slopes $\beta$ over a wide range in both redshift and
luminosity.  This is the first time this has been possible to do so
with such precision, and we use these $UV$-continuum slope $\beta$
distributions to make inferences about how the dust properties of
galaxies likely vary with both luminosity and redshift.  We then use
these results to derive a SFR versus stellar mass sequence for
galaxies at $z\sim4$, to intepret the evolution of specific star
formation rate with cosmic time, and to compare the SFR density
results at $z\sim4$-8 with that inferred from the stellar mass
density.

Here are our primary findings:
\begin{itemize}
\item{Galaxies at high redshift lie along a well-defined
  ($\sigma_{\beta}\sim0.34$) $UV$-color versus magnitude sequence
  at all redshifts under study ($z\sim4$-7).  Previously Labb{\'e} et
  al.\ (2007) and Bouwens et al.\ (2009) presented evidence for
  similar sequences at $z\sim1$-3 and $z\sim2.5$-4.}
\item{The biweight mean $UV$-continuum slope $\beta$ shows an
  approximately linear relationship with $UV$ luminosity in all four
  redshift intervals we examine, i.e., $z\sim4$, $z\sim5$, $z\sim6$,
  and $z\sim7$. The mean $\beta$ of higher luminosity galaxies is
  redder than that found for lower luminosity galaxies in all four
  samples.  We demonstrate that this trend is not an artifact of
  source selection in our analysis, contrary to the suggestion of
  Dunlop et al.\ (2012).  See \S4.6 and Appendix B.  Similar trends
  were found by Labb{\'e} et al. (2007), Overzier et al.\ (2008),
  Bouwens et al.\ (2009), Bouwens et al.\ (2010a), Finkelstein et
  al.\ (2010), Lee et al.\ (2011), and Wilkins et al.\ (2011: \S4.4).}
\item{No statistically-significant evolution in the slope of
  $UV$-continuum slope $\beta$ - luminosity sequence is found over the
  entire redshift range $z\sim4$-7 (see Figure~\ref{fig:bestfit}).
  Particularly striking are the slopes of the $\beta$ - luminosity
  relationship at $z\sim4$, $z\sim5$, and $z\sim6$ where the slopes
  are $-$0.11$\pm$0.01, $-$0.16$\pm$0.03, and $-$0.15$\pm$0.04,
  respectively.  The slope of this relationship at $z\sim7$, i.e.,
  $-$0.21$\pm$0.07, is consistent with that at lower redshift (later
  times).  The derived slopes to the $\beta$-luminosity relationship
  are in good agreement with previous results at $z\sim3$-5 (e.g.,
  Bouwens et al.\ 2009; Overzier et al.\ 2008; Wilkins et al.\ 2011).}
\item{We observe an evolution in the intercept to the $UV$-slope
  $\beta$ vs. luminosity relationship with cosmic time, in the sense
  that higher redshift galaxies of a given $UV$ luminosity are bluer
  than lower redshift galaxies of the same $UV$ luminosity (see
  Figure~\ref{fig:bestfit}).  While such an evolution in colors had
  already been found over the redshift range $z\sim5$-6 to $z\sim3$-4
  (Lehnert \& Bremer 2003; Stanway et al.\ 2005; Bouwens et al.\ 2006,
  2009, 2010a), this confirms this result at much higher confidence.
  We remark that this evolution in the colors might have been expected
  based on the evolution seen in the dust extinction vs. bolometric
  luminosity relationship from $z\sim2$ to $z\sim0$ (e.g., Reddy et
  al.\ 2006b, 2010; Buat et al.\ 2007) and also the evolution seen in
  the mass-metallicity relationship from $z\sim3.5$ to $z\sim0$ (e.g.,
  Tremonti et al.\ 2004; Erb et al.\ 2006a; Maiolino et al.\ 2008;
  Mannucci et al.\ 2009; Laskar et al.\ 2011: see \S5.4).}
\item{We observe similarly red $UV$-continuum slopes $\beta$, i.e.,
  $-1.6$ to $-1.9$, for the most luminous galaxies (and presumably
  most massive) in all four redshift samples examined here (similarly
  red values were also found by Lee et al.\ 2011 and Willott et
  al.\ 2012).  The similarity of the observed $\beta$'s for the most
  luminous galaxies (and presumably the most massive) underscores the
  potential importance of mass in setting the properties of individual
  galaxies (see also Finkelstein et al.\ 2012).}
\item{We find further evidence that the mean $UV$-continuum slope
  $\beta$ of faint $z\sim6$-7 galaxies is very blue, i.e.,
  $\langle\beta\rangle \lesssim-2.5$ (\S4.8).  After correcting for
  the relevant biases (Appendix B), the biweight mean $UV$-continuum
  slope $\beta$ we measure for $\sim-18$ AB mag sources is
  $-2.5\pm0.2$ and $-2.7\pm0.2$ at $z\sim6$ and $z\sim7$,
  respectively.  While very blue overall (relative to the typical
  galaxy at low redshift), these $\beta$'s are not inconsistent with
  what one can achieve with conventional stellar population modelling
  and therefore do not require exotic stellar populations to explain.
  Instead, the blue $\beta$'s appear to be as expected given the
  observed trends in $\beta$ versus both redshift and luminosity.  Our
  measured $\beta$'s confirm earlier results by Bouwens et
  al.\ (2010a), Bouwens et al.\ (2010b), Oesch et al.\ (2010a), Bunker
  et al.\ (2010), and Finkelstein et al.\ (2010).}
\item{We can use the intrinsic scatter in the $UV$-continuum slope
  $\beta$ distribution (Table~\ref{tab:uvslope}) to set limits on
  variations in the dust extinction or instantaneous SFR of galaxies
  on the ``star-forming'' sequence (\S5.7).  The inferred scatter of
  $\sigma_{\beta}\sim0.34$ in the $UV$-continuum slope $\beta$
  distribution corresponds to a maximum scatter of 0.3 dex in the dust
  extinction of galaxies and 0.9 dex in the instantaneous SFRs.}
\item{We argue that changes in $UV$-continuum slope $\beta$ as a
  function of redshift and luminosity are primarily driven by changes
  in the mean dust extinction of galaxies (\S5.2, \S5.4: see e.g.
  \S4.5 of Bouwens et al.\ 2009).  Using the current determinations of
  the $UV$-continuum slope distribution at $z\sim4$-7 and the
  IRX-$\beta$ relationship at $z\sim0$ (Meurer et al.\ 1999; Overzier
  et al.\ 2011), we estimate the approximate dust extinction of
  galaxies as a function of luminosity at $z\sim4$, 5, 6, and 7 (see
  Figure~\ref{fig:dust}: see \S5.3).  We find that the dust extinction
  for galaxies at lower luminosities and high redshift is essentially
  zero (Figure~\ref{fig:dust} and \S5.3: e.g., see also Bouwens et
  al.\ 2009).}
\item{We find good agreement between the SFR density inferred from
  stellar mass density estimates and that inferred from the
  dust-corrected UV observations over the redshift range $z\sim4$-7
  (Figure~\ref{fig:sfzmass}: \S6.3).  The agreement is better than
  that found by Bouwens et al.\ (2011b) using the Bouwens et
  al. (2009) estimates of the dust correction.}
\item{We have used our new estimates of the dust extinction and
  mass-to-light estimates from the literature (e.g., Stark et
  al.\ 2009; Gonz{\'a}lez et al.\ 2011) to reexamine the relationship
  between the SFR and stellar mass $M_{*}$ of galaxies (\S5.5).  We
  find that the SFR is proportional to $M_{*} ^{0.73\pm0.32}$
  (Figure~\ref{fig:sfrmass}: see also Labb{\'e} et al. 2010a).  The
  exponent to this relationship is much closer to 1.0 including a
  correction for dust extinction than not including it.  This
  relationship is therefore plausibly close to the simple
  proportionality, i.e., SFR $\propto$ $M_{*}$, expected in many
  cosmological hydrodynamical simulations (e.g., Finlator et
  al.\ 2011).}
\item{Our new estimates of the dust extinction imply higher values of
  the SSFR at $z\sim4$-7 for the intermediate mass ($\sim5\times10^9$
  $M_{\odot}$) galaxies where this dependence was quantified (\S5.6).
  The implied change in the SSFRs at $z\gtrsim4$ is approximately a
  factor of $\sim$2-3 higher than before.  With the implied changes
  to the SSFR, the SSFR at $z\sim5$-7 is therefore plausibly higher
  than that at $z\lesssim2$ (Figure~\ref{fig:ssfr}) and hence
  plausibly evolves from $z\sim4$-7 to $z\sim2$.  The observed
  evolution is therefore in a similar sense to what is expected in
  many theoretical models (e.g., Bouch{\'e} et al.\ 2010; Dav{\'e}
  2010; Dutton et al.\ 2010; Weinmann et al.\ 2011; Krumholz \& Dekel
  2011).}
\end{itemize}
The star-forming (or mass-metallicity) sequence identified here from
the well-defined, redshift-independent $\beta$ vs. luminosity relation
can provide us with powerful constraints on the build-up and evolution
of galaxies at early times.  The modest scatter and uniform slope for
this sequence suggest a scenario in which galaxies build up and evolve
in a uniform manner for most of early cosmic time.  Luminous galaxies
on this sequence are redder than lower luminosity galaxies, due to
their larger dust extinction.  Lower redshift galaxies are also
redder, at all luminosities again most likely due to their higher dust
content.

Given the limited size and S/N of $z\sim7$ samples, the $UV$-continuum
slope distribution at $z\sim7$ is more poorly defined than at $z\leq6$
and would benefit from even deeper observations than are currently
available.  Such observations would be valuable not only for reducing
the size of current photometric errors, but also for extending our
samples to fainter magnitudes at high S/N.

In addition, we look forward to more WFC3/IR observations over the
$\sim$800 arcmin$^2$ CANDELS fields (Grogin et al.\ 2011; Koekemoer et
al.\ 2011).  These wide-area observations will allow us to accurately
establish the $UV$-continuum slope distribution to high luminosities
where the rarity of sources has made such determinations difficult in
the past.

\acknowledgements

We are happy to thank Romeel Dav{\'e}, Pratika Dayal, Harry Ferguson,
Kristian Finlator, Steve Finkelstein, Mauro Giavalisco, Roderik
Overzier, Casey Papovich, Naveen Reddy, Daniel Schaerer, Daniel Stark,
and Simone Weinmann for useful discussions.  We are greatly
appreciative to Kristian Finlator for analyzing the UV-continuum slope
distribution in his cosmological hydrodynamical simulations and
providing us with the quantitative results.  Massimo Stiavelli and
Kristian Finlator provided us with helpful feedback on our manuscript.
Extensive feedback from our referee Naveen Reddy was very useful to us
in improving the overall clarity and rigor of this manuscript.  We are
grateful to all those at NASA, STScI and throughout the community who
have worked so diligently to make Hubble the remarkable observatory
that it is today.  We acknowledge the support of NASA grant NAG5-7697,
NASA grant HST-GO-11563, and ERC grant HIGHZ \#227749; PO acknowledges
support from NASA through a Hubble Fellowship grant \#51278.01 awarded
by the Space Telescope Science Institute.

\appendix

\section{A.  Comparisons with $UV$-continuum slope measurements based on a single $UV$ color}

In the present study, we determine $UV$-continuum slopes $\beta$ from
a fit to all available flux observations in the rest-frame $UV$.  Only
passbands sufficiently redward of the Lyman Break are included to
avoid contamination from Ly$\alpha$ emission or IGM absorption (below
$\sim$1216 \AA).  The goal is to use all available flux information to
constrain the $UV$-continuum slope and to adopt the most extended
possible wavelength baseline.  This allows us to minimize the errors
on our $UV$-continuum slope measurements.

One potential difficulty with this approach is that the wavelength
baseline used to estimate the $UV$-continuum slope varies somewhat
depending on the sample.  The wavelength baseline adopted for $z\sim4$
galaxies, for example, is different from the wavelength baseline used
for $z\sim6$ galaxies (e.g., see Figure~\ref{fig:examp}) and similarly
for galaxies at $z\sim5$ and $z\sim7$ (see also
Table~\ref{tab:pivotbands}).  Because of these different wavelength
baselines, we could potentially measure different $UV$-continuum
slopes for galaxies at different redshifts even if there is no
intrinsic evolution in the underlying galaxy populations themselves.

To determine the extent to which our varying wavelength baseline
biases our $UV$-continuum slope measurements, we also estimated
$UV$-continuum slopes $\beta$ for sources in our samples using the
wavelength baseline 1600\AA$\,\,$to 2200\AA$\,\,$.  All galaxies in
our $z\sim4$-7 samples have coverage at these wavelengths and so the
$UV$-continuum slope measurements can be made in a consistent way
across all four samples.  Because of the very limited wavelength
ranges involved here, we derive the $UV$-continuum slopes from 2-3
rest-frame $UV$ bands.  We have derived the following formula to
convert the observed colors to the equivalent $UV$-continuum slope
$\beta$.  The formula are as follows:
\begin{eqnarray}
\beta = 2.94 (i_{775}-Y_{105}) - 2.00~~&\textrm{($z\sim3.8$)} \label{eq:betab}\\
\beta = 2.53 (i_{775}-(2Y_{098}+J_{125})/3) - 2.00~~&\textrm{($z\sim3.8$)} \\
\beta = 2.93 (z_{850}-J_{125}) - 2.00~~&\textrm{($z\sim5$)} \\
\beta = 3.09 ((Y_{105}+J_{125})/2 - H_{160}) - 2.00~~&\textrm{($z\sim6$)} \\
\beta = 2.79 ((Y_{098}+J_{125})/2 - H_{160}) - 2.00~~&\textrm{($z\sim6$)} \\
\beta = 4.29 (J_{125}-H_{160}) - 2.00~~&\textrm{($z\sim7$)} 
\label{eq:beta2}
\end{eqnarray}
These formulae are similar to those already presented in Bouwens et
al.\ (2010a), Dunlop et al.\ (2012), and Wilkins et al.\ (2011).

To quantify possible systematics, we compared the $UV$-continuum slopes
$\beta$ derived using our fiducial approach (using the full flux
information from the $UV$-continuum) and that estimated from our
Lyman-Break samples using the above formula.  We found that the median
$UV$-continuum slope $\beta$ using our fiducial approach is offset by
just $0.03$, $-0.15$, $0.06$, and 0.0 relative to that determined by
Eqs.~\ref{eq:betab}-\ref{eq:beta2} for our $z\sim4$, $z\sim5$,
$z\sim6$, and $z\sim7$ samples, respectively.  These offsets are
comparable in size than the $\Delta \beta\sim0.10$-0.28 uncertainties
we estimate to be present in the derived $UV$-continuum slopes $\beta$
based upon potential systematics in the photometry.  \textit{We would
  therefore expect no large biases in our $UV$-continuum slope
  measurements as a result of our use of a variable wavelength
  baseline.}  Our measurements of the $UV$-continuum slope $\beta$
should thus be both reliable and have smaller uncertainties.

\begin{figure}
\epsscale{1.1}
\plotone{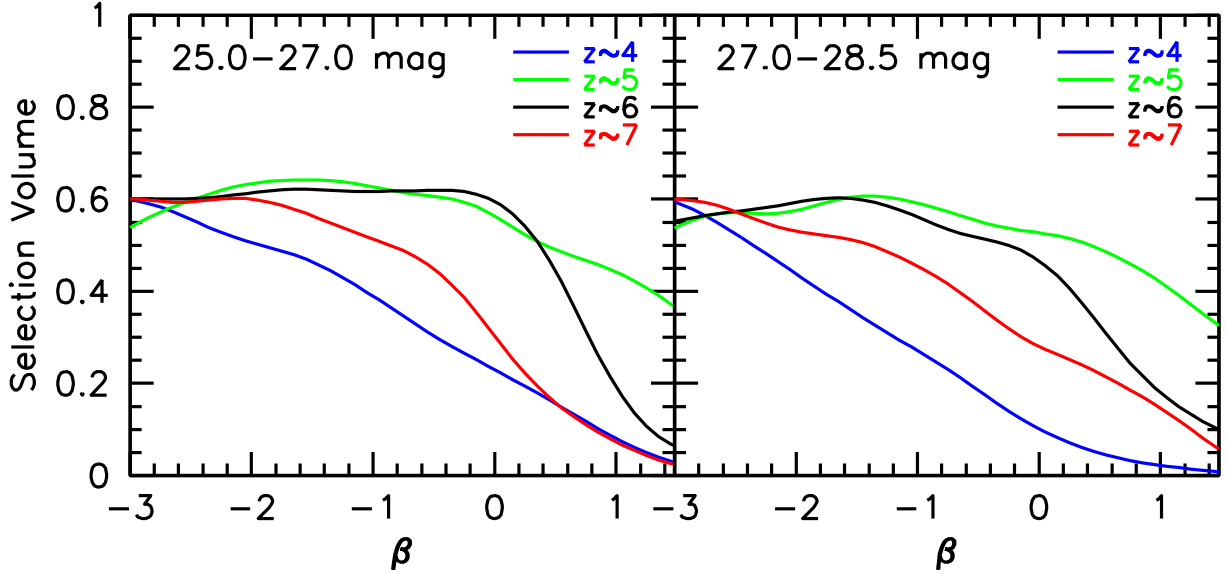}
\caption{Relative volumes available for selecting $z\sim4$, $z\sim5$,
  $z\sim6$, and $z\sim7$ galaxies versus $UV$-continuum slope $\beta$,
  given our Lyman-Break selection criteria (Figure~\ref{fig:selcrit}).
  This selection volume is calculated for the HUDF09 fields over the
  magnitude ranges 25.0-27.0 mag (\textit{left panel}) and 27.0-28.5
  mag (\textit{right panel}).  While our selections are more efficient
  at selecting galaxies with bluer $UV$-continuum slopes $\beta$,
  these selections are also effective in identifying galaxies to quite
  red $UV$-continuum slopes $\beta$.\label{fig:selvol}}
\end{figure}

\begin{figure}
\epsscale{0.7}
\plotone{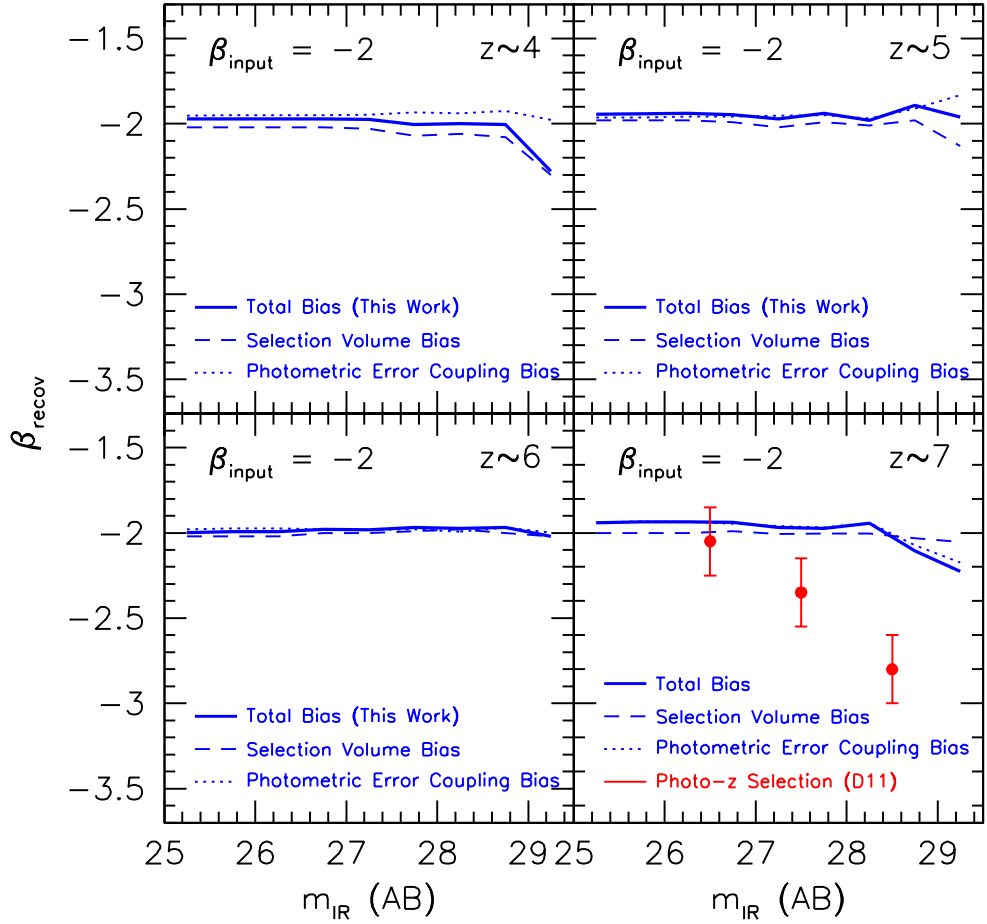}
\caption{Biases in the mean $UV$-continuum slope $\beta$ distribution
  expected to arise from $\beta$-dependent selection effects
  (\textit{dashed lines}: Appendix B.1.1) and from any coupling of the
  photometric errors in the measurement process to source selection
  (\textit{dotted lines}: Appendix B.1.2).  The panels present the
  mean $UV$-continuum slope $\beta$ we would expect to find from our
  observed Lyman-Break samples given an input $UV$-continuum slope
  $\beta$ distribution with a mean $\beta$ of $-2$.  The results are
  presented as a function of the rest-frame $UV$ magnitudes of the
  sources.  The biases expected using the ``robust'' photometric
  redshift selection of Dunlop et al.\ (2012: D11) are also included
  (\textit{red circles}: from their figure 8).  It is remarkable how
  poorly the photometric redshift techniques of Dunlop et al.\ (2012)
  perform relative to the present approach.  See
  Figure~\ref{fig:biasdunlop}, \S4.6, Appendix B.1.2, and Appendix D
  for a brief explanation for the differences between the present
  approach and that employed by Dunlop et al.\ (2012).  In general, we
  expect extremely small biases in the mean $UV$-continuum slopes
  $\beta$ derived from our samples, i.e., $\Delta \beta \lesssim 0.1$,
  relative to the actual values.  Of course, at very faint magnitudes,
  the biases are a little larger, but they are still much smaller than
  the biases suffered by photometric redshift techniques where source
  selection and $\beta$ measurements are tightly
  coupled.\label{fig:bias0c}}
\end{figure}

\begin{figure*}
\epsscale{1.15}
\plotone{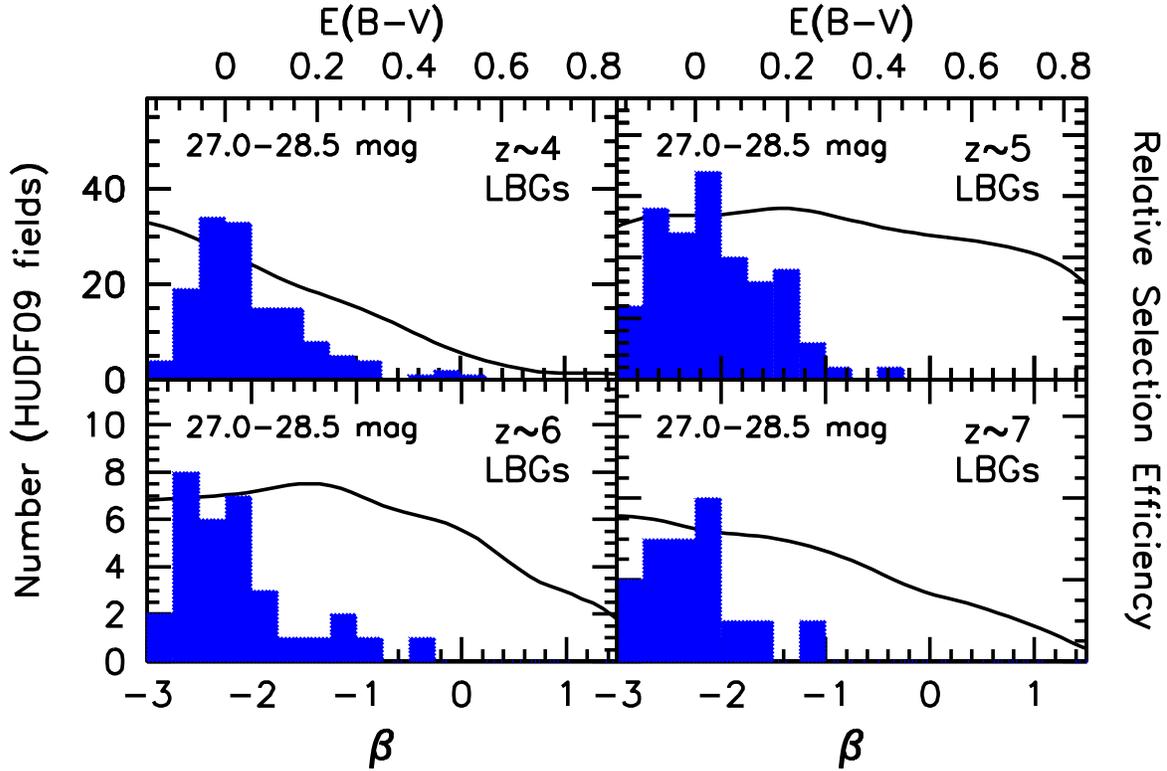}
\caption{Distribution of $UV$-continuum slopes $\beta$ for faint
  (27.0-28.5 mag) galaxies in the three HUDF09 fields (\textit{blue
    histogram}).  For comparison, the selection volume available to
  identify galaxies versus $UV$-continuum slope is also included as
  the black lines (see also Figure~\ref{fig:selvol}).  The selection volumes
  presented here are normalized to match the number of galaxies found
  near the peak of distribution.  This figure is similar to Figure 5
  of Bouwens et al.\ (2009).  A clear deficit of star-forming galaxies
  with very red $UV$-continuum slopes is observed.  This figure
  demonstrates that this deficit is almost certainly real and is not
  simply the result of selection biases.\label{fig:betadist}}
\end{figure*}

\begin{figure}
\epsscale{0.65}
\plotone{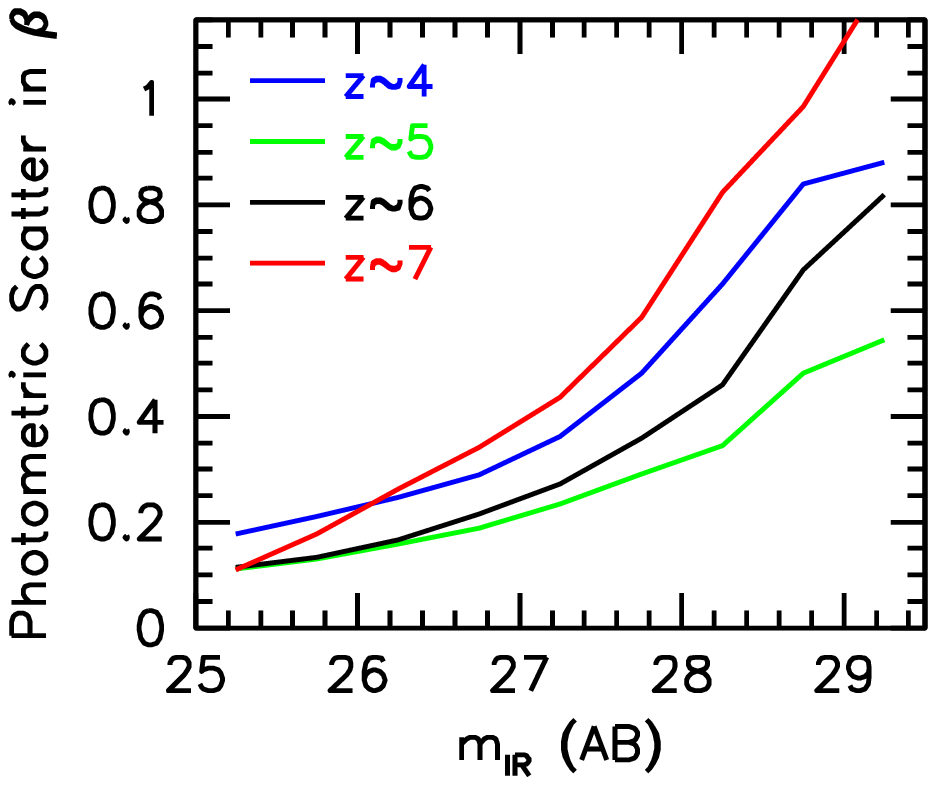}
\caption{Scatter in the observed $UV$-continuum slope $\beta$
  distribution expected to originate from photometric errors
  (``noise'').  The plotted results are for $UV$-continuum slope
  $\beta$ determinations over the HUDF.  The scatter is determined by
  adding artificial sources to the observations and then recovering
  these sources and measuring their $UV$-continuum slopes.
  Differences between the input $UV$-continuum slopes $\beta$ and that
  recovered from the observations are used to quantify the approximate
  scatter in the $UV$-continuum slope distribution introduced by
  noise.  The scatter we find in $\beta$ is much lower for lower
  redshift samples than for our highest redshift sample.  We have the
  longer wavelength baseline and larger number of broadband filters
  available to establish the $UV$-continuum slope $\beta$ at lower
  redshift.  The scatter in our $UV$-continuum slope $\beta$
  measurements for our $z\sim4$-5 samples is $\sim$1.5$\times$
  smaller using the full flux information in the rest-frame $UV$ SED
  (i.e., $i_{775} z_{850} Y_{105} J_{125}$ for our $z\sim4$ sample and
  $z_{850}Y_{105}J_{125}H_{160}$ for our $z\sim5$ sample) than using
  information in a single $UV$ color.\label{fig:bias1}}
\end{figure}

\section{B.  Corrections to the $UV$-continuum slope distribution}

Our measurements of the $UV$-continuum slopes $\beta$ are, of course,
subject to a variety of selection and measurement biases (e.g.,
Bouwens et al.\ 2009; Dunlop et al.\ 2012).  These biases include a
preferential selection for sources with very blue $UV$-continuum
slopes $\beta$ and the effect of noise in artificially increasing the
scatter in the measured $UV$-continuum slopes.  Since these effects
can have a significant effect on the derived $\beta$ distribution, we
cannot draw scientifically useful conclusions unless we determine the
magnitude of these biases on the $UV$-continuum slope $\beta$
distribution and apply a correction.

In this appendix, we quantify the approximate effect that source
selection and photometric error coupling bias have on the
$UV$-continuum slope distribution (Appendix B.1) and then derive a
correction for these biases (Appendix B.2).  In Appendix B.3, we
quantify the effect of photometric errors in increasing the overall
scatter in the $UV$-continuum slope distribution.  Finally, in
Appendix B.4, we compare the $UV$-continuum slopes $\beta$
distribution we derive from ultra-deep and wide-area data sets to
ensure that our results show a basic self consistency.  This is
important for verifying that all the relevant biases are understood
and correctly treated.

\begin{figure*}
\epsscale{1.15} \plotone{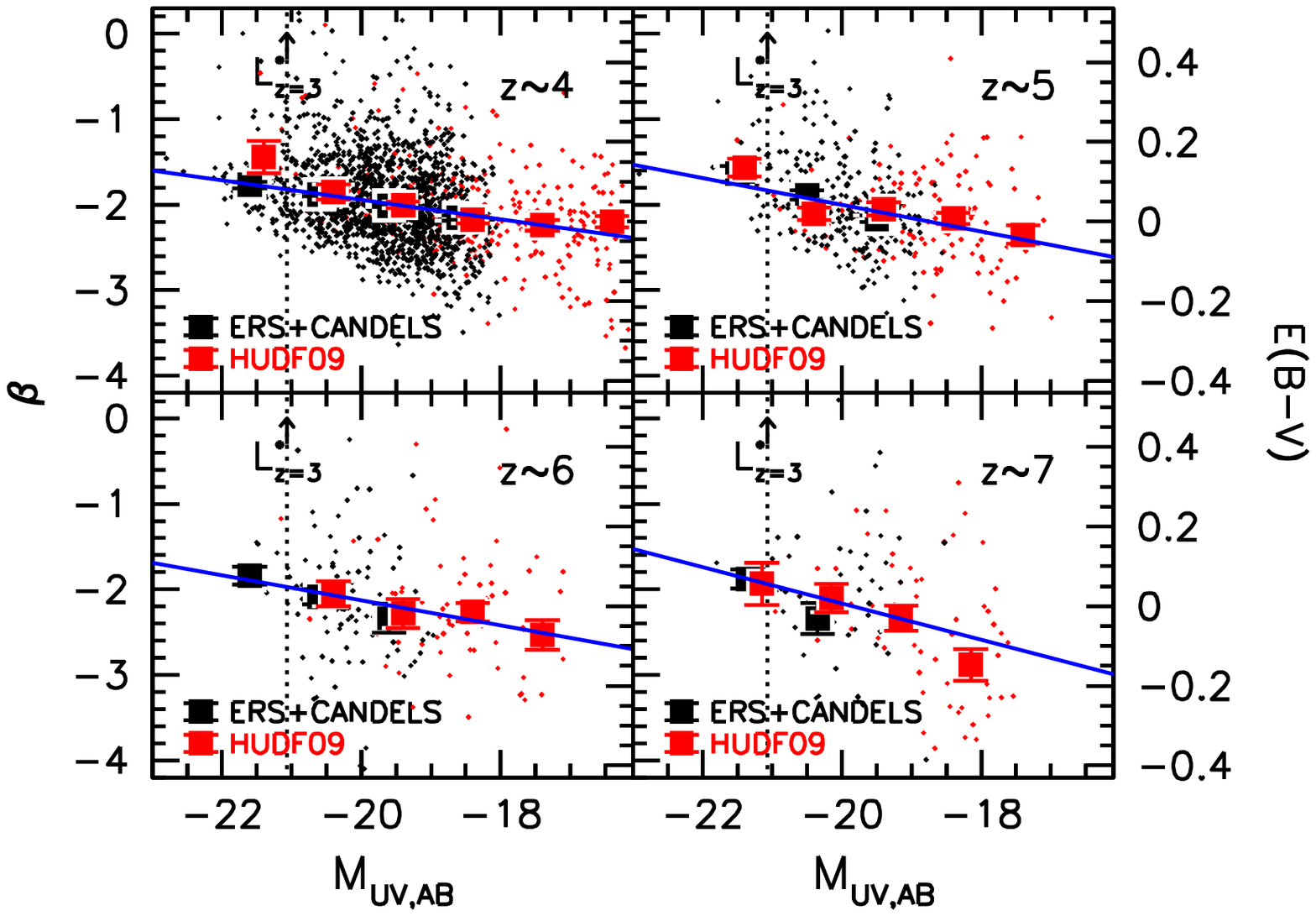}
\caption{$UV$-continuum slope $\beta$ versus $UV$ luminosity for our
  $z\sim4$, $z\sim5$, $z\sim6$, and $z\sim7$ samples.  The red points
  are sources from our HUDF09 samples and the black points are from
  our ERS+CANDELS samples.  The large solid squares show the biweight
  mean $UV$-continuum slope measured in 1.0 mag intervals.  The error
  bars give the $1\sigma$ uncertainties on the biweight mean.  The red
  and black squares are horizontally offset for clarity (to avoid
  direct overlap with each other).  The blue lines show the best-fit
  relationship between $UV$-continuum slope $\beta$ and luminosity
  presented (\S3.6).  Excellent agreement is observed between the mean
  $\beta$ determinations in our ultra-deep HUDF09 and those in our
  wide-area ERS+CANDELS selections in the luminosity range where the
  determinations overlap.  This suggests that the $\beta$ distribution
  we derive from our shallower data sets is fair and any possible
  residual biases (not accounted for in our calculated corrections:
  Appendix B.1-Appendix B.3) are small.\label{fig:colmagq}}
\end{figure*}

\subsection{B.1. Potential Biases in the Mean $\beta$}

We would expect small biases in the mean $UV$-continuum slope $\beta$
we measure from our observed samples.  These biases arise whenever
galaxies with specific properties -- intrinsic or observed -- are
easier to select than others.  We discuss two such biases: (1) one
based on the fact that galaxies with certain intrinsic colors are
easier to select than others (Appendix B.1.1) and (2) one arising from
the fact that the information used to measure $\beta$ is not always
independent from that used to select the sources (Appendix B.1.2).

The bias described in Appendix B.1.1 arises because not all intrinsic
values for $\beta$ are selected with the same efficiency, and the
bias described in Appendix B.1.2 arises because the average measured
value for $\beta$ is not always equal to the intrinsic value.

The purpose of Appendix B.1 is to provide a description of the
relevant biases and their approximate size for the most simplistic
cases (i.e., if the $\beta$ distribution is a delta function).  In
Appendix B.2, we calculate these biases assuming a more realistic
input distribution for $\beta$.

\subsubsection{B.1.1  ``Selection Volume Bias'': Biases Related to the Intrinsic Selectability 
of Sources}

In general, we would expect our high-redshift selections to be more
effective in identifying galaxies with certain $UV$-continuum slopes
than others -- resulting in small biases in the mean $UV$-continuum
slope $\beta$ derived from our samples.  Since bluer galaxies are
almost always easier to select than redder galaxies in Lyman-break
galaxy selections, this bias would be towards bluer colors.  This
effect is often referred to as ``template bias'' in the literature.

To quantify the extent to which object selection affects the
distribution of $UV$-continuum slopes recovered from the observations,
we run extensive Monte-Carlo simulations.  In these Monte-Carlo
simulations, we insert artificial galaxies into the real data with a
wide variety of luminosities and colors and then reselect these
galaxies using the same method as we used on the real sources.  We
then use these simulations to determine how the selection efficiency
and hence selection volume depends upon the intrinsic $UV$-continuum
slope $\beta$.  A detailed description of these simulations is given
in Appendix C.

Figure~\ref{fig:selvol} provides an illustration of the basic results
from these simulations in the magnitude ranges 25.0 - 27.0 mag and
27.0 - 28.5 mag.  Indeed we find the expected trends in the
simulations.  The selection efficiency is largest for galaxies with
the bluest $UV$-continuum slopes.  As in other Lyman-break selections,
the selection volumes show a sizeable decrease towards redder
$UV$-continuum slopes, especially for $\beta\gtrsim0.0$.

The effect of these biases on the mean $UV$-continuum slope $\beta$
derived from the observations however are relatively modest for
typical samples.  For example, assuming the intrinsic distribution of
$UV$-continuum slopes $\beta$ has a mean $\beta$ of $-1.5$ and
$1\sigma$ scatter of 0.3 (see Table~\ref{tab:uvslope}), the bias in
the mean $\beta$ is just $\Delta\beta\sim0.1$ for a $\sim$28 mag
galaxy in our $z\sim4$ selection.  We estimate the size of this bias
by starting with an initial distribution of $\beta$'s, weighting this
distribution by the effective volume at each $\beta$ (for a given
magnitude), and then computing the shift in the biweight mean $\beta$.

Figure~\ref{fig:bias0c} shows the effect of these selection biases on
the mean $UV$-continuum slope $\beta$ found in our $z\sim4$-7
selections.  The biases shown in Figure~\ref{fig:bias0c} assume that
the intrinsic $\beta$ distribution has a mean value of $-2.0$ and
$1\sigma$ scatter of 0.3.  While we note a bias towards bluer
$UV$-continuum slopes $\beta$, these biases are not extraordinarily
large.  Similarly small selection biases are found in our other data
sets, i.e., the two other HUDF09 fields, the ERS field, and the
CDF-South CANDELS field.

Comparing the number of sources observed with a given $\beta$ with the
computed selection volumes for this $\beta$ gives us another means of
assessing the importance of this bias.  As can be seen in
Figure~\ref{fig:betadist} for our HUDF selections, we can select
sources to much redder $\beta$'s than generally observed in our
selections.  As in our direct estimates of the bias
(Figure~\ref{fig:bias0c}), this suggests that the selection effects
are only having a modest effect on our derived $\beta$'s.

We emphasize that the biases presented in this subsection and the next
are simply intended to be illustrative; more realistic estimates of
the bias (and those values of the bias we will use to correct the
observations) are presented in Appendix B.2.

\subsubsection{B.1.2.  Photometric Error Coupling Bias}

The intrinsic selectability of sources is not the only cause for
biases in the mean $UV$-continuum slope measurements.  For many
samples, biases can arise as a result of the fact that source
selection is done on the basis of the same photometric information as
used for the $UV$-continuum slope $\beta$ measurements.  This can be
mildly problematic, since the $UV$-continuum slope $\beta$
measurements are then coupled to the selectability of a source.  If
errors in the photometry of a source make the source easier to select,
any effect these same photometric errors have on the measured $\beta$
results in biases.  Dunlop et al.\ (2012) show how significant this
effect can be using their own photometric redshift selection as a
reference (Figure~\ref{fig:biasdunlop}).

The size of this effect can be determined using the same simulations
as described above.  The measured values of $\beta$ (after selection)
are compared with the input values of $\beta$.  Differences between
the mean $UV$-continuum slope $\beta$ measured for the selected
galaxies and that input into the simulations is the bias.  Throughout
we will refer to this bias as the ``photometric error coupling bias.''

In Figure~\ref{fig:bias0c}, we show the mean $UV$-continuum slopes
$\beta$ we recover for our $z\sim4$-7 selections in the HUDF09 field
assuming an input $UV$-continuum slope $\beta$ of $-2$.  Strikingly,
the output distribution of $UV$-continuum slopes $\beta$ we find is
almost exactly the same as our input distribution, so it is
immediately clear that biases resulting from noise are small.
Nonetheless, we do note a small bias in the recovered $UV$-continuum
slope distribution towards \textit{redder} slopes.

The explanation for both of these effects is simple.  The bias is
small because we use a completely different region of the $UV$ SED
(including the two bluest rest-frame $UV$ bands) for selecting
galaxies than we use to measure the $UV$-continuum slope (the second
bluest $UV$ band and redder).  The effect of photometric scatter on
the selection of sources is therefore largely independent of the
effects of this scatter on $UV$-continuum slope $\beta$ estimates.

The bias towards redder $UV$-continuum slopes occurs as a result of
the small overlap between the bands used for selection and those used
for measurement of the $UV$-continuum slopes.  Since this overlapping
band is the \textit{long} wavelength anchor for the selection, any
scatter towards fainter values makes the source bluer in the
$UV$-continuum and therefore easier to select.  However, since this
overlapping band is also the \textit{short} wavelength anchor for
$UV$-continuum slope $\beta$ estimates, this same scatter towards
fainter values results in \textit{redder} values for the
$UV$-continuum slope $\beta$.

\subsection{B.2.  Corrections to the observed $UV$-continuum slope 
distribution}

In Appendix B.1, we estimated the approximate bias in the mean
$UV$-continuum slope $\beta$ we would determine from the observations.
We began by estimating the bias that would result from the fact that
galaxies with bluer $UV$-continuum slopes $\beta$ are easier to select
than galaxies with redder $UV$-continuum slopes (Appendix B.1.1).  We
then estimated the bias we would expect in the measured values for
$\beta$ as a result of the slight coupling that occurs between source
selection and the measurement process (Appendix B.1.2).  Of course,
both biases affect the mean $UV$-continuum slope $\beta$, so we need
to add the above biases together to obtain the total bias in the
$UV$-continuum slope $\beta$.  The results are presented in
Figure~\ref{fig:bias0c} (\textit{solid lines}).

Remarkably enough, the approximate bias in the mean $UV$-continuum
slope $\beta$ is almost zero.  The bias towards bluer slopes (from the
intrinsic selection biases: B.1.1) largely offsets the bias towards
redder slopes (from the slight coupling between source selection and
$\beta$ measurements: B.1.2), resulting in a very small bias overall.
Given the very small size of the biases expected here for our faint
samples, it is certainly striking how large these same biases are for
the ``robust'' photometric redshift selection of Dunlop et
al.\ (2012).  Figures~\ref{fig:biasdunlop} and \ref{fig:bias0c}
provide a rather dramatic illustration of the differences.  The
contrast in biases is especially noteworthy, especially given the
large number of candidates in our faint samples.

While we can calculate the approximate effect of the aforementioned
biases on $\beta$ given some input distribution, for the simulations
to be accurate we must use the true underlying $\beta$ distribution
for these inputs.  Since we are not able to establish the underlying
$\beta$ distribution without first establishing the biases (so we can
apply a correction to the observed distribution of $\beta$'s), it was
necessary to follow an iterative approach.  The observed $\beta$
served as a starting point for the computation of the biases.  In each
iteration, we used the previous best estimate of the $UV$-continuum
slope $\beta$ distribution to establish the relevant corrections until
we obtained convergence.

Biases in the mean $UV$-continuum slope were estimated in a very
similar way for each of our samples ($z\sim4$, $z\sim5$, $z\sim6$, and
$z\sim7$).

\subsection{B.3. Biases in the $1\sigma$ Scatter Measured for the $\beta$ Distribution}

Flux measurement errors act to significantly broaden the
$UV$-continuum slope $\beta$ distribution.  The added scatter can be
substantial, particularly near the selection limits for our samples.
Typical uncertainties on individual $UV$-continuum slope $\beta$
measurements range from 0.4 to 1.0, with the largest uncertainties
being relevant for $z\sim7$ galaxies where only two passbands
($J_{125},H_{160}$) are available to make the measurement and the
wavelength baseline is relatively short.  An illustration of the
extent to which these errors can increase the apparent scatter in the
$\beta$ distribution can be found in Bouwens et al.\ (2009: Figure 4).

To quantify the extent to which flux measurement error (``noise'')
increases the spread in the $UV$-continuum slope $\beta$ distribution,
we again rely on the simulations described in Appendix B.1 (see also
Appendix C).  For these simulations, we assume that all galaxies have
exactly the same $UV$-continuum slope $\beta$, with no scatter, and
then measure the scatter in the recovered distribution.  The derived
$1\sigma$ scatter is shown in Figure~\ref{fig:bias1} for $z\sim4$-7
galaxies identified in the HUDF observations.  Scatter is largest for
faint sources in each of our $z\sim4$-7 samples and also in our
$z\sim6$-7 samples.  Scatter in our lower redshift samples is smaller
because of the much larger number of passbands and longer wavelength
baselines over which we use to measure the $UV$-continuum slopes
$\beta$ (Table~\ref{tab:pivotbands} and \S3.3).

Another benefit of having run these simulations is that they allow us
to quantify the extent to which we can reduce the overall
uncertainties in $\beta$ using the full flux information in the $UV$
continuum, rather than using just a single color (\S3.3 and Appendix
A).  These gains are most dramatic for $z\sim4$ and $z\sim5$ galaxies
in our fields where we use the information in four passbands
(Table~\ref{tab:pivotbands}) rather than just two (Appendix A).  In
the $z\sim4$ or $z\sim5$ cases, for example, we find a factor of
$\sim$1.5 reduction in the scatter.

\subsection{B.4.  Comparisons between the HUDF09 and ERS+CANDELS 
Determinations}

In the previous sections, we outlined the procedure we use to
determine the distribution of $UV$-continuum slopes $\beta$ for
high-redshift galaxy samples, after correcting for various selection
and measurement biases.

An important check on these results is to compare the distributions we
derive at different depths.  How well does the distribution of
$UV$-continuum slopes $\beta$ obtained from the ultra-deep HUDF09 data
set agree with that obtained from the wide-area ERS+CANDELS
observations?  The results are shown in Figure~\ref{fig:colmagq} for
all four of our Lyman-Break selections.  While the agreement is
excellent overall, this should perhaps not come as a great surprise --
given the very small corrections that need to be made to each of our
selections (Figure~\ref{fig:bias0c}).

\section{C.  Simulation Procedure Used to Assess Biases}

The $UV$-continuum slope $\beta$ distributions we find are sensitive
to a wide variety of object selection and measurement issues.  To
determine the approximate size of these effects, we ran an extensive
set of Monte-Carlo simulations where we inserted artificial galaxies
into the real data, reselected them and measured their properties, all
in the same way as done for our real samples (\S3).  Galaxies are
randomly included in the simulated images over the full redshift
window where we might conceivably select them, i.e., $z\sim2.9$-5.0
for our $z\sim4$ samples, $z\sim4.0$-6.0 for our $z\sim5$ samples,
$z\sim5.0$-7.0 for our $z\sim6$ samples, and $z\sim6.0$-8.0 for our
$z\sim7$ samples.  For these simulations, we start with real
pixel-by-pixel images of similar luminosity galaxies from the $z\sim4$
HUDF sample of Bouwens et al.\ (2007).  The $z\sim4$ HUDF sample of
Bouwens et al.\ (2007) implicitly defines the size and morphology
distribution used in our simulations for $z\sim4$ galaxies, both in
terms of the absolute sizes, the size-luminosity relation, and also
the width of the size distribution.  Sizes of the sources are scaled
as $(1+z)^{-1}$ to match the observed size-redshift trends (e.g.,
Bouwens et al.\ 2004; Ferguson et al.\ 2004; Buitrago et al.\ 2008).
Such a size scaling results in a close reproduction of the actual size
distribution of galaxies in the $z\sim7$-8 observations (Oesch et
al.\ 2010b; Appendix A of Bouwens et al.\ 2011b).  We have utilized
such techniques extensively in the last decade to establish the
reliability of our selections, to quantify biases in measured
magnitudes and colors, and to establish the needed corrections (e.g.,
Bouwens et al.\ 2003, 2004, 2006, 2007, 2008, 2009, 2011b).

\begin{figure*}
\epsscale{0.6}
\plotone{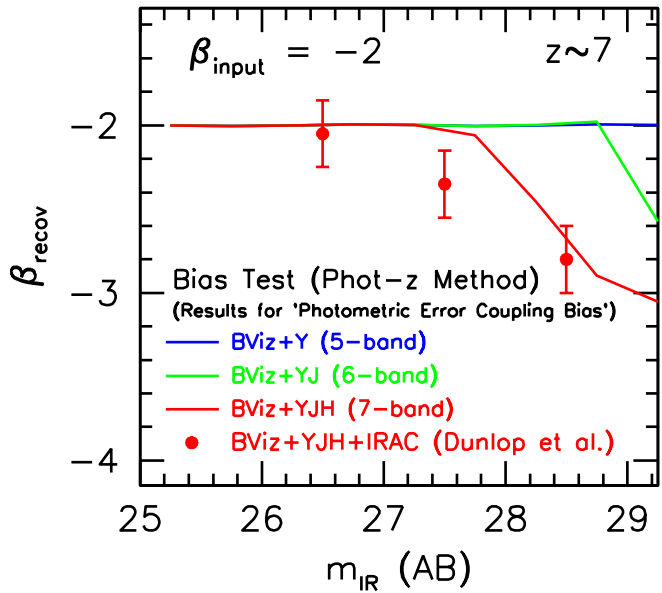}
\caption{Mean $UV$-continuum slope $\beta$ we recover versus $IR$
  magnitude at $z\sim7$ using a photometric redshift selection
  procedure and input $UV$-continuum slopes $\beta$ of $-2$ (see
  Appendix D).  The procedure we use to select galaxies and measure
  their $UV$-continuum slopes $\beta$ here is designed to be similar
  to that used by Dunlop et al.\ (2012).  Shown are the results
  selecting sources using the 5-band $BVizY$ photometry
  (\textit{blue}), 6-band $BVizYJ$ photometry (\textit{green}), and
  the full 7-band $BVizYJH$ photometry (\textit{red}).  Our results
  for the full 7-band photometry seem very similar to that obtained by
  Dunlop et al.\ (2012: \textit{solid red circles}).  The mean
  $UV$-continuum slope $\beta$ we recover using five or six-band
  photometry is much less biased than what we recover using the full
  seven-band photometry.  The reason a 7-band selection results in
  biased measures of the $UV$-continuum slope $\beta$ is that similar
  information (i.e., the $J$ and $H$-band fluxes) is used (1) to measure their
  $UV$-continuum slopes $\beta$ and (2) to select the sources.
  Because of this coupling, the 7-band photometric redshift selection
  excludes as unreliable precisely those sources which are reddest in
  their observed $UV$ slopes -- biasing the slope results.  This is
  the same as the ``photometric error coupling bias'' discussed in
  Appendix B.1.2, and it becomes very large at low S/N.  Significantly
  less biased measures of the mean $UV$-continuum slope $\beta$ can be
  obtained using five or six-band photometric redshift selections (due
  to the independence of the information being used for source
  selection and the measurement of $\beta$).  These simulations
  effectively show that $UV$-continuum slopes $\beta$ can be recovered
  accurately to very faint flux levels, given the appropriate
  technique.\label{fig:dunlop}}
\end{figure*}

\section{D.  Biases in measuring $\beta$ using a photometric redshift procedure?}

While we offer one approach to establishing the $UV$-continuum slope
$\beta$ distribution, a complementary approach was put forward by
Dunlop et al.\ (2012).  In this approach, high-redshift sources are
selected using a photometric redshift procedure and their
$UV$-continuum slopes $\beta$ are estimated directly from their
observed colors.  Using this methodology, Dunlop et al.\ (2012) found
no dependence of $\beta$ on luminosity or redshift -- which is in
contrast to the present results and most results in the literature
(\S4.3-\S4.4).

One of the most interesting claims Dunlop et al.\ (2012) made using
their approach was that measurements of the $UV$-continuum slope
$\beta$ at faint magnitudes were likely unreliable, and hence it was
not possible to determine the mean $UV$-continuum slope $\beta$ to
very faint magnitudes.  Dunlop et al.\ (2012) demonstrated these
significant biases by inserting high-redshift sources into their
catalogs with specific $UV$-continuum slopes $\beta$, adding noise to
the photometry, and then deriving the mean $UV$-continuum slope
$\beta$ for the selected sources.  Dunlop et al.\ (2012) found that
galaxies at fainter magnitude levels, particularly $>$28 mag, were
biased to much bluer $UV$-continuum slopes $\beta$ than the sources
they put into their simulations.  These biases resulted from the
effect of noise in perturbing the colors of sources in their catalogs
and these same colors being used for both source selection and $\beta$
measurements.  This is the same photometric error coupling bias we
discussed in Appendix B.1.2.

Are such biases really as unavoidable as Dunlop et al.\ (2012)
suggest?  To investigate this, we wrote software to approximately
replicate the procedure by Dunlop et al.\ (2012).  We then ran
Monte-Carlo simulations involving more than $6\times10^4$ sources.
Each source was assigned a random $H_{160}$-band magnitude between 25
AB mag and 30 AB mag, keeping the intrinsic $\beta$ fixed for all
sources in a simulation run.  Fluxes for individual sources were
computed based on the input $\beta$'s assuming power-law SEDs.  Noise
was added to the fluxes of sources in the simulations assuming
HUDF09-depth data.  Finally, sources were selected using a similar
photometric redshift procedure to that employed in Dunlop et
al.\ (2012) and McLure et al.\ (2011).  This involved comparing the
measured fluxes for each source with the fluxes from various template
SEDs to determine the minimum $\chi^2$ at each redshift and then
calculating the redshift likelihood function $P(z)$.

For SED templates, we used the four templates from Coleman et
al.\ (1980) and the two bluest starburst templates from Kinney et
al.\ (1996), linearly interpolating between adjacent templates.  While
this provides us with slightly less freedom in fitting arbitrary SEDs
than is available to Dunlop et al.\ (2012) -- who utilising the
Bruzual \& Charlot (2003) spectral synthesis code consider a wide
range of ages, metallicities, and dust extinctions -- the SED template
set we consider covers a sufficient range to illustrate the relevant
biases in $\beta$.  No prior was assumed.  Sources were selected if
(1) the primary redshift solution is preferred at $>$95\% confidence
over the secondary redshift solution (i.e., $\chi^2 \textrm{(best-fit,
  high-z)} < \chi^2 \textrm{(best-fit, low-z)} + 4$, (2) $>$50\% of
the integrated redshift likelihood $\int P(z)$ is at $z>6$, and (3)
the best-fit $\chi^2$ is statistically acceptable (typically
$\chi^2<10$ given the number of flux constraints and free parameters
in the SED modeling).  Following Dunlop et al.\ (2012), we estimate
the $UV$-continuum slope $\beta$ for sources in the simulations from
the observed $J_{125}-H_{160}$ colors, using the expression
$\beta=4.43(J_{125}-H_{160})-2.00$.

The results of our simulations are shown in Figure~\ref{fig:dunlop}
for initial $\beta$ values of $-2.0$ and $-2.5$.  Not surprisingly, we
obtain similar results with our simulations ({\em red line}) as what
Dunlop et al. (2012) obtained ({\em solid red circles}).  The
simulations indicate a strong bias in the recovered $UV$-continuum
slopes ($\beta$) at faint magnitudes towards bluer slopes.  This is
particularly true at $>28$\,mag where the recovered $\beta$ is
$\delta\beta\simeq 0.8$ bluer than the intrinsic value.  As we show
below, these biases depend critically on the passbands used to select
sources.  This bias likely originates from a close coupling of the
information used to select sources and that used to measure $\beta$
(Appendix B.1.2).  This coupling is problematic because any effect
that noise has in making the apparent UV slopes bluer would also cause
objects to be more amenable to selection via the Dunlop et al. (2012)
procedure.

Is it possible for us to recover the $UV$-continuum slope $\beta$ in a
less biased fashion?  The answer is yes, but it requires that we use
independent information to select the sources from what we use to
measure the $UV$-continuum slopes -- thereby decoupling the effect of
noise on the two processes.  We can do this by repeating the above
photometric redshift selection, but only using the flux information in
the five passbands that do not contribute to our $UV$-continuum slope
estimates, i.e., the $B_{435}$, $V_{606}$, $i_{775}$, $z_{850}$, and
$Y_{105}$ bands.  We also consider a selection using the five
passbands that do not contribute plus the bluest remaining band, i.e.,
the $B_{435}$, $V_{606}$, $i_{775}$, $z_{850}$, $Y_{105}$, and
$J_{125}$ bands.  The results are shown in Figure~\ref{fig:dunlop},
and it is clear that the mean $UV$-continuum slope $\beta$ we measure
is less biased, especially in the five passband case.  Indeed, the
mean $UV$-continuum slope $\beta$ can be successfully recovered to
$\sim$29 AB mag, even in the six passband case.  \textit{This provides
  a clear demonstration that it is possible to measure the mean
  $UV$-continuum slope $\beta$ to very faint flux levels, given the
  appropriate technique.}

We minimize the biases by excluding those passbands used for the
$\beta$ measurements from the selection process (i.e. excluding the
reddest one or two passbands from the photometric redshift
determinations).  Are there any disadvantages to this approach?  While
one might argue that this results in some loss of ability to control
for contamination, such losses are small -- as one can see from the
following considerations: (1) The only secure characteristic we know
about high redshift star-forming galaxies is that they disappear at
wavelengths bluer than $\sim$1216\AA$\,\,$and are detected blueward of
that in the rest-frame $UV$. Use of five or six-band photometry to
compute photometric redshifts allows us to verify that sources have
this characteristic. (2) Selection of high-redshift sources can be
improved by excluding those sources with the reddest colors redward of
the break.  Only one color is necessary for this task.  For a $z\sim7$
selection, this means a 6-band $BVizYJ$ photometric redshift selection
(the same bands are used for our $z\sim7$ Lyman-Break
selection). Adding a 7th band to the photometric redshift selection
might appear to help, but actually does not, as this results in
significantly biased $UV$ slope $\beta$ measurements (see
Figure~\ref{fig:dunlop}).  (4) Even if a small number of contaminants
make it into one's samples one can mitigate the effect they have on
the mean $UV$-continuum slope $\beta$ through the use of robust
statistics like the biweight mean (see \S3.5).  Thus there appears to
be few disadvantages to using an approach that minimizes the biases.
The gains in measurement reliability greatly offset any potential
concerns.

\section{E.  $UV$-continuum Slope $\beta$ Measurements for Individual Sources}

In the interests of transparency and to facilitate comparisons with
other high-redshift studies, we include a complete list of the
$UV$-continuum slopes $\beta$ that we derive for all 2518 individual
sources in our $z\sim4$, $z\sim5$, $z\sim6$, and $z\sim7$ samples in
Table~\ref{tab:betalist}.

We note that a small fraction ($\lesssim$1\%) of the sources in our
samples have very red UV-continuum slopes $\beta$, i.e., $> 0.5$, and
therefore may not be at high redshifts (since this would imply that
the apparent break at $\sim 1$ micron in their SEDs may not be due to
absorption by neutral hydrogen, but due to their overall spectral
shape).  We should emphasize, however, that these sources have
essentially no effect, i.e., $d\beta\lesssim 0.01$, on the
biweight-mean $\beta$'s we report.

\begin{deluxetable*}{cccccc}
\tablecaption{A complete list of the UV-continuum slopes $\beta$ we 
       measure for sources in our $z\sim4$, $z\sim5$, $z\sim6$, and $z\sim7$ 
       samples from the HUDF09+ERS+CANDELS data sets\tablenotemark{a}\label{tab:betalist}}
\tablehead{\colhead{R.A.} & \colhead{Declination} & \colhead{$M_{UV,AB}$} & \colhead{$\beta$} & \colhead{$<z>$} & \colhead{Data Set\tablenotemark{b}}}
\startdata
03:32:38.49 & -27:48:21.41 & $-$17.82 & $-$1.92$\pm$0.24 & 4 & 1  \\
03:32:38.42 & -27:48:18.68 & $-$16.46 & $-$1.61$\pm$0.57 & 4 & 1  \\
03:32:37.67 & -27:48:16.87 & $-$17.49 & $-$2.19$\pm$0.37 & 4 & 1  \\
03:32:37.31 & -27:48:13.74 & $-$18.41 & $-$2.42$\pm$0.19 & 4 & 1  \\
03:32:38.14 & -27:48:12.75 & $-$17.82 & $-$1.35$\pm$0.34 & 4 & 1
\enddata
\tablenotetext{a}{Table~\ref{tab:betalist} is published in its entirety in the electronic edition of the Astrophysical Journal.  A portion is shown here for guidance regarding its form and content.}
\tablenotetext{b}{The data set from which the source was selected and in
          which its UV-continuum slope beta derived (1 = HUDF09,
	  2 = HUDF09-1, 3 = HUDF09-2, 4 = CANDELS/ERS)}
\end{deluxetable*}

\end{document}